\documentclass[a4paper,fleqn,usenatbib]{mnras}

\usepackage[T1]{fontenc}
\usepackage{ae,aecompl}

\usepackage{graphicx}	
\usepackage{amsmath}	
\usepackage{amssymb}	
\usepackage{multirow}
\usepackage{threeparttable}
\usepackage{color}

\title[DEEM and FRD mesurement]{DEEM, a versatile platform of FRD measurement for highly multiplexed fibre systems in astronomy}

\author[Y. Yan et al.]{
Yunxiang Yan$^{1,2}$\thanks{E-mail: yxyan@nao.cas.cn (Y. Yan)},
Qi Yan$^{3}$,
Gang Wang$^{1}$,
Weimin Sun$^{3}$,
A-Li Luo$^{1}$,
Zhenyu Ma$^{3}$,
\newauthor
Qiong Zhang$^{3}$,
Jian Li$^{1}$
and Shuqing Wang$^{1}$
\\
$^{1}$Key Laboratory of Optical Astronomy, National Astronomical Observatories, Chinese Academy of Sciences, Beijing 100012, China \\
$^{2}$University of Chinese Academy of Sciences, Beijing 100049, China \\
$^{3}$Key Lab of In-fiber Integrated Optics, Ministry Education of China, Harbin Engineering University, Harbin 150001, China \\
}

\date{Accepted XXX. Received YYY; in original form ZZZ}

\pubyear{2017}

\begin{document}
\label{firstpage}
\pagerange{\pageref{firstpage}--\pageref{lastpage}}
\maketitle

\begin{abstract}
We present a new method of DEEM, the direct energy encircling method, for characterising the performance of fibres in most astronomical spectroscopic applications. It's a versatile platform to measure focal ratio degradation (FRD), throughput, and point spread function (PSF). The principle of DEEM and the relation between the encircled energy (EE) and the spot size were derived and simulated based on the power distribution model (PDM). We analysed the errors of DEEM and pointed out the major error source for better understanding and optimisation. The validation of DEEM has been confirmed by comparing the results with conventional method which shows that DEEM has good robustness with high accuracy in both stable and complex experiment environments. Applications on the integral field unit (IFU) show that the FRD of 50$\mu$m core fibre is substandard for the requirement which requires the output focal ratio to be slower than 4.5. The homogeneity of throughput is acceptable and higher than 85 per cent. The prototype IFU of the first generation helps to find out the imperfections to optimise the new design of the next generation based on the staggered structure with 35$\mu$m core fibres of $N.A.$=0.12, which can improve the FRD performance. The FRD dependence on wavelength and core size is revealed that higher output focal ratio occurs at shorter wavelengths for large core fibres, which is in agreement with the prediction of PDM. But the dependence of the observed data is weaker than the prediction.
\end{abstract}




\section{Introduction}
Optical fibres are now widely used in astronomy. Introducing fibres into the large scale telescope makes it more efficient for multi-object survey with large field of view (FOV) \citep{Angel1977A}, such as SDSS \citep{york2000sloan}, AAT \citep{Lewis2002The}, LAMOST \citep{cui2012large} , DESI \citep{Flaugher2014The} and Subaru \citep{Sugai2015Prime}. But the non-conservation of the optical etendue that enlarges the solid angle of the output spot in the output end, which is known as focal ratio degradation (FRD), and the transmission efficiency limit the energy utilisation and decrease the signal-to-noise ratio (SNR) of spectra especially in high-resolution spectrographs and highly multiplexed fibre systems \citep{Grupp2003The, Feger2012A, Wang2013LAMOST}. Point spread function (PSF) and the reconstruction of the far-field also affect the precision of the spectra \citep{Baudrand2001Modal,Lemke2011Modal,Rawson1983Modal}. FRD cannot be completely eliminated in practical applications but it is important and promising to minimize the influence of FRD to improve the efficiency \citep{Xue2013LAMOST}. Various models have been proposed to describe the FRD performance with many influential factors like wavelength, bending, fibre length, core diameter, etc \citep{Ramsey1988Focal, Clayton1989The, Avila1998Results, Schmoll1998FRD, Crause2008Investigation, Bryant2010Hexabundles, Bryant2011Characterization, Poppett2010The}. Especially the FRD dependence on wavelength is different in the results of many groups. Some researchers \citep{Carrasco1994A,Poppett2007Fibre} found out a decreasing trend with increasing wavelength as predicted by PDM \citep{Gloge1972Derivation}. While some groups \citep{Crause2008Investigation,Pazder2014The,Matthew2004SparsePak} showed no significant dependence on wavelength. On the other hand, a weak opposite trend was reported by \citet{Murphy2008Focal} which was verified experimentally. Throughput is not only affected by the materials and the fabricating process but also relates to the application environment and FRD. However, the experimental results are not always consistent with predictions of the theoretic models because of the limitation of the experimental accuracy and one must repeat many different experiments to calibrate the parameters to satisfy the design requirements for different instruments.

There are two common methods for determining FRD according to the geometric type of the light source. One method is using a collimated beam or a very thin annual beam at a specific angle to represent a particular input focal ratio \citep{Ferwana2004All, Haynes2004New, Haynes2008AAOmicron, Haynes2011Relative}. The FRD is determined by the radial full width at half-maximum (FWHM) of the output annual spot \citep{Mattias2009Air}. The other is using a filled cone beam or a cone beam with a centre obscuration to simulate the behaviours of fibres in a telescope \citep{Lee2002Properties, Oliveira2004Studying, Oliveira2011FRD, Santos2014Studying, Pazder2014The}. The FRD will measure the encircled energy (EE) within a certain focal ratio.

Both of the conventional methods with CCD need the imaging process to record the output spots \citep{Murphy2008Focal, Murphy2013The} and they have some disadvantages in the real-time measurement. For example, the collimated beam method can hardly measure the energy utilisation and the cone beam method usually costs too much time in data recording and processing for each test and it is too sensitive to the environment light \citep{Finstad2016Collimated}. In the cone beam technique, a very stable light source is needed, and the recorded images with different exposure time or under different illumination condition will bias the diameters of output spots if the subtraction of the background is not completely done, which is tested and discussed in this paper (see Section 3). As the CCD camera is moving in different positions, the output power distribution varies in different images and brings uncertainties into the determination of the diameters.

To conquer the problems, a modified platform of direct energy encircling method (DEEM) is proposed based on the cone beam technique. This method is convenient for measuring FRD, throughput, EE ratio and PSF at the same time, especially for highly multiplexed fibre system like the integral field unit (IFU). The stability of DEEM is improved by the design of the two-arm detectors which can record the reflective and the refractive light at the same time to conduct a closed loop system. DEEM skips the imaging process to acquire the EE ratio of the output spot directly to enable rapid measurement. To confirm the validation of DEEM, comparison between DEEM and the conventional method was made to ensure the feasibility and the precision. Two methods have different optical construction in the output end, so the error analysis including different error sources has been discussed separately. An important process is the subtraction of the background which is one of the major errors that brings offset in FRD measurements. We proposed three types of noises to correct the background subtraction, with which the results showed a good improvement in the precision in the conventional method. Finally the FRD dependence on wavelength and core size was tested at the wavelength range from 400nm to 900nm and we found out that the measured results agreed with the predicted trend of FRD dependence by PDM. The results indicated that the dependence was not only dominated by the wavelength of input light but also affected by the coherence of the light source.

In recent years, the degree of multiplexed instruments in astronomy is increasing fast, more and more surveys need the support of IFU \citep{Hill2008Design}, such as SAMI \citep{Bryant2015The} with 13 fibre-based IFUs and HETDEX with 33000 fibres \citep{Kelz2014VIRUS}. In Section 4, we also presented the applications of DEEM on the first generation of the prototype IFU for the Fiber Arrayed Solar Optical telescope (FASOT) \citep{Qu2011A} to rapidly measure the performance of FRD, throughput, position arrangement and so forth. A time-saving and accurate method is required for the quality assurance of fibres.

\section{DEEM, direct energy encircling method}
Both of the integral field spectroscopy (IFS) and the multi-object survey require the highly multiplexed fibre system. The IFUs for FASOT contain 8192 fibres in the two segments including 4096 fibres for each. The wavelength ranges from 400nm to 900nm and the first priority waveband is 515nm$\sim$526nm with resolution power of 110,000. The final input focal ratio is required to be slower than 4.5 and the transmission efficiency should be larger than 75 per cent for the whole range of the wavelength and better than 80 per cent for 500nm$\sim$660nm. In the initial design the fibres were multi-mode fibres of 125$\mu$m core size, but fibres with small core size from 70$\mu$m down to 50$\mu$m were considered to compact the structure and increase the filling ratio of IFU in the later requirements. And the first generation of prototype IFU is assembled in Harbin Engineering University with fibres of 50$\mu$m core size. Currently, LAMOST combines 4000 fibres of 320$\mu$m core size in the focal plane within the diameter of 2m. To improve the light throughput and the observing efficiency, the upgrade is planned to reduce the size of focal plane down to 1.0m and enlarge the population of fibres up to 5000 with smaller core size of 200$\mu$m or even smaller depending on the environment of the new site (A candidate Ali site in Tibet on the altitude of 5100m), including the light pollution, seeing, etc. Measuring the FRD performance and the transmission property would be a giant project in these telescopes. A general platform of DEEM is proposed to meet the demands with the ability to measure many types of fibres of different core sizes, numerical apertures ($N.A.$) and input focal ratios (larger than 2.0).

The conventional cone beam method for testing FRD records a series of images with a CCD camera by moving it away from the fibre end to different positions. The output focal ratio is determined by the focal distances and the diameters of the output spots within a certain EE ratio. It is intuitive in visual aspects, but the operation costs too much time and the precision of the results is very sensitive to the experiment environment and the alignment precision. The input condition including alignment and incident position on the fibre end should be carefully performed to ensure the results are valid \citep{Yan2017A}. A stable light source is important because the output spots captured by CCD with different intensity would bias the diameter if the subtraction of the noise is not completely done.

The platform of DEEM begins with the energy usage of the output spot. The closed loop provides the feedback of the changes of the light source to improve the robustness of the system. DEEM tries to unify the relation between the spot size and the energy utilisation which is usually presented by EE based on PDM. It mainly concentrates on the power distribution and the PSF of the output light from the fibre. These two factors determine the FRD and the energy usage which contribute to the observation efficiency of a telescope as well as the SNR of the spectra quality. DEEM can also take the advantage of the two-arm measurement design to reconstruct the 3D position distribution for highly multiplexed fibre devices.

Both of the two methods aim to encircle the integral energy within a certain EE ratio. So precisely determining the spot size and the EE ratio is of great importance to measure the FRD performance and the throughput. In this section we will analyse the relation between the EE ratio and the diameter of output spots with the model of PDM to construct the fundamental principle of DEEM.

\subsection{Power distribution model (PDM) and encircled energy (EE)}
Geometric optics and wave optics are commonly used in waveguide analysis. And the mode theory in wave optics is more appropriate for describing mode transmission or energy transmission. PDM is one of the wave theories to characterise the energy transmission from the input end to the output end through an optical fibre. The mode field or the power distribution directly affects the PSF and the energy usage. Since the important factor of the diameter of the output spot is determined by power distribution within a certain EE, FRD can also be described by PDM. The model of PDM was firstly proposed by \citet{Gloge1972Derivation} and later adapted by \citet{Gambling1975Mode}. It uses the far-field distribution to represents a direct image of the distribution of different guided modes with different power. The analytic expression of equation (\ref{eq1}) describes the distribution of power $P$ in a fibre of length $L$ under the input condition of a beam with the axial angle of incidence $\theta _{in}$ and the output beam with output angle $\theta _{out}$.
\begin{equation}\label{eq1}
\begin{split}
& \frac{{\partial P\left( {{\theta _{out}},{\theta _{in}}} \right)}}{{\partial L}} = - A\theta _{in}^2P\left( {{\theta _{out}},{\theta _{in}}} \right) \\
& \qquad \qquad \qquad + \frac{D}{{{\theta _{in}}}}\frac{\partial }{{\partial {\theta _{in}}}}\left( {\theta _{in} \frac{{\partial P\left( {{\theta _{out}},{\theta _{in}}} \right)}}{{\partial {\theta _{in}}}}} \right)
\end{split}
\end{equation}
where $A$ is the absorption coefficient, and $\theta _{in}$ and $\theta _{out}$ represent the input and output angles, respectively. And $D$ is a parameter that depends on the constant ${d_0}$ that characterises micro bending:
\begin{equation}\label{eq2}
D = {\left( {\frac{\lambda }{{2a{n_1}}}} \right)^2}{d_0}
\end{equation}
where $\lambda $ is the wavelength of light, $a$ is the core diameter and ${n_1}$ is the refraction index of the core. The modified measurement of the constant $d_0$ was proposed by \citet{Yan2017A}.

In 1975, \citeauthor{Gambling1975Mode} solved the equation with the constraint condition of a collimated input beam at an angle of incidence ${\theta _{in}}$ :
\begin{equation}\label{eq3}
\begin{split}
& P\left( {{\theta _{out}},{\theta _{in}}} \right) =  \left[ {\frac{{\exp \left( {{{ - bL} \mathord{\left/ {\vphantom {{ - bL} 2}} \right. \kern-\nulldelimiterspace} 2}} \right)}}{{1 - \exp \left( { - bL} \right)}}} \right] \\
& \qquad \times \exp \left\{ { - \left( {\frac{{{\chi _{out}} + {\chi _{in}}}}{2}} \right)\left[ {\frac{{1 + \exp \left( { - bL} \right)}}{{1 - \exp \left( { - bL} \right)}}} \right]} \right\}\\
& \qquad \times {I_0}\left[ {\frac{{{{\left( {4{\chi _{out}}{\chi _{in}}} \right)}^{{1 \mathord{\left/
 {\vphantom {1 2}} \right.
 \kern-\nulldelimiterspace} 2}}}\exp \left( {{{ - bL} \mathord{\left/
 {\vphantom {{ - bL} 2}} \right.
 \kern-\nulldelimiterspace} 2}} \right)}}{{1 - \exp \left( { - bL} \right)}}} \right]
\end{split}
\end{equation}
where $\chi  = {\left( {{A \mathord{\left/
 {\vphantom {A D}} \right.
 \kern-\nulldelimiterspace} D}} \right)^{{1 \mathord{\left/
 {\vphantom {1 2}} \right.
 \kern-\nulldelimiterspace} 2}}}{\theta ^2}$, $b = 4{\left( {AD} \right)^{{1 \mathord{\left/
 {\vphantom {1 2}} \right.
 \kern-\nulldelimiterspace} 2}}}$, and ${I_0}$ is the modified Bessel function of the zeroth order.

Equation (\ref{eq3}) describes the output intensity for the case of a collimated input beam, and for other situations, assuming that $G\left( {{\theta _{in}},\varphi ,{\theta _m}} \right)$ represents angular distribution of the input light, where ${\theta _{in}}$ is the incident angle with respect to the optical axis, $\varphi $ is the azimuthal angle and ${\theta _m}$ is the solid angle of the incident light cone with respect to the incident angle, the output profile can be derived from:
\begin{equation}\label{eq4}
\begin{split}
& F\left( {{\theta _{out}},{\theta _{in}}} \right) = \\
& \qquad \int_0^{2\pi } {\int_0^\pi  {G\left( {{\theta _{in}},\varphi ,{\theta _m}} \right)P\left( {{\theta _{out}},{\theta _{in}}} \right)} } \sin {\theta _{in}}d{\theta _{in}}d\varphi
\end{split}
\end{equation}

Equation (\ref{eq4}) is a more general expression. In the practical applications, there are some particular cases when if the input light is symmetrical with respect to the fibre axis, the function $G$ will be simplified to be only a function of $\theta _m$ and there is no dependence on $\varphi $. Furthermore, according to the distribution of the profile of the output beam, we can obtain the total light energy within the area of a cone with half angle of $\theta _0$ by integrating the function $F$ as follows:
\begin{equation}\label{eq5}
T\left( {{\theta _0}} \right) = \int_0^{{\theta _0}} {F\left( \theta  \right) \cdot 2\pi \theta d\theta }
\end{equation}
The function $T(\theta )$ increases monotonically and describes the integral energy distribution within an aperture characterised by the angle $\theta _0$.

Generally, the order of magnitude of parameter $D$ is about $10^{-4}$ and absorption coefficient $A < 1$, so the expression of $bL \ll 1$ is valid for a short fibre unless the fibre reaches to hundreds or thousands meters or even longer. Therefore, we can make the approximation of $\exp ( { - bL} ) \approx 1 - bL$, and the equation (\ref{eq3}) can be written as
\begin{equation}\label{eq6}
P\left( {{\theta _{out}},{\theta _{in}}} \right) \approx \frac{1}{{bL}}\exp \left( { - \frac{{{\theta _{out}}^2 + {\theta _{in}}^2}}{{4DL}}} \right){I_0}\left( {\frac{{{\theta _{out}}{\theta _{in}}}}{{2DL}}} \right)
\end{equation}
The maximum of the input and the output angles $\theta _{in}$, $\theta _{out}$ are limited by $N.A.$ as follows:
\begin{equation}\label{eq7}
\sin {\theta _{\max }} = N.A.
\end{equation}
In general, $N.A.$ is smaller than 0.22, so we can consider applying small angle approximation on the output angle. According to the geometrical relationship between the output angle and the diameter of the spot, we can deduce the approximate expression,
\begin{equation}\label{eq8}
\theta  \approx  \sin \theta \approx \tan \theta  = \frac{r}{f}
\end{equation}
where $r$ is the radius of the spot and $f$ is the focal distance away from the fibre end. Substituting for ${\theta _{out}}$  from equation (\ref{eq6}), we get
\begin{equation}\label{eq9}
\begin{split}
& P\left( {{r_i},{f_i},{\theta _{in}}} \right) \approx \\
& \qquad \frac{1}{{bL}}\exp \left( { - \frac{{{r_i}^2 + {{\left( {{\theta _{in}}{f_i}} \right)}^2}}}{{4DL{f_i}^2}}} \right){I_0}\left( {\frac{{{r_i}{\theta _{in}}}}{{2DL{f_i}}}} \right)
\end{split}
\end{equation}
Equation (\ref{eq9}) indicates that the output spot is either a Gaussian centred spot or a ring spot depending on the incident angle ${\theta _{in}}$. And the Gaussian cross section and width is
\begin{equation}\label{eq10}
{\sigma _i} = {\left( {2DL} \right)^{{1 \mathord{\left/
 {\vphantom {1 2}} \right.
 \kern-\nulldelimiterspace} 2}}}{f_i}
\end{equation}
As the output profile is a Gaussian-like distribution, it is uneasy to determine the diameter without a clear boundary. For a Gaussian beam, a common practice is to relate the spot size to the FWHM according to:
\begin{equation}\label{eq11}
w\left( z \right) = \frac{{FWHM}}{{\sqrt {2\ln 2} }} = \frac{1}{2}\sigma
\end{equation}
where $z$ is in the propagation direction from the centre axis of the beam, 2$w(z)$ is the spot size at the position in a distance of $z$ away from the beam's focus where is equivalent to the fibre end, so 2$w(f)$ is the spot size on the focal plane and 2$w(0)$ on the end face of the fibre the same as mode field diameter (MFD).

\begin{figure}
   \centering
   \includegraphics[width=\hsize]{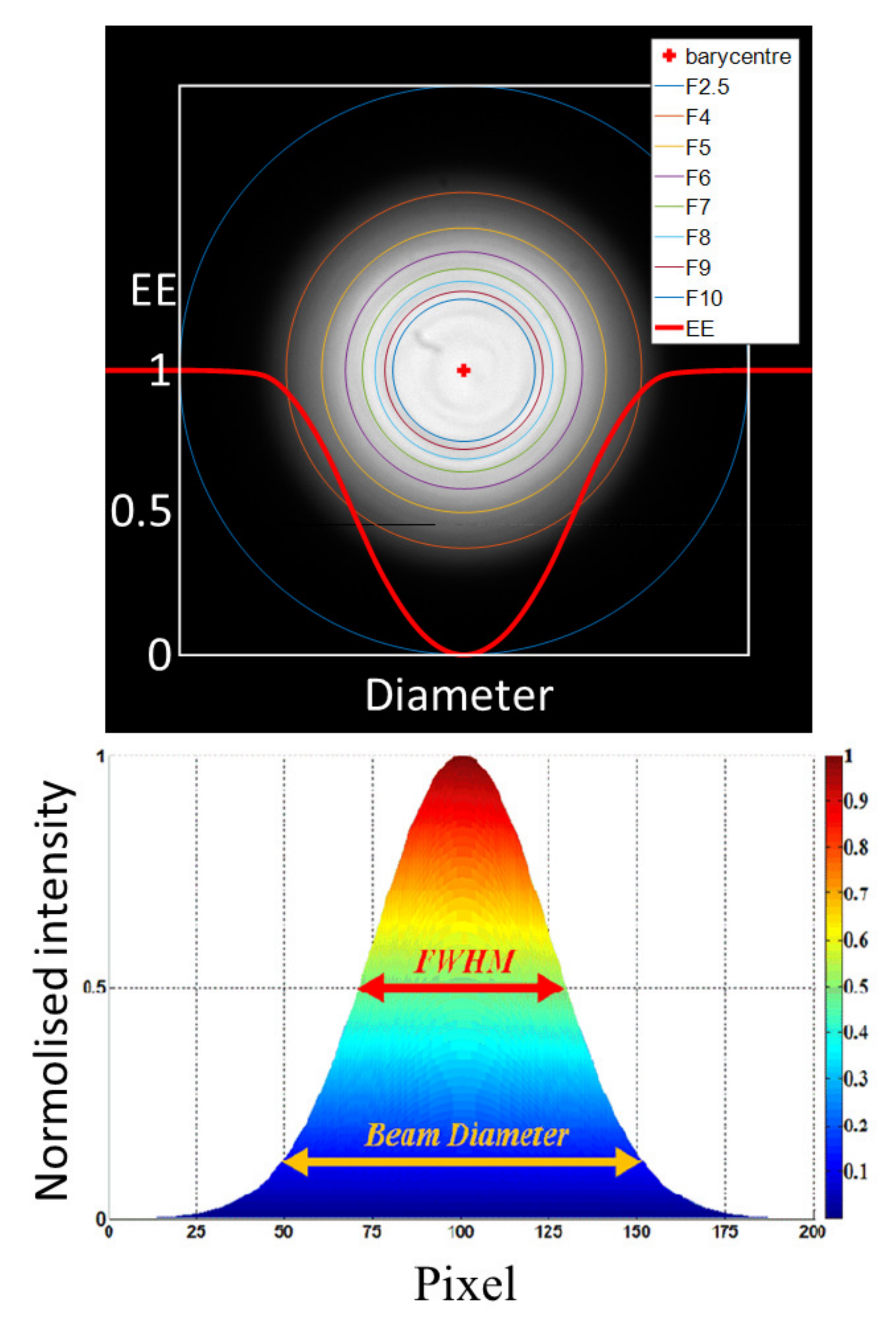}
      \caption{Mathematics used to obtain the spot size. The EE ratio is controlled by the rings in the image. To acquire the total energy, a circle determined by $N.A.$ is to collect the whole power within the spot as the blue ring for $F$=2.5 close to $N.A.$=0.2. The thick red curve is the EE ratio. For a Gaussian beam, the diameter is usually determined by FWHM when the intensity decreases down to $1/e^2$ of the maximum.}
         \label{fig:1}
   \end{figure}

The mainly dominated factor is EE ratio for the spot size of $w(f)$ or MFD. The way to determine EE is to estimate the proportion of the sum of encircled energy within a certain aperture with respect to the total energy integrated in an area limited by $N.A.$ as shown in Fig.\ref{fig:1}.

For an aperture of size 2$w(z)$, only part of the light transmits through the circle and still there is some light that cannot be collected. A standard Gaussian function has a diameter (2$w(z)$ as used in the text and about 1.7 times larger than the FWHM) which is determined by the boundary where the intensity decreases to $1/e^2$ of the maximum. For this circle, the fraction of power transmitted through the aperture is about 86.5 per cent. Similarly, about 90 per cent of the beam's power will flow through a circle of radius $r = 1.07w(z)$, 95 per cent through a circle of radius $r = 1.224w(z)$, and 99 per cent through a circle of radius $r = 1.52w(z)$. This method is usually applied in an annual spot circumstance to predict the divergence of a collimated beam during the transmission to investigate the FRD performance.

\begin{figure*}
   \centering
   \includegraphics[width=\hsize]{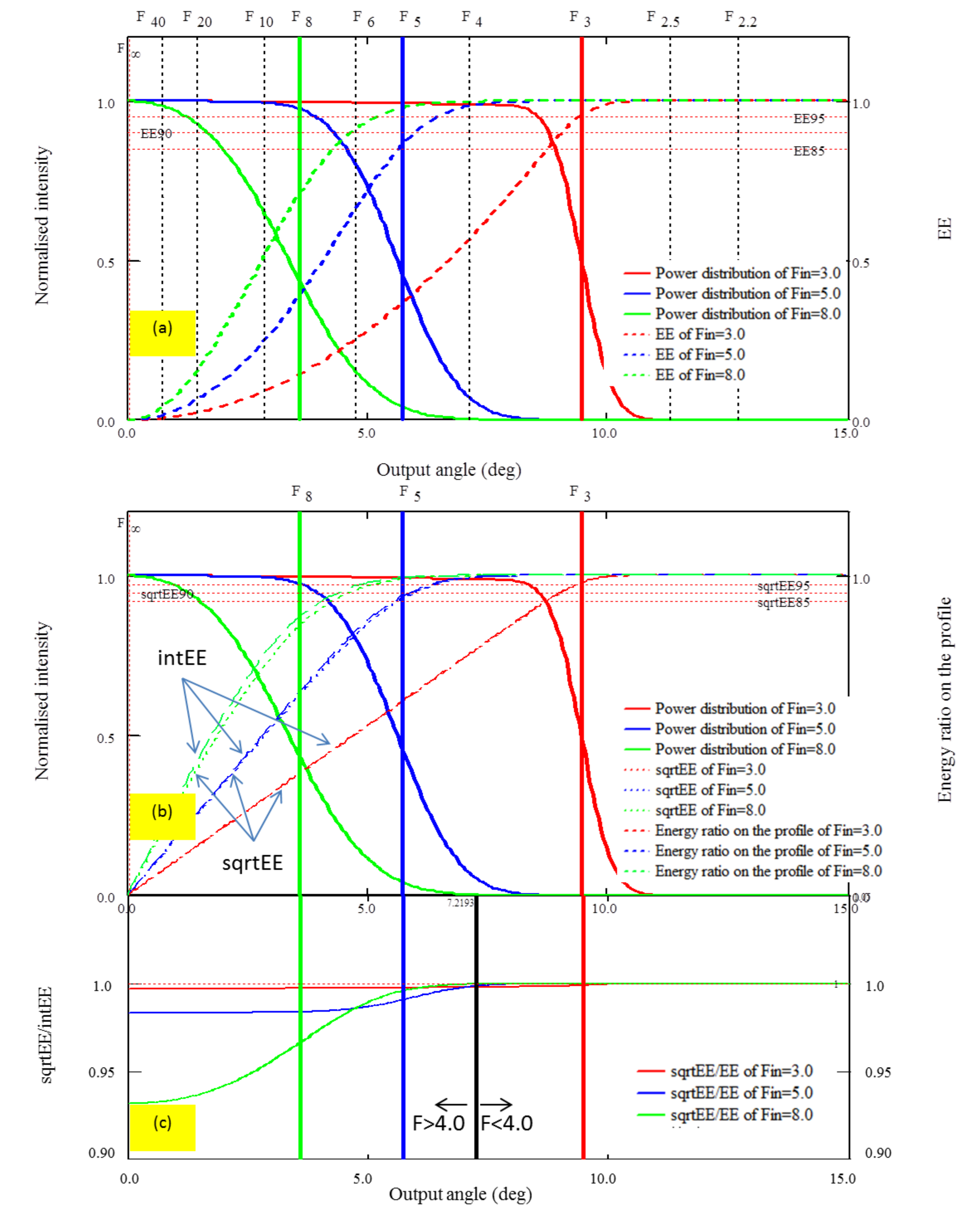}
      \caption{The profile and EE ratio of simulation results. (a): Profile of the normalised intensity cut and the EE ratio within an aperture of a specific angle. A lower EE ratio occurs within the same aperture size of the input focal ratio for a slower input beam. (b),(c): Energy ratio on the intensity cut in arbitrary direction (intEE) across the barycentre and the square root of the EE within an aperture (sqrtEE). The ratio of sqrtEE/intEE reaches to 0.999 when the input focal ratio is smaller than 4.0. While the difference is larger than 7.0 per cent for smaller input focal ratio. So the diameter from intEE is larger than that of sqrtEE in the determination of the spot size because the diffusion increases with the decreasing input focal ratio.}
         \label{fig:2}
   \end{figure*}

Considering an input cone beam of focal ratio $F_{in}$=3.0, 5.0 and 8.0, respectively, the simulation results of the output power distribution and EE ratio by PDM are shown in Fig.\ref{fig:2}. Fig.\ref{fig:2}(a) shows the profile of the normalised intensity cut across the energy barycentre and the EE ratio is radially integrated within an aperture limited by a certain angle. Fig.\ref{fig:2}(b) shows the energy ratio (intEE) on the intensity cut across the energy barycentre within a certain angle rather than the encircled energy profile. As the power distribution is symmetry in three dimensional space, the energy ratio on one dimension should be the square root of EE (sqrtEE). For example, the EE ratio of EE90 measured in radial integration of the encircled energy can be converted to 95 per cent ($ 95 per cent \approx \sqrt {90 per cent }$, sqrtEE90) of the energy in one dimension, which can reveal the difference of energy proportion calculated from the intensity cut and the radial integration of the encircled energy. The ratio of sqrtEE/intEE is derived to indicate the influence of the two kinds of methods on the energy ratio in different input focal ratios as shown in Fig.\ref{fig:2}(c). In the simulation results of Fig.\ref{fig:2}(a), 86.3 per cent of energy is encircled in the aperture of the same focal ratio $F_{out}$=5.0 and the relative difference of output angle is 5.8 per cent between EE85 and EE90 and 7.5 per cent between EE90 and EE95, but it becomes larger than 14 per cent from EE95 to EE99. From this point of view, choosing EE85$\sim$EE95 as the common interval in the determination of the diameter is better for its good robustness and relatively low bias.

According to the simulation results in Fig.\ref{fig:2}(a), if the output focal ratio is the same with the input focal ratio, the intensity on the boundary will become lower and lower with the increasing input focal ratio, which indicates that the FRD would be worse for a larger input focal ratio. Fig.\ref{fig:2}(c) shows the difference between the two kinds of ratios: intEE and sqrtEE. Compared the two kinds of energy ratios, we find that they are consistent with each other when the output focal ratio is smaller than 4.0, and in this case the value of sqrtEE/intEE is larger than 0.999. When the input focal ratio is $F_{in}$=8.0, the difference of energy ratio between the two methods can be more than 5 per cent, which means the throughput will be different within the same output focal ratio determined by the power distribution in the 3D space (sqrtEE) and by the profile of the intensity cut (intEE). While the difference in throughput is much smaller when the input focal ratio is $F_{in}$=3.0. This infers that when the input light is a flat function, the diffusion of the output spot increases with the decreasing solid angle or the increasing input focal ratio. The different output power distribution established in different input focal ratio will affect the estimation of the diameter. The better way to acquire a realistic spot size is to encircle the energy in three dimensional space rather than on the profile cut.

\subsection{Experiment setup of DEEM}
The PDM model shows that the image of the output power distribution is a Gaussian function which is a boundaryless spot. Generally, the diameter of the spot size is determined within a certain EE ratio in conventional methods. The images of the output spots are easily contaminated by the ambient light of the background. The noise is subtracted by deducting the corresponding dark image, but it is not easy to handle the precision in the determination of the diameter because the measurement is an open loop system without the feedback to revise the subtraction of the background. DEEM consists of an incident system and a two-arm measurement system. The incident system controls the input condition of the intensity and the input focal ratio. The two-arm measurement system, including a reference arm and a testing arm, supports the feedback of noise during each measurement of the diameter and the energy. And the noise of the ambient light or the background can be corrected by the feedback from the reference arm. It skips the imaging process to measure the spot size by making use of the output power distribution within a certain EE ratio directly from the diaphragm.

\begin{figure*}
   \centering
   \includegraphics[width=\hsize]{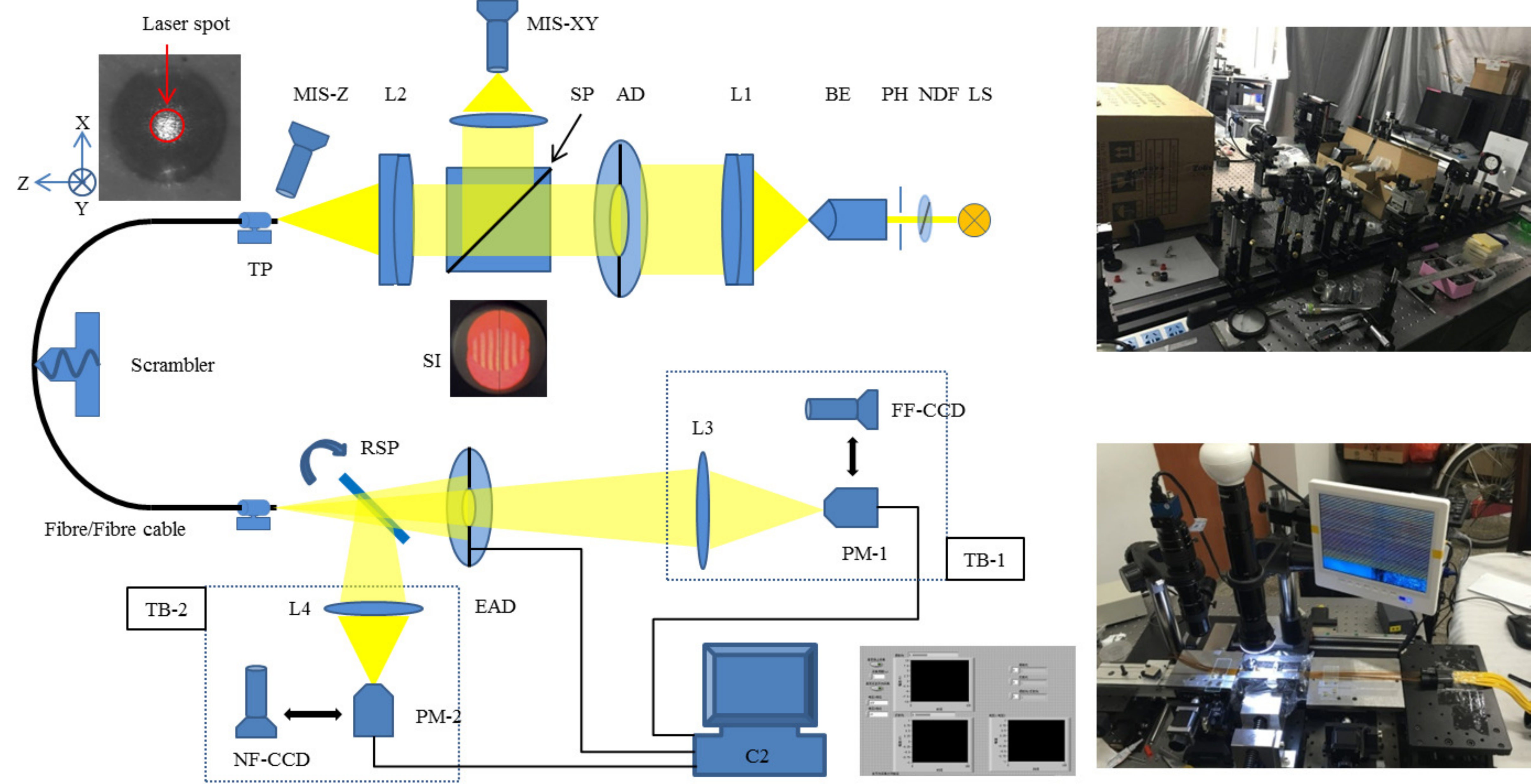}
      \caption{Schematic diagram of DEEM. LS: light source (laser or white light); NDF: neutral density filter to change the input power; PH: pinhole; BE: beam expander. L1$\sim$L4: apochromatic lens; AD: adjustable diaphragm to control the input focal ratio; SP: beam splitter; MIS: microscopic imaging system; SI: shearing interferometer; TP: transfer platform combined with six-axis translation stage (precision less than 1$\mu$m); RSP: rotatable beam splitter; EAD: electric-driven adjustable diaphragm; NF-CCD\&FF-CCD: near- and far- field CCD; PM: power meter; TB: convertible testbed; C2: control console to control and record the EE ratio, diameter of the diaphragm and the output spot. The light source can be laser or LED and the scrambler is mounted on the fibre or fibre cable to suppress speckle to smooth the output spots for laser illumination system.}
         \label{fig:3}
   \end{figure*}

The experiment setup is shown in Fig.\ref{fig:3}. The incident system contains the light source, the collimating optics and the input focal ratio controlling system. The shearing interferometer (SI) is to ensure the light is collimated from the lens L1. The three dimensional monitor system consists of the microscopes MIS-XY and MIS-Z can inspect the input position on the fibre end face. In the output end, the system can directly limit the aperture of the output spot to form a specific EE ratio with the electric-driven adjustable diaphragm (EAD) and the diameter of the spot is the same size as EAD. The output image is recorded simultaneously by near- and far- field CCD (NF-CCD and FF-CCD) if needed. The bi-detector design can eliminate the influence of the ambient light, so DEEM can reduce the sensitivity to the unstable light source. The focal length of the lens reaches to 150mm with the diameter of 75mm (lens L3 in Fig.\ref{fig:3}), which can ensure the total energy of the output spot is collected and supports a long operation distance of more than 100mm for the electric-driven diaphragm to move backward and forward.

The light source can be laser or white light (broadband light). For large core fibres, 320$\mu$m for instance, the incident light is limited by a pinhole of 200$\mu$m. As for small core fibres, the input light can be replaced by a fibre with the same size or smaller core to decrease the input spot size. At the same time, the preparation of the fibre end face (used as the light source waveguide) should be well cleaved or polished to suppress scattering to ensure the output power distribution of the incident light is smooth. The alignment in both input and output ends is important for measuring FRD accurately. In DEEM system, the reverse incidence method (RIM) based on the principle of reversibility of light is proposed to ensure the status of alignment by injecting the light from the input and output ends of the fibre as in lens L1 and L4 in Fig.\ref{fig:3}, respectively. And the shearing interferometer is to inspect the collimated light between the lenses L1 and L2 in both positive and inverse incident directions. Two main steps of RIM are shown as follows. First, the expanded light from the pinhole passes through the lens L1 to be collimated light and focuses on the focal point of the lens L2. Move lens L1 slightly in x- and y- directions, and fine-tune the angle of the lens L1 till the interference fringe on the shearing interferometer is parallel and stable. Thus the normal of lens L1 is on the optical axis of the incident light. Second, the testing fibre is required to be fixed on the focal point of the lens L2. According to the reversibility of light, we inject the light from the lens L4 into the testing fibre, and the output light also should be collimated when it passes through the lens L2. Similarly, fine-tune the position of the fibre end and the angle of lens L2 and inspect the collimated light through the shearing interferometer. The three dimensional monitor system of MIS-XY and MIS-Z can check the light spot on the fibre end face. With the help of RIM, the alignment of the incident system can be ensured within the level of 0.01mm in z-direction and 0.001mm in x- and y- directions.

\subsection{Principle of DEEM}
\begin{figure}
   \centering
   \includegraphics[width=\hsize]{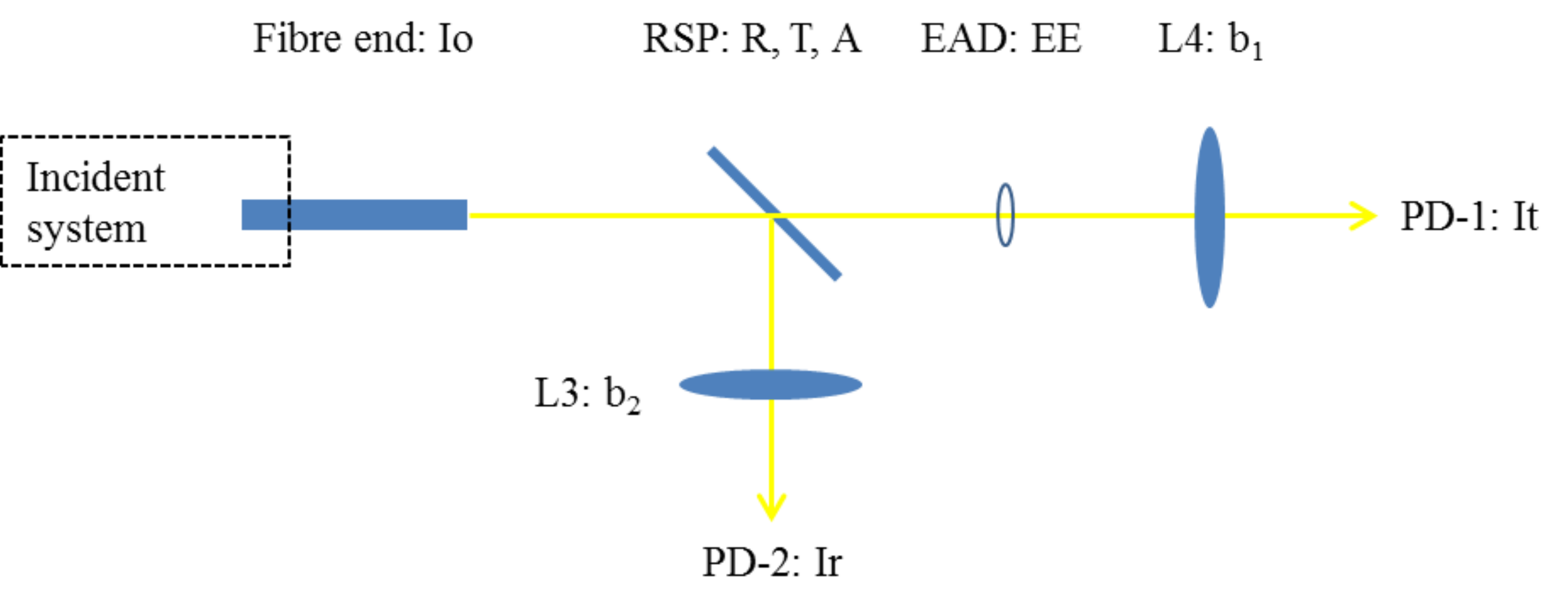}
      \caption{The simplified output system. $I_o$: the intensity of the output spot. $I_r$: reflective light. $I_t$: transmission light. $b_1$, $b_2$: effective transmission coefficient of the lens.}
         \label{fig:5}
   \end{figure}

The proposal of DEEM aims to improve the FRD measurements in two aspects: one is the measurement efficiency to shorten the test time in FRD and throughput; the other is to improve the precision and the stability in complex experiment environment, such as the unstable light source, the variable ambient light and so forth. In DEEM system, the convenience is that we only need to measure the intensity of $I_r$ and $I_t$ or the ratio of $I_t/I_r$ as shown in the simplified output system Fig.\ref{fig:5} to acquire the output focal ratio which enables the rapid measurement. The output beam from the fibre is divided into two beams by a rotatable beam splitter (RSP), of which one beam passes through the electric-driven adjustable diaphragm (EAD) directly to the testbed TB-1 (including the power meter (PM-1) and far-field CCD (FF-CCD)) and the other one goes to the second testbed TB-2 (including the power meter (PM-2) and near-field CCD (NF-CCD)).

For the beam splitter, the effective reflection coefficient ($R_{eff}$) and the effective transmission coefficient ($T_{eff}$) are a function of the reflection ($R$), transmission ($T$) and absorption ($A$).
\begin{equation}\label{eq12}
\left\{ {\begin{array}{*{20}{c}}
{T + R + A = 1}\\
{T:R:A = \alpha :\beta :\gamma }
\end{array}} \right.
\end{equation}
where $\alpha $, $\beta $ and $\gamma $ are constants for a specific splitter. Then the effective reflective and transmission coefficients $R_{eff}$, $T_{eff}$ can be written in the form:
\begin{equation}\label{eq13}
\left\{ {\begin{array}{*{20}{c}}
{{R_{eff}} = f\left( {T,R,A} \right)}\\
{{T_{eff}} = g\left( {T,R,A} \right)}
\end{array}} \right.
\end{equation}
Assuming that the output intensity is $I_o$, the power $I_r$ and $I_t$ can be determined as follows:
\begin{equation}\label{eq14}
{I_r} = {I_o} \cdot f\left( {T,R,A} \right) \cdot {b_2}
\end{equation}
\begin{equation}\label{eq15}
{I_t} = {I_o} \cdot g\left( {T,R,A} \right) \cdot EE \cdot {b_1}
\end{equation}
where $b_1$ and $b_2$ are the effective transmission coefficients of lenses L3 and L4, respectively.
Substituting $I_o$ from equation (\ref{eq14}) into equation (\ref{eq15}), we get
\begin{equation}\label{eq16}
\frac{{{I_t}}}{{{I_r}}} = \frac{{g\left( {T,R,A} \right)}}{{f\left( {T,R,A} \right)}} \cdot EE \cdot \frac{{{b_1}}}{{{b_2}}}
\end{equation}
Let $\frac{{g\left( {T,R,A} \right)}}{{f\left( {T,R,A} \right)}} \cdot \frac{{{b_1}}}{{{b_2}}} = C$ , we may write equation (\ref{eq16}) as
\begin{equation}\label{eq17}
EE = \frac{{{I_t}}}{{{I_r}}} \cdot \frac{1}{C}
\end{equation}
where $C$ is a constant for a specific beam splitter. Thus we only need to measure the ratio of $I_t/I_r$ to calculate the output focal ratio within a certain EE ratio.
Let $k = {{{I_t}} \mathord{\left/
 {\vphantom {{{I_t}} {{I_r}}}} \right.
 \kern-\nulldelimiterspace} {{I_r}}}$ , substituting into equation (\ref{eq17}), then we get
\begin{equation}\label{eq18}
k = \frac{{{I_t}}}{{{I_r}}} = C \cdot EE
\end{equation}
The factor $k$ is the ratio of the reflective light and the transmission light in the real-time measurement. Once the beam splitter is selected, the constant $C$ is known for sure and $I_r$ is unchanged. We set the $k$ on the console panel and the DEEM system will regulate the transmission light $I_t$ to satisfy the requirement.

The intensity of $I_t$ is controlled by the adjustable diaphragm EAD and the aperture is the diameter $D$ of the output spot located in a pre-set distance $f$ away from the fibre end. Then the focal ratio is determined by
\begin{equation}\label{eq19}
F_{out}  = \frac{f}{D} = \frac{{\Delta f}}{{\Delta D}}
\end{equation}
And the definition of FRD is given in equation (\ref{eq:FRD}):
\begin{equation}\label{eq:FRD}
\Delta FRD = \left( {\frac{{{F_{in}} - {F_{out}}}}{{{F_{in}}}}} \right) \times 100 per cent
\end{equation}

\begin{figure}
   \centering
   \includegraphics[width=\hsize]{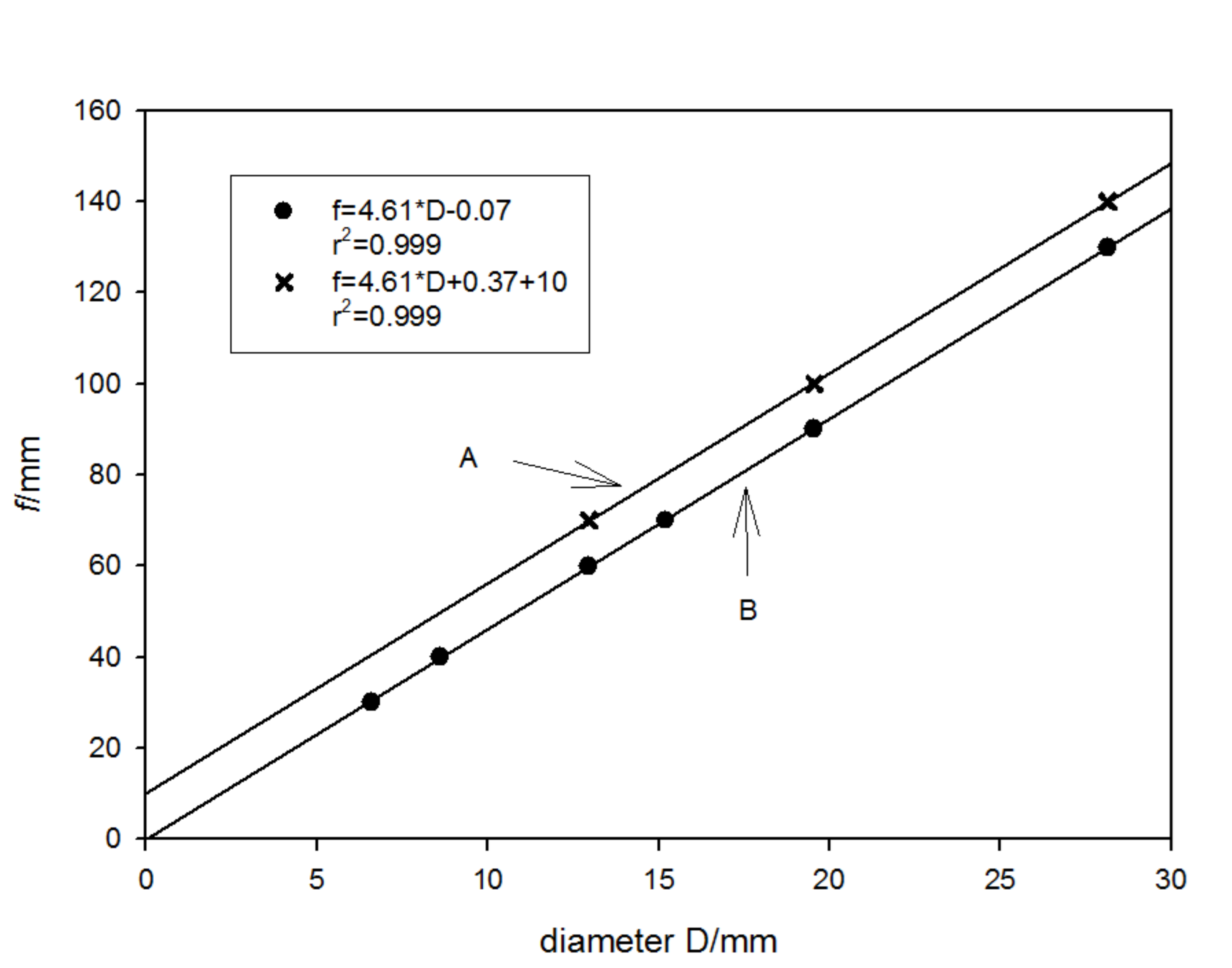}
      \caption{The regression fitting curves of two arrays. The line A with crosses is lifted up with 10mm for clarity. It plots the three measured positions. The other line B with full-filled circles is fitted by differential multiplexing method DMM. The slope of the curve is the focal ratio: 4.61 for both fitting curves. The intercept on $f$-axis indicates the fitted zero point of the fibre end, which is the smaller the better. DMM gives a smaller $f$-intercept which is closer to the fibre end.}
         \label{fig:6}
   \end{figure}

In the measurement process, we record three different diameters $(d_1,d_2,d_3)$ of the diaphragm EAD in three specific distances $(f_1,f_2,f_3)$ away from the fibre end. In the data processing, a differential multiplexing method (DMM) is implemented to raise the data density. As the positions of the diaphragm is well defined (the precision is less than 0.01mm), an array of six ordered pairs \[\left[ {\begin{array}{*{20}{c}}
{\left( {{f_1},{d_1}} \right)}&{\left( {{f_2},{d_2}} \right)}&{\left( {{f_3},{d_3}} \right)}\\
{\left( {\Delta {f_{12}},\Delta {d_{12}}} \right)}&{\left( {\Delta {f_{13}},\Delta {d_{13}}} \right)}&{\left( {\Delta {f_{23}},\Delta {d_{23}}} \right)}
\end{array}} \right]\] can be derived (where $\Delta {f_{ij}} = f_i - {f_j}$ and $\Delta {d_{ij}} = d{_i} - {d_j}$, assuming that $f{_i} > {f_j}$ and $d{_i} > {d_j}$.), with which a linear regression fitting curve is plotted to calculate the focal ratio as shown in Fig.\ref{fig:6}. In this way, we take the advantages of the data to enhance the accuracy with the least times of measurements.

The regression linearity is very high in both fitted lines and reaches to $r^2$=0.999. The intercepts on $f$-axis of two lines are 0.07mm for line B and 0.37mm for line A excluding the additional 10mm. The value of the intercept infers the fitting error on the position of the output fibre end. In the ideal situation, it should be zero. Considering the precision of the experimental environment, the fitting results will be better and more accurate with a smaller intercept. According to equation (\ref{eq19}), the slope of the fitting curve is the focal ratio. And the output focal ratio is 4.61 in the results of Fig.\ref{fig:6}.

\subsection{CCD-IM based on conventional cone beam technique}
\begin{figure}
   \centering
   \includegraphics[width=\hsize]{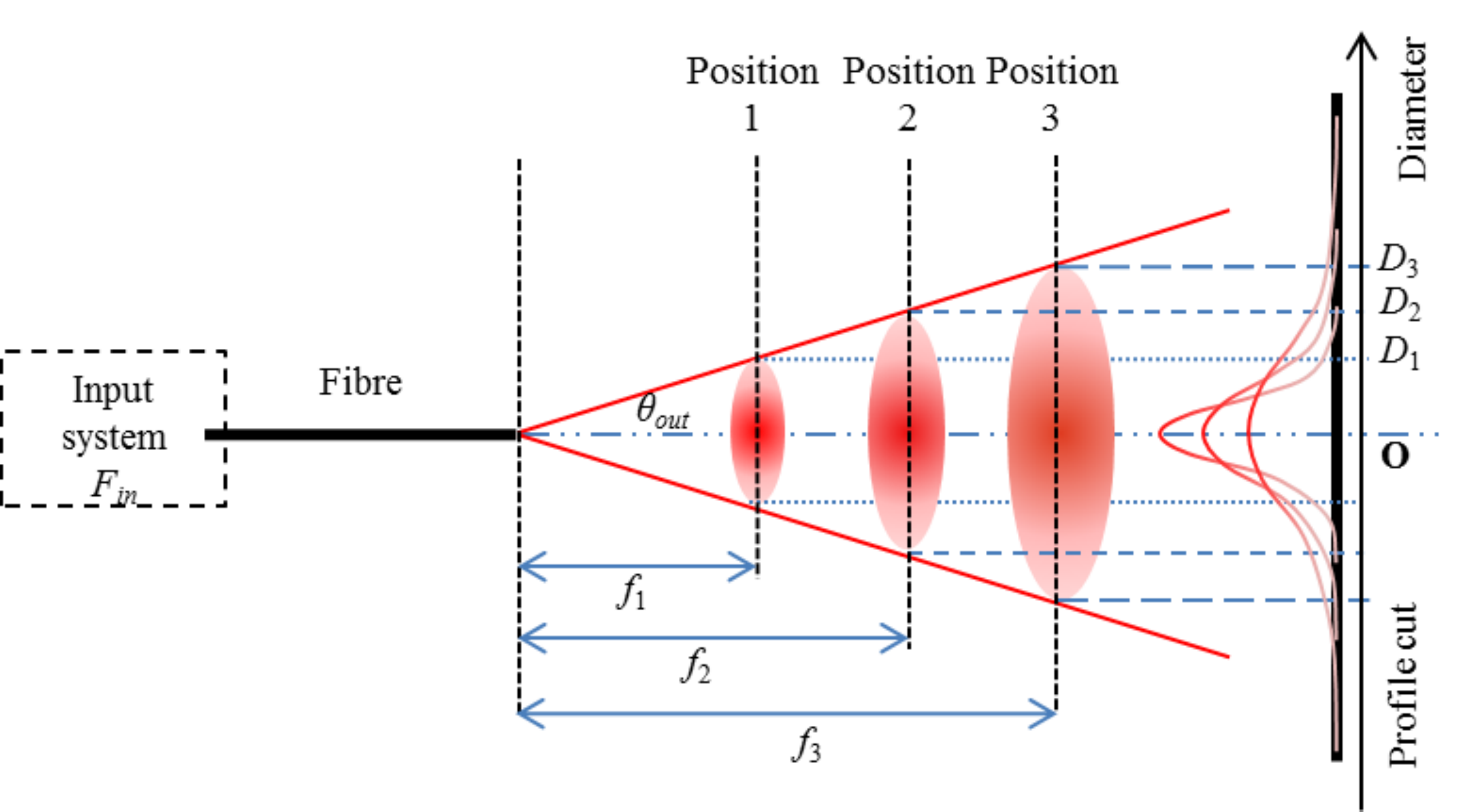}
      \caption{The conventional cone beam method for measuring FRD. The input system is the same as in DEEM. A CCD camera is placed behind the fibre end to record a series of images in different positions. Then fit the distance $f$ and spot size $D$ with linear regression to compute the output focal ratio.}
         \label{fig:CCD-IM}
   \end{figure}

The input system is the same in both of DEEM system and the conventional imaging method with CCD (donated as CCD-IM) based on the cone beam technique. The procedure of the FRD measurement in CCD-IM is shown in Fig.\ref{fig:CCD-IM}. A CCD camera is placed in several positions of different distances ($f$) away from the fibre end to record the output spots. Then the diameter ($D$) of each spot is estimated within a certain EE ratio. Similarly, the linear regression is applied to fit the curve of $f$ and $D$ to measure the output focal ratio.

\begin{figure}
   \centering
   \includegraphics[width=\hsize]{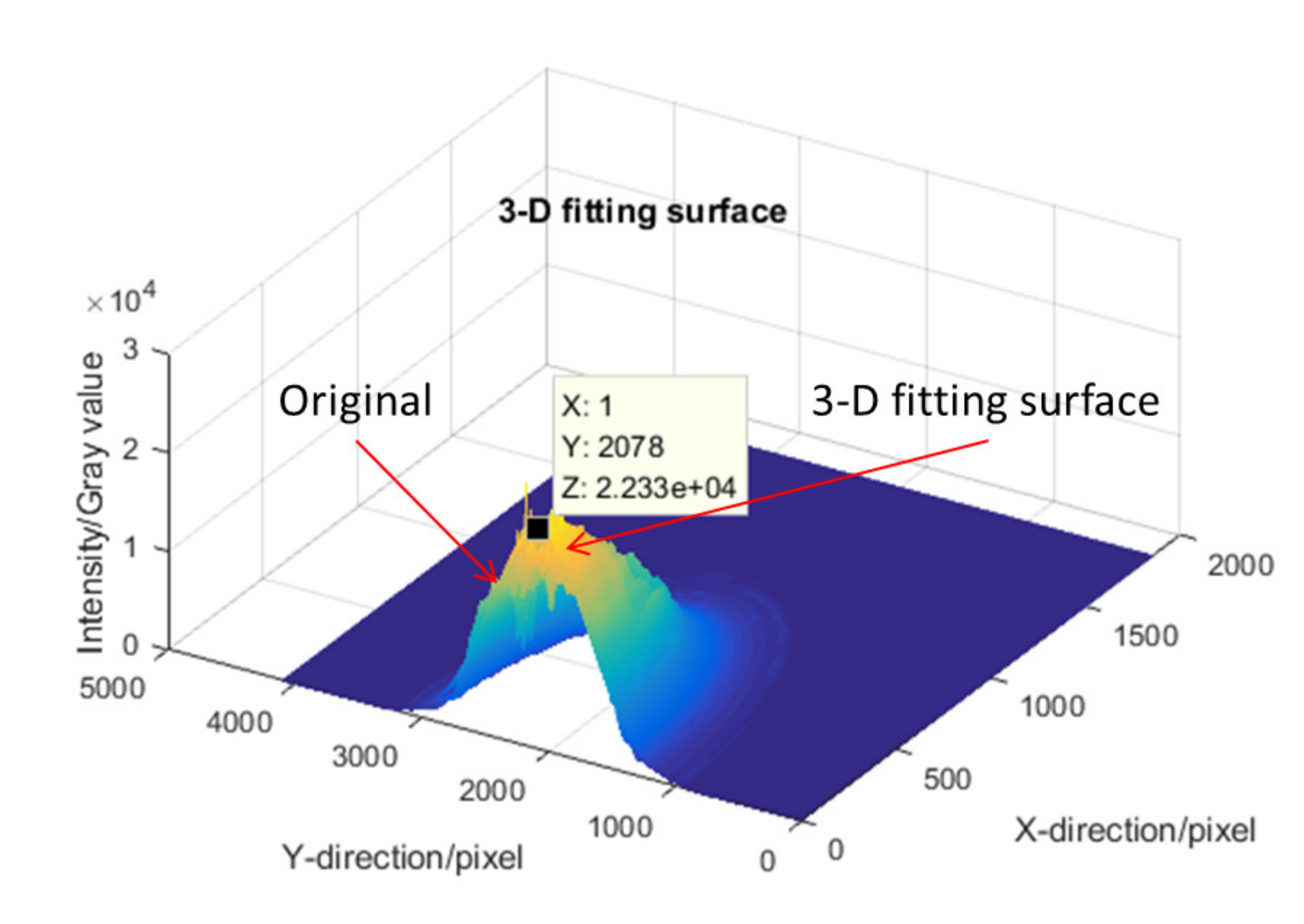}
      \caption{The 3D fitting surface in CCD-IM. The image only shows half of the spot for clarity. The 3D fitting method can make full use of the whole image with every pixel, which can eliminate the uncertainties from the selection effect compared with a profile cut. It also suppresses the influence of laser speckle. Even though a scrambling device is applied in the optical system, the fluctuations of laser speckle still can be seen in the original distribution. It must be pointed out that the 3D fitting method is to assist to determine the centre point of the spot and the EE ratio is integrated and calculated by using the original distribution. The reason is that not all the output distributions are Gaussian spots and some of them are top-hat functions. So it is inaccurate to encircle the energy on the fitting surface (See Fig.\ref{fig:15}).}
         \label{fig:threedimensionalfitting}
   \end{figure}

The barycentre is important for the determination of the diameter of the spot. The two dimensional fitting curves on the profile cut of the image on different directions will lead to different barycentres which bring a major impact on the error in the measured EE ratios. A viable way shown in Fig.\ref{fig:threedimensionalfitting} is to fit the whole spot in three dimensional space to reduce the influence caused by speckle patterns, especially for the laser source, and to eliminate the random errors induced by selection effect.

The alignment should also be considered in CCD-IM. If the normal of the CCD camera is not parallel to the optical axis, the recorded image of the spot will be elliptical. Two parameters $\sigma _x$ and $\sigma _y$ in the three dimensional Gaussian fitting method implemented in the experiments indicate the standard deviation in two directions. The threshold of the ratio of abs($\sigma _x/\sigma _y$-1)=0.02 (<0.5 per cent in FRD) is set to determine the validity of the recorded spots. Those images under the threshold are regarded as circular spots and others should be applied with a factor of cos($\sigma _x/\sigma _y$-1) to correct the spot size in the computing process.

The dark image is recorded in advance to perform background subtraction and the detailed process will be discussed in the comparison tests. The centre point of the Gaussian fitting surface is the barycentre. The EE value controls the diameter of the spot. In the program code, when the EE ratio reaches to the pre-set value (usually the first time it gets larger than the pre-set value, for example, the measured EE of 90.04 per cent>90 per cent), the corresponding radius $r$ (the pixel number) is saved. At the same time, more five pairs of values of $r-3, r-2, r-1, r+1, r+2$ and the corresponding EE ratios $EE_{-3}, EE_{-2}, EE_{-1}, EE_{+1}, EE_{+2}$ are recorded to construct a matrix:
\[\left[ {\begin{array}{*{20}{c}}
{{r_{ - 3}}}&{{r_{ - 2}}}&{{r_{ - 1}}}&{{r_0}}&{{r_1}}&{{r_2}}\\
{E{E_{ - 3}}}&{E{E_{ - 2}}}&{E{E_{ - 1}}}&{E{E_0}}&{E{E_1}}&{E{E_2}}
\end{array}} \right]\]
Then the diameter is derived by weighted averaging as follows:
\begin{equation}\label{eq:rave}
{r_{ave}} = \frac{1}{n} \cdot \frac{{\sum\limits_{j \ne i} {\left( {E{E_j} - EE90} \right){r_i}} }}{{\sum {\left( {E{E_i} - EE90} \right)} }}
\end{equation}
Generally, once the CCD camera is installed in the system, the alignment status changes very slightly. Thus the EE ratio is more sensitive to the speckle patterns. The speckle patterns bring fluctuations on the profile, which will bring offset on the barycentre that might cause a major error on the EE ratio. So a scrambling device is usually used to suppress the speckle. A common method of scrambling is to exchange the near and far fields by means of a lens relay as a double scrambler, but it has to split the fibre \citep{Avila1998Fiber,Spronck2013Fiber,Avila2008Optical}. The alternative way is to squeeze the fibre with mechanical stress to generate bending effect \citep{Avila2006Photometrical}. But it increases FRD and affects the longevity of the fibres. Both of the methods can guarantee high scrambling gain especially in the condition of incoherent light. But they can hardly smooth the speckle patterns in the far field of laser beam in a static situation. Mechanical agitation and the moving diffuser are popular approaches to suppress the fluctuations of the speckle \citep{Reynolds2014A,Roy2014Scrambling,Mahadevan2014Suppression,Mccoy2012Optical,Yang2014Laser,Goodman2010Very}. Applying the diffuser in scrambling has the advantage of non-contact operation on fibres, but it also complicates the optical construction to remain the same input condition. A vibrating scrambler at the frequency of 65Hz with low-amplitude vibration of 0.5cm is used in our experiments to suppress the laser speckle \citep{Wang2016Phase}. With the mechanical agitation, the contrast of the output spot is reduced to provide a stable energy barycentre in the 3D fitting surface. In the program code, other eight points around the barycentre determined by the 3D fitting surface are chosen as new centres to calculate the diameters in order to suppress the influence of speckle patterns. The final size of the spot is averaged with the weighting factor $w$:
\[w = \left[ {\begin{array}{*{20}{c}}
1&2&1\\
2&4&2\\
1&2&1
\end{array}} \right] \cdot \frac{1}{{16}}\]

Except the fitting method, another common practice is using double integral method to position the barycentre ($X_m,Y_m$)as in equation (\ref{eq:barycentre}):
\begin{equation}\label{eq:barycentre}
\left\{ {\begin{array}{*{20}{c}}
{{X_m} = \frac{{\sum {{m_i}{x_i}} }}{{\sum {{m_i}} }}}\\
{{Y_m} = \frac{{\sum {{m_i}{y_i}} }}{{\sum {{m_i}} }}}
\end{array}} \right.
\end{equation}
where $m_i$ is the intensity (grey value) on each pixel and $x_i$, $y_i$ the pixel position. For a well-scrambled spot in our experiments, the difference of diameters between the two methods is small within 5 pixels (0.045mm, E$N.A.$<0.0003). While the difference can be more than 30 pixels (0.27mm) for the same input condition without scrambling. Since this method is sensitive to the fluctuations on the power distribution, the diameter is determined by the 3D fitting method.

\section{Validation tests of DEEM}
We first should estimate the constant $C$ in equation (\ref{eq18}). A very simple way to measure $C$ is to regulate the diaphragm EAD to its maximum and let all the light pass through the aperture to the detector, in which the EE equals to 100 per cent, and the constant $C$ is determined by
\begin{equation}\label{eq20}
C = \frac{{{I_t}}}{{{I_r}}} \cdot \frac{1}{{EE}} = \frac{{{I_t}}}{{{I_r}}}
\end{equation}
Another problem needing to pay attention is the stability and the linearity of both separated beams. Let $f\left( {T,R,A} \right) \cdot {b_1} = {C_1}$  and  $g\left( {T,R,A} \right) \cdot {b_2} = {C_2}$ , and the partial differential with respect to time $t$ in equation (\ref{eq14}) and (\ref{eq15}), we get:
\begin{equation}\label{eq21}
\frac{{{I_r}\left( t \right)}}{{{I_o}\left( t \right)}} = {C_1} \Rightarrow \frac{{d{{{I_r}\left( t \right)} \mathord{\left/
 {\vphantom {{{I_r}\left( t \right)} {{I_o}\left( t \right)}}} \right.
 \kern-\nulldelimiterspace} {{I_o}\left( t \right)}}}}{{dt}} = 0
\end{equation}
\begin{equation}\label{eq22}
\frac{{{I_t}\left( t \right)}}{{{I_o}\left( t \right)}} = {C_2} \Rightarrow \frac{{d{{{I_t}\left( t \right)} \mathord{\left/
 {\vphantom {{{I_t}\left( t \right)} {{I_o}\left( t \right)}}} \right.
 \kern-\nulldelimiterspace} {{I_o}\left( t \right)}}}}{{dt}} = 0
\end{equation}
Equation (\ref{eq21}) and (\ref{eq22}) suggest that the ratio of $I_r/I_o$ and $I_t/I_o$ are independent of time $t$, though the reflective light and transmitted light change with time if the light source is not rigorously stable. Therefore, if $I_r$ and $I_t$ have good linearity with respect to $I_o$, the measurement of constant $C = {{{I_t}} \mathord{\left/
 {\vphantom {{{I_t}} {{I_r}}}} \right.
 \kern-\nulldelimiterspace} {{I_r}}}$  will be a fixed constant. In this way, equation (\ref{eq18})$\sim$(\ref{eq20}) construct the fundamental of DEEM. As the constant $C$ has been measured, we only need to set the value of $k$ according to equation (\ref{eq23}) to measure the focal ratio within a certain EE ratio as follows:
\begin{equation}\label{eq23}
Given:EE \Rightarrow k = C \cdot EE
\end{equation}

\begin{figure*}
   \centering
   \includegraphics[width=\hsize]{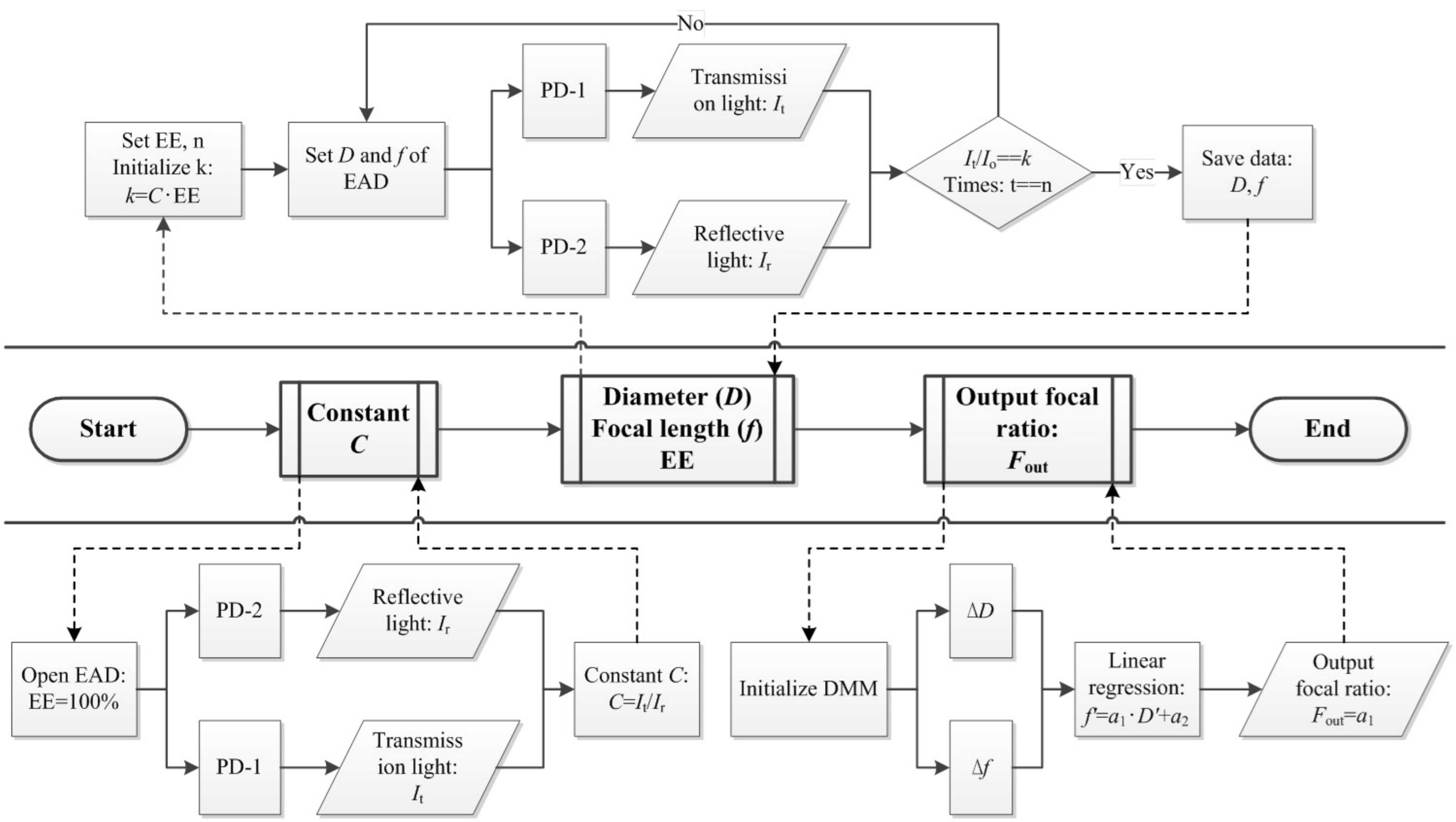}
      \caption{The flowchart of DEEM.}
         \label{fig:7}
   \end{figure*}

The basic procedure is shown in the flowchart of Fig.\ref{fig:7}. In the experiments, we test the performance of DEEM with two different beam splitters (SP1 and SP2) under a stable laser illumination system and a variable light source system.

\subsection{Measurements of the constant $C$}
\begin{figure}
   \centering
   \includegraphics[width=\hsize]{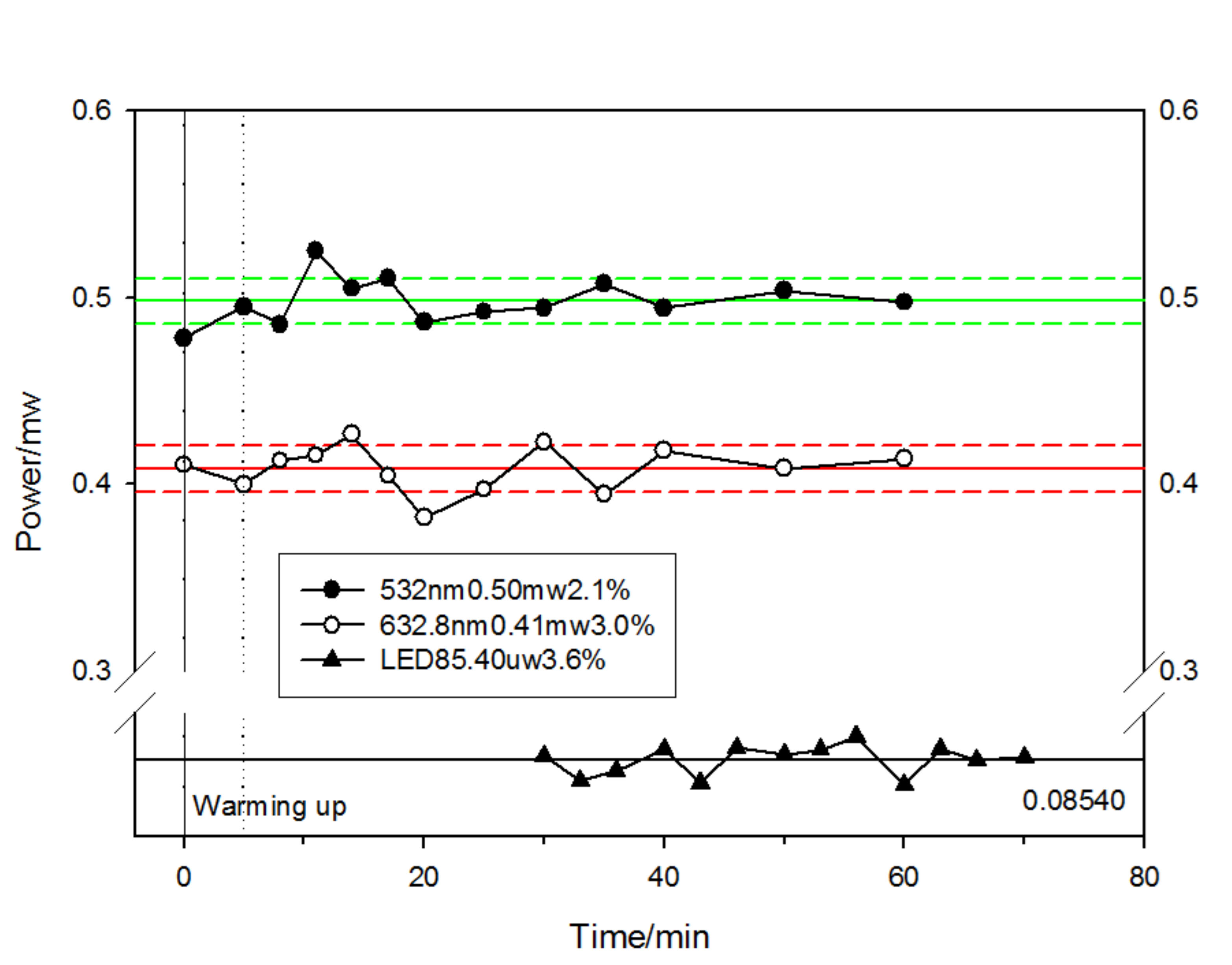}
      \caption{The stability test of the light sources. The stability is not the major error source in FRD measurement in DEEM. But it can affect the throughput test in both DEEM and the conventional method. It also influences the subtraction of noise in the conventional method (See Section 3.4).}
         \label{fig:8}
   \end{figure}

Two stable laser sources ($\lambda $ = 532nm and 632.8nm) are implemented and a stability test during 60 minutes shows that the variation is 2.1 per cent for the green laser and 3.0 per cent for the red laser as shown in Fig.\ref{fig:8}. The intensity is recorded behind the neutral density filter (NDF) and the period of the first five minutes is for warming up. LED is sensitive to the temperature, so the warming time is longer for 30 minutes. The variation of LED is 3.6 per cent.

In fact, the measurement of the constant $C$ is not susceptible to the stability of the light source. According to the principle of DEEM, we should rotate the density filter to change the input light power to record a group of $I_r$ and $I_t$ to make linear regression. The two power detectors record the power of $I_r$ and $I_t$ simultaneously so that even if the light power is changing with time, the relative proportion is unchanged, which can be seen in equation (\ref{eq21}) and (\ref{eq22}) and the constant $C$ should be stable. And such kind of property provides us a convenient way to evaluate the value of the constant $C$. The movement direction of the electric-driven diaphragm is in one-way, either moving away from the output fibre end or moving close to the fibre end. And the aperture of the diaphragm is either enlarging or shrinking with no returning to avoid the deadpath error.

According to equation (\ref{eq20}), we set the diaphragm EAD to its maximum (meaning EE=100 per cent) and let all the light go through the aperture to be collected by the detector PM-1. Then regulate the density filter to change the input light power and record the $I_o$, $I_t$ and $I_r$ in Table \ref{tab:1}. All the tests were carried out till the light source was stable and the fibre was chosen as a multi-mode fibre with the core size of 320$\mu$m.

\begin{table*}
\caption{Measurement of the constant $C$. The measurements of the power ($\mu$W) of $I_o$, $I_r$ and $I_t$ repeat five times to compute the average value. Then the constant $C$ is derived from the linear regression of $I_t$/$I_r$.}             
\label{tab:1}      
\centering                          
\begin{tabular}{c c c c c c c c c}        
\hline                 
 \multicolumn{3}{c}{Power of SP1$@$632.8nm} & \multicolumn{3}{c}{Power of SP2$@$632.8nm} & \multicolumn{3}{c}{Power of SP1$@$532nm} \\    
$I_o$ & $I_r $ & $I_t$ & $I_o$ & $I_r $ & $I_t$ & $I_o$ & $I_r $ & $I_t$ \\
\hline
   105.81 & 39.81 & 47.33 & 174.01 & 71.51 & 77.62 & 73.22 & 26.20 & 32.94 \\
   107.80 & 40.42 & 48.67 & 176.04 & 72.30 & 79.30 & 23.36 & 8.32 & 10.46 \\
   25.72 & 9.62 & 11.51 & 113.78 & 46.67 & 52.31 & 8.36 & 2.94 & 3.70 \\
   54.94 & 20.33 & 24.73 & 60.67 & 24.89 & 27.22 & 146.24 & 52.44 & 65.86 \\
   96.02 & 35.52 & 42.90 & 45.60 & 18.73 & 20.30 & 83.61 & 30.02 & 37.62 \\
   13.33 & 4.96 & 5.98 & 34.81 & 14.27 & 15.55 & 112.22 & 40.20 & 50.54 \\
   6.36 & 2.36 & 2.57 & 23.50 & 9.63 & 10.58 & 54.03 & 19.32 & 24.28 \\
   9.43 & 3.50 & 4.23 & 17.31 & 7.08 & 7.74 & 132.56 & 47.54 & 59.74 \\
   39.51 & 14.62 & 17.84 & 12.12 & 4.95 & 5.38 & & & \\
   79.63 & 28.20 & 34.01 & & & & & & \\
   \\
   \multicolumn{2}{c}{Slope($I_t/I_r$)} & 1.202 & \multicolumn{2}{c}{Slope($I_t/I_r$)} & 1.095 & \multicolumn{2}{c}{Slope($I_t/I_r$)} & 1.257 \\
   \multicolumn{2}{c}{$r^2$} & 0.999 & \multicolumn{2}{c}{$r^2$} & 0.999 & \multicolumn{2}{c}{$r^2$} & 0.999 \\
   \multicolumn{2}{c}{$C$} & 1.20 & \multicolumn{2}{c}{$C$} & 1.10 & \multicolumn{2}{c}{$C$} & 1.26 \\
\hline                                   
\end{tabular}
\end{table*}

\begin{figure*}
   \centering
   \includegraphics[width=\hsize]{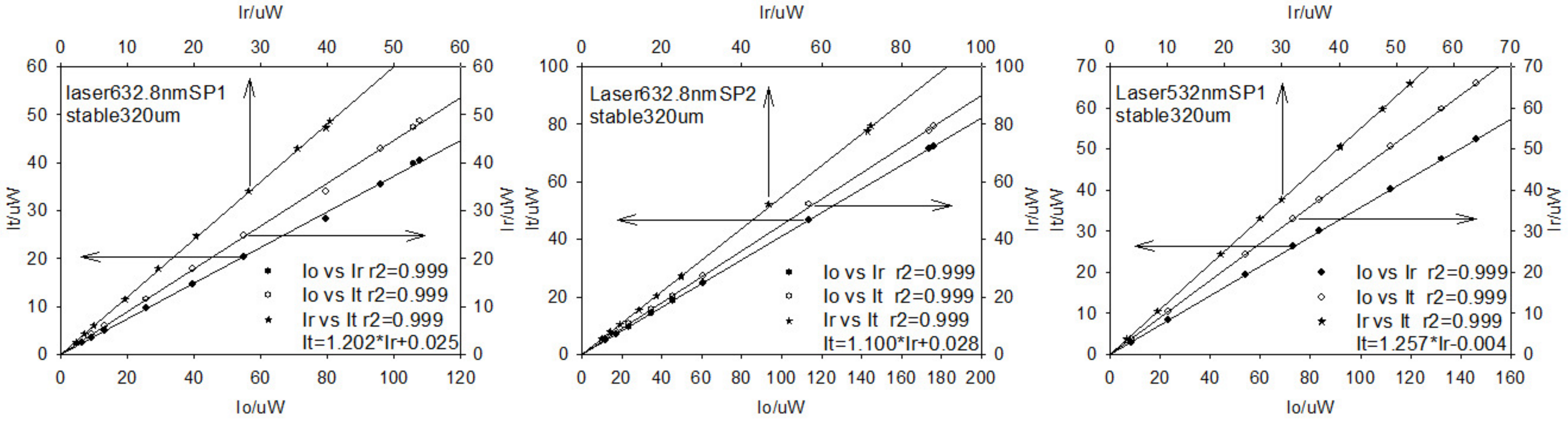}
      \caption{The fitted curves of $I_r$, $I_t$ and $I_o$. The slope of the linear regression $I_t$/$I_r$ is the measured constant $C$. The intercept on $I_t$ axis indicates the offset of the transmission light and the relative error is less than 0.05 per cent (much smaller than the variation of the light source), which means the response of the power meter is synchronised.}
         \label{fig:9}
   \end{figure*}

The constant $C$ is the slope of the linear regression of $I_t/I_r$. And the fitting curve of $I_r/I_o$ and $I_t/I_o$ can be used as indicators to evaluate the accuracy and the confidence level of the measurements of the constant $C$ as shown in Fig.\ref{fig:9}. Here we introduce some new parameters $I_{ro}$-intercept ($I_{ro}$), $I_{to}$-intercept ($I_{to}$) and $I_{tr}$-intercept ($I_{tr}$) to represent the intercepts on the vertical axis of the fitting curves. The three new parameters describe the bias in power measurements and the error analysis will be discussed in Section 3.2. The results indicate that the constant $C$ of different splitters is not the same at different wavelengths but it will not affect the results in DEEM because the constant $C$ is an intermediate variable according to the principle of DEEM. From this point of view, a rigorous stable light source is not essential for DEEM system because even if the input light power varies with time, the constant $C$ remains unchanged. Then the ratio of $k$ is still a fixed constant for a given EE ratio and the computing process will always satisfy the condition of equation (\ref{eq16}).

The slopes of the fitting curves are 1.202, 1.095 and 1.257. In the experiment setup, the precision of the power detector is limited to two decimal, so the constant $C$ is corrected to 1.20 for SP1 at 632.8nm, 1.10 for SP2 at 632.8nm and 1.26 for SP1 at 532nm. The intercepts of $I_{ro}$, $I_{to}$ and $I_{tr}$ rely on the deviation degree of the stability of the light source and the splitter. In the tests, the value of the intercepts should be the smaller the better.

\subsection{Error analysis}
In this section we will study the error calculation and the error sources and discuss the method to reduce the uncertainties efficiently. First, the misalignment in both input and output ends of the fibre will contribute errors to the measurement. Second, the diffraction around the aperture of the diaphragm, especially for laser source, will affect the power readings of the power meter. Finally, the uncertainties of the two key parameters, the constant $C$ and the EE ratio, also determine the accuracy of the DEEM system. In addition, FRD is sensitive to the profile of the fibre end face. Fibre ends prepared by polishing or cleaving have different roughness and stress on the end face which is one of the important effects that dominate FRD \citep{Allington-Smith2013End}.

Due to the non-linearity between the focal ratio and the half cone angle of the spot, the influences on FRD from the variation of the diameters are not the same for different input focal ratios. For example, if the focal distance is 100mm, then an increment of 1mm in the diameter of spot will result in the percentage changes in $\Delta FRD$ to be 4.8 per cent and 9.1 per cent for focal ratios of F/5.0 and F/10.0, respectively. Considering the range of F-ratios being tested in the experiment are limited to be less than the cone angles of the critical $N.A.$ of the fibres. Therefore, an approximated linearity relationship between the inverse of focal ratio and the sin$\theta$ of the half cone angle is valid as in equation (\ref{eq:sinf}):
\begin{equation}\label{eq:sinf}
\sin \theta  \approx  \frac{D}{{2 \cdot f}} = \frac{1}{2} \cdot \frac{1}{F}
\end{equation}
To better quantify the errors in FRD measurement, we convert the variations in the alignment into the uncertainties in $N.A.$ (denoted as E$N.A.$) to characterise the changes in the diameters or the angles of the spot. And one can easily translate the errors from E$N.A.$ to the changes in F-ratio in different input focal ratios according to equation (\ref{eq:sinf}).

The stability of the light source is a potential error source. While in DEEM, it can be suppressed in the two-arm measurement system. Another important error source is the noise, including the background of the environment and the dark current and so forth. If the background can be well characterised and subtracted, the influence can also be minimized. In both of DEEM and CCD-IM, the subtraction of noise must be done, but there are some differences in the subtraction of the two methods. The approach of these two error sources will be discussed separately in Section 3.4.

\subsubsection{Alignment}
For alignment issue, RIM is applied to ensure the input light is normally injected into the fibre end as described in Section 2.2. For a bare fibre, a CCD camera is placed between the diaphragm EAD and the lens L3 to inspect the projection image of the output light. Then record a series of output spots while moving the diaphragm EAD forward and backward along the parallel rails and fine-tuning the position and the angle of the testing fibre end. If the light source is laser, diffraction around the aperture of the diaphragm is apparent to be noticed. In this case, enlarge the aperture of the diaphragm to let the entire light pass through the circle and the diffraction patterns just disappear. Then we repeat these steps above, the output fibre end and the diaphragm can be well controlled on the optical axis. The shift of fitting surface of the output spot on the CCD is limited in 10 pixels and it results in error of E$N.A.$=0.0003.

The measurement uncertainties in the input optics also exist in the determination of the input focal ratio which is affected by the controlling devices of the stage, the diaphragm and the positioning holder. With the help of RIM and the monitor system, the alignment error of the collimated light from the light source can be ignored compared to the precision of the travel stage (<0.01mm) and the diaphragm (<0.05mm), which results in E$N.A.$<0.00033. The uncertainty of the centre position of the diaphragm on the optical axis is less than 0.045mm and it leads to the error in $N.A.$ of E$N.A.$<0.0003. On the other hand, when we measure the FRD performance of an IFU, the monitor system can help to select the fibre and inspect the incident position of the input light spot on the fibre end face as shown in Fig.\ref{fig:3} (top left). The repositioning of each fibre on the IFU head to the input beam introduces an uncertainty of <0.005mm (E$N.A.$<0.00007) depending on the reference scale of the microscope.

In the output end, the alignment also contributes to the uncertainties in the FRD measurements. For a bare fibre, the output angle is well controlled by RIM. While the V-groove plate of the IFU is fixed on the optical axis with some slight angle differences among the fibres depending on the angle variations caused by end face termination and how the fibre is taped in the V-grooves. First, the image of the distribution of the fibres in the V-grooves is recorded on the focus plane to demarcate the baseline. Then scan the fibre in the input end with the same input beam to image the output spots. Finally, move the CCD backward and forward to measure the average shift of centroid of each fibre core to compare with the baseline. The results of the measured angle uncertainty with respect to the optical axis are smaller than 0.011rad and the angle error of 95 percent of the fibres (77 out of 81 fibres) is less than 0.0019rad. Considering the magnification factor of $\delta$=75/200 of the imaging system, the errors in E$N.A.$ are corrected to 0.0041 and 0.0007. And the uncertainties of the centre position of the fibres in the V-grooves is up to E$N.A.$=0.00011. The alignment of the detectors including the CCD camera in CCD-IM and the power meter in DEEM also affect the measurement error. For CCD-IM, the angle error exist in the travel of the camera on the stage. The uncertainty is controlled less than 1.0$^\circ$ and the correction of this error is applied in the computing process (see Section 2.4). In the DEEM system, the power meter is fixed in a specific position and there is no need to move the detector. So the uncertainty comes from the response of power readings (<0.33 per cent, see Section 3.2.2) and the precision of the diaphragm which is less than E$N.A.$ of 0.00033.

Another important error source is the error in the focus of the fibre. For large core multimode fibres, 320$\mu$m for instance, the core size is big enough and different modes focus in different positions. This type of error highly depends on the fibre parameters and it can be characterised in the fitting curve of linear regression as shown in Fig.\ref{fig:6}.

We repeat the measurements of repositioning and end cleaving to test the accuracy of the alignment as shown in Table \ref{tab:alignment} and the relative error of output focal ratio is less than 0.8 per cent. Throughput was tested in different coupling positions ($\pm$5$\sim\pm$40$\mu$m from the centre point) and the input spot was totally injected into the fibre core. The beam size of the input light was <198.0$\mu$m at 10 per cent of the peak and <211.2$\mu$m at 1 per cent for 320$\mu$m fibre, <74.8$\mu$m at 1 per cent for 125$\mu$m fibre and <35.2$\mu$m at 1 per cent for 50$\mu$m fibre. The variation in throughput was consistent within the difference of 2.4 per cent compared with that of the centre position for fibres of 320$\mu$m and 125$\mu$m core, while the difference increased to 8.2 per cent for 50$\mu$m core fibre when the offset was $\pm$10$\mu$m, where the input beam reached to the edge of the fibre core. And the throughput uncertainty can be reduced to smaller than 2.8 per cent as the same level as the light source by aligning the beam to the fibre centre with the monitor system.

In the experiment, the electric-driven diaphragm was fixed in the optical path to measure the diameter of the output spot. This approach can eliminate the uncertainty of the distance away from the fibre end when the diaphragm is moved to different positions. Then the relative error of output focal ratio is dominated only by repositioning and cleaving in both input and output end and it can be derived from the uncertainties of the diameter of the output spot as follow:
\begin{equation}\label{eq:reFRD}
\Delta FRD = \frac{{{f \mathord{\left/
 {\vphantom {f {{D_i}}}} \right.
 \kern-\nulldelimiterspace} {{D_i}}} - {f \mathord{\left/
 {\vphantom {f {{D_j}}}} \right.
 \kern-\nulldelimiterspace} {{D_j}}}}}{{{f \mathord{\left/
 {\vphantom {f {{D_i}}}} \right.
 \kern-\nulldelimiterspace} {{D_i}}}}} = 1 - \frac{{{D_i}}}{{{D_j}}}
\end{equation}

\begin{figure}
   \centering
   \includegraphics[width=\hsize]{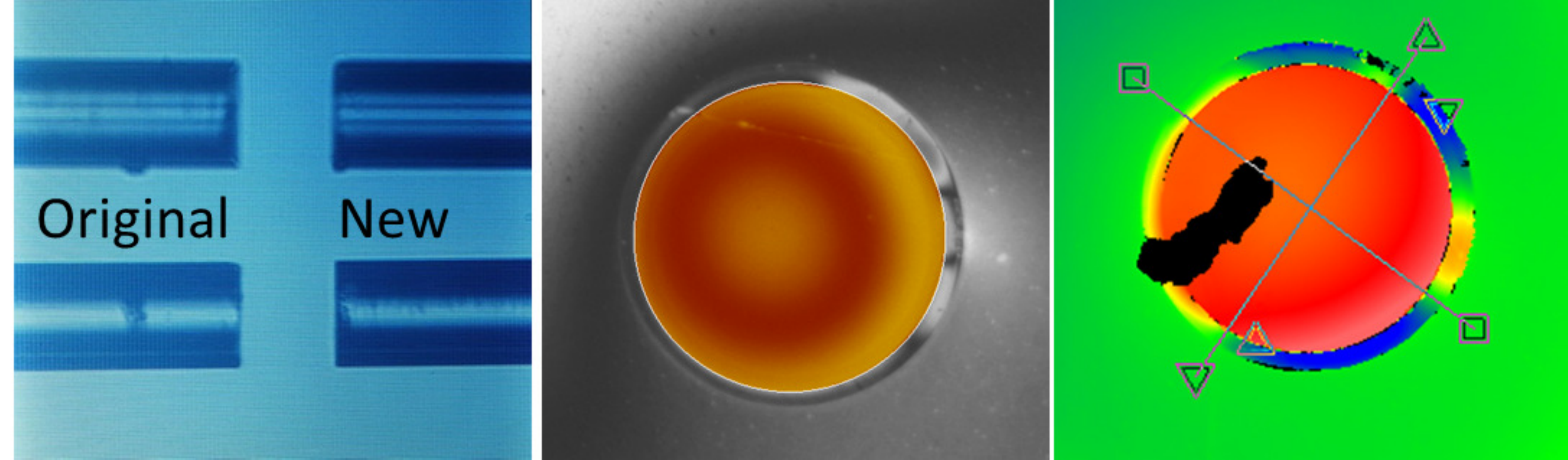}
      \caption{The end face in the microscope. The surface roughness is an important influence to dominate FRD. So those fibres with serious pits on the end face and the defects near the cleave point spreading into the core are rejected. The angle should be inspected between the original and new end face as in left image to ensure the input beam is normally incident into the fibre. The middle image shows the acceptable end face. Contamination sometimes exists on the surface and it must be cleaned by dehydrated alcohol or ether.}
         \label{fig:endface}
   \end{figure}

The end face termination with the surface angle less than 1$^\circ$ by cleaving should be inspected by a microscope to ensure the quality of fibre end without serious defects, rejecting the fibres with bad surface roughness or breaks near the incision to the fibre core. Sometimes the fibre end is contaminated by some ashes as shown in Fig.\ref{fig:endface} because of the static electricity and it can be cleaned by dehydrated alcohol or ether. In the experiments of several times of cleaving, no significant changes occur in the output spot sizes. So the well-controlled end finish is not the major influence on the output focal ratio.

\begin{table*}
\caption{Diameters (mm) of the output spots acquired by DEEM in the repeated measurements of repositioning and cleaving.}             
\label{tab:alignment}      
\centering                          
\begin{tabular}{c c c c c c c c c c}        
\hline                 
    Wavelength & \multicolumn{3}{c}{320$\mu$m core fibre} & \multicolumn{3}{c}{125$\mu$m core fibre} & \multicolumn{3}{c}{50$\mu$m core fibre} \\
   & Diameter & $\Delta FRD$ & Throughput & Diameter & $\Delta FRD$ & Throughput & Diameter & $\Delta FRD$ & Throughput \\
   & & (per cent) & (per cent) & & (per cent) & (per cent) & & (per cent) & (per cent) \\
\hline
   532nm & 18.03$\pm$0.12 & 0.7  & 91.3$\pm$2.0 & 18.99$\pm$0.10 & 0.5 & 90.8$\pm$1.8 & 20.61$\pm$0.12 & 0.6 & 84.5$\pm$1.6 \\
   632.8nm & 18.18$\pm$0.12 & 0.7 & 93.1$\pm$1.6 & 19.06$\pm$0.11 & 0.6 & 92.6$\pm$2.5 & 20.86$\pm$0.15 & 0.7 & 86.3$\pm$2.1 \\
   LED & 18.10$\pm$0.14 & 0.8 & 92.2$\pm$2.2 & 19.13$\pm$0.12 & 0.6 & 90.9$\pm$1.3 & 20.82$\pm$0.14 & 0.7 & 87.1$\pm$1.9 \\
\hline                                   
\end{tabular}
\end{table*}

\subsubsection{Diffraction effects on power readings}
\begin{figure}
   \centering
   \includegraphics[width=\hsize]{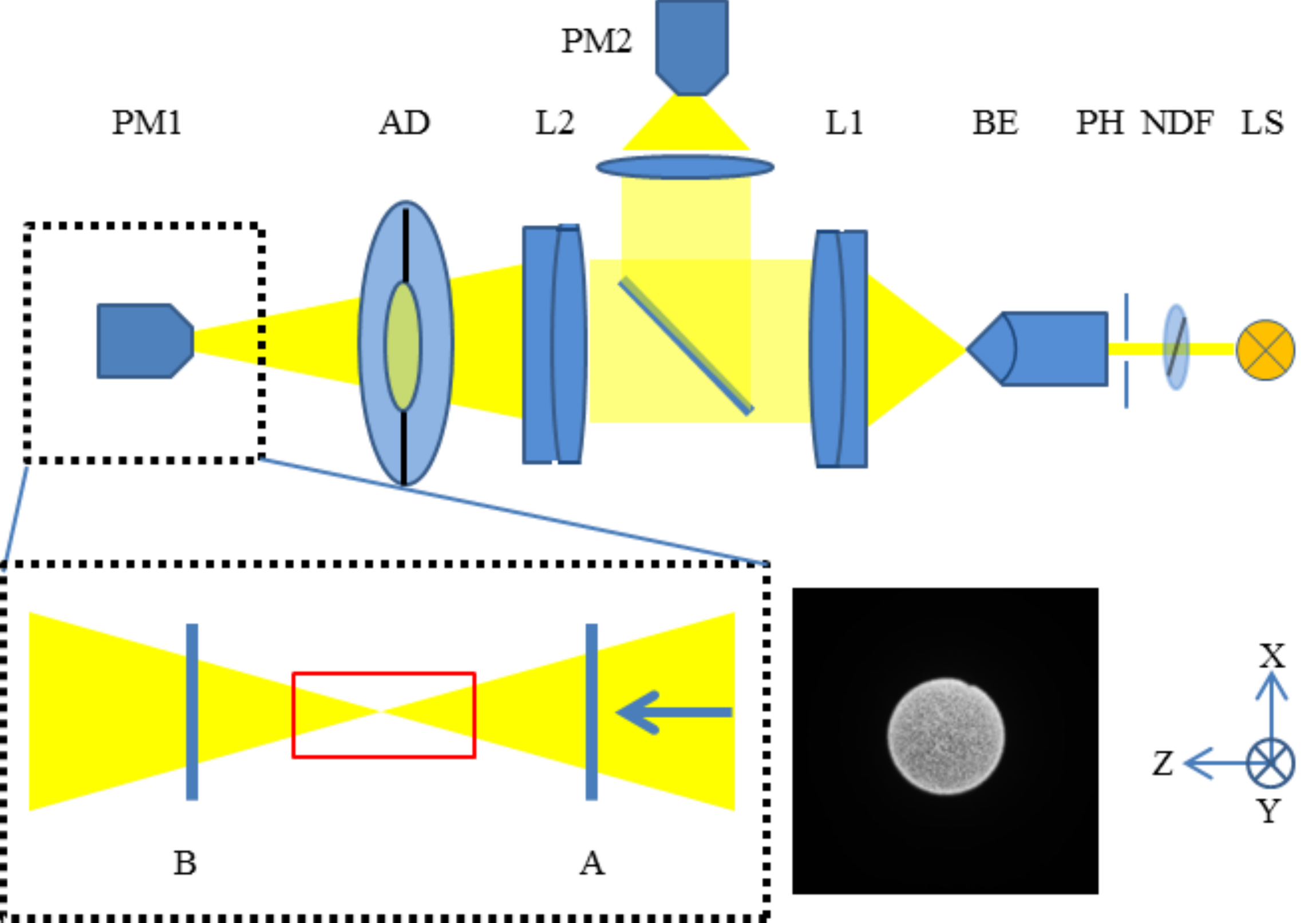}
      \caption{Power reading test. The diaphragm is placed between the power meter PM1 and the lens L2 to construct the same optical situation as in the output end. Removing the test fibre is to exclude the uncertainties of the transmission light in the fibre. Power meters PM1 and PM2 record the total energy and the power within the aperture to compare the variations caused by diffraction and the stability of the light source. The power meter PM1 should be placed around position A and B to avoid the damage to the detector caused by highly focused laser in the focal point (in red frame).}
         \label{fig:diffractiontestbed}
   \end{figure}

The EE ratio in DEEM is acquired by regulating the electric-driven adjustable diaphragm EAD to encircle the output light within a specific aperture. It saves much time to measure the FRD performance, but the diffraction around the aperture occurs inevitably. In CCD-IM, no such aperture exists, so there is no need to concern about the diffraction in the images recorded by CCD. To evaluate the influence on the power readings, the light source of laser was chosen to test the uncertainties as shown in Fig.\ref{fig:diffractiontestbed}. In the experiment, the power detector was placed behind the lens L2 to record the input light power rather than the power out from the testing fibre to avoid the potential energy variation caused by the transmission in the fibre, which might conceal the influence of diffraction. The adjustable diaphragm was placed between the lens L2 and the detector in order to simulate the diffraction situation just the same as in the output end.

The diameter of the detector window is 9.5mm. To avoid the potential damage by highly focused laser, the detector is placed near the position A or B or other positions far away from the focal point. Table.\ref{tab:diffraction} shows the power readings of the detector fixed in six positions (three of which in A side and others in B side). The focal ratio of the light was $F$=5.0 limited by the diaphragm. The power readings are stable within the relative error less than 2.3 per cent which is the same level of the stability of the light source. So we believe that the slight variation in the power reading is from the changes of the light source. If the difference is mainly caused from the uncertainty of the light source, the EE ratio ($EE = \frac{{{{{P_1}} \mathord{\left/ {\vphantom {{{P_1}} {{P_2}}}} \right. \kern-\nulldelimiterspace} {{P_2}}}}}{{BS}}$, where $BS$ is a constant of the split ratio of the beam splitter) will not change since the two detectors record the refractive light and the transmitted light simultaneously and then the influence of the diffraction is negligible which is confirmed in the test that the relative difference of PM1/PM2 is less than 0.33 per cent. And it also shows that the DEEM system can improve the stability even in an unstable input condition.

\begin{table*}
\caption{Power readings ($\mu$W) of diffraction test. And the relative difference of the ratio PM1/PM2 is smaller than 0.33 per cent, which indicates the error of the measured EE ratio is less than 0.33 per cent. The variation in different positions is 2.3 per cent within the stability of the light source. So we believe that the difference is dominated by the uncertainty of the light source.}             
\label{tab:diffraction}      
\centering                          
\begin{tabular}{c c c c c c c c c c c c c}        
\hline                 
   Light source & Diameter of & PM & \multicolumn{6}{c}{Power in six positions} & Average & $\Delta P$ & $P_1/P_2$ \\
   Wavelength & diaphragm (mm) & & A1 & A2 & A3 & B1 & B2 & B3 & & (per cent) & \\
\hline
   532nm & 12.0 & PM1 & 429.91 & 422.63 & 434.23 & 428.55 & 442.52 & 427.31 & 430.86$\pm$6.84 & 1.6 & 7.93 \\
   & & PM2 & 54.26 & 53.46 & 54.86 & 53.93 & 55.58 & 53.59 & 54.28$\pm$0.81 & 1.5 & \\
   & 18.0 & PM1 & 430.98 & 431.56 & 428.35 & 433.13 & 423.98 & 422.38 & 428.40$\pm$4.35 & 1.0 & 7.87 \\
   & & PM2 & 54.61 & 54.90 & 54.32 & 54.96 & 54.17 & 53.85 & 54.47$\pm$0.43 & 0.8 & \\
   & 26.0 & PM1 & 426.03 & 426.60 & 427.03 & 429.36 & 428.84 & 426.58 & 427.41$\pm$1.36 & 0.3 & 7.87 \\
   & & PM2 & 53.58 & 54.22 & 54.32 & 55.46 & 54.56 & 53.60 & 54.29$\pm$0.70 & 1.3 & \\
   632.8nm & 12.0 & PM1 & 342.84 & 330.57 & 339.49 & 332.64 & 343.76 & 339.24 & 338.09$\pm$5.37 & 1.6 & 7.93 \\
   & & PM2 & 43.15 & 41.44 & 42.99 & 41.85 & 44.10 & 42.13 & 42.61$\pm$0.98 & 2.3 & \\
   & 18.0 & PM1 & 341.21 & 336.87 & 338.81 & 334.83 & 330.34 & 331.69 & 335.63$\pm$4.17 & 1.2 & 7.92 \\
   & & PM2 & 43.23 & 42.67 & 42.82 & 42.17 & 41.68 & 41.76 & 42.39$\pm$0.62 & 1.5 & \\
   & 26.0 & PM1 & 334.27 & 336.65 & 335.61 & 329.11 & 330.94 & 327.49 & 332.35$\pm$3.71 & 1.1 & 7.89 \\
   & & PM2 & 42.48 & 42.91 & 42.55 & 41.44 & 42.16 & 41.28 & 42.14$\pm$0.65 & 1.5 & \\
   ave$P_1/P_2$ & & & & & & & & & & & 7.90$\pm$0.026 \\
\hline                                   
\end{tabular}
\end{table*}

\subsubsection{Precision of the constant $C$}
\begin{figure}
   \centering
   \includegraphics[width=\hsize]{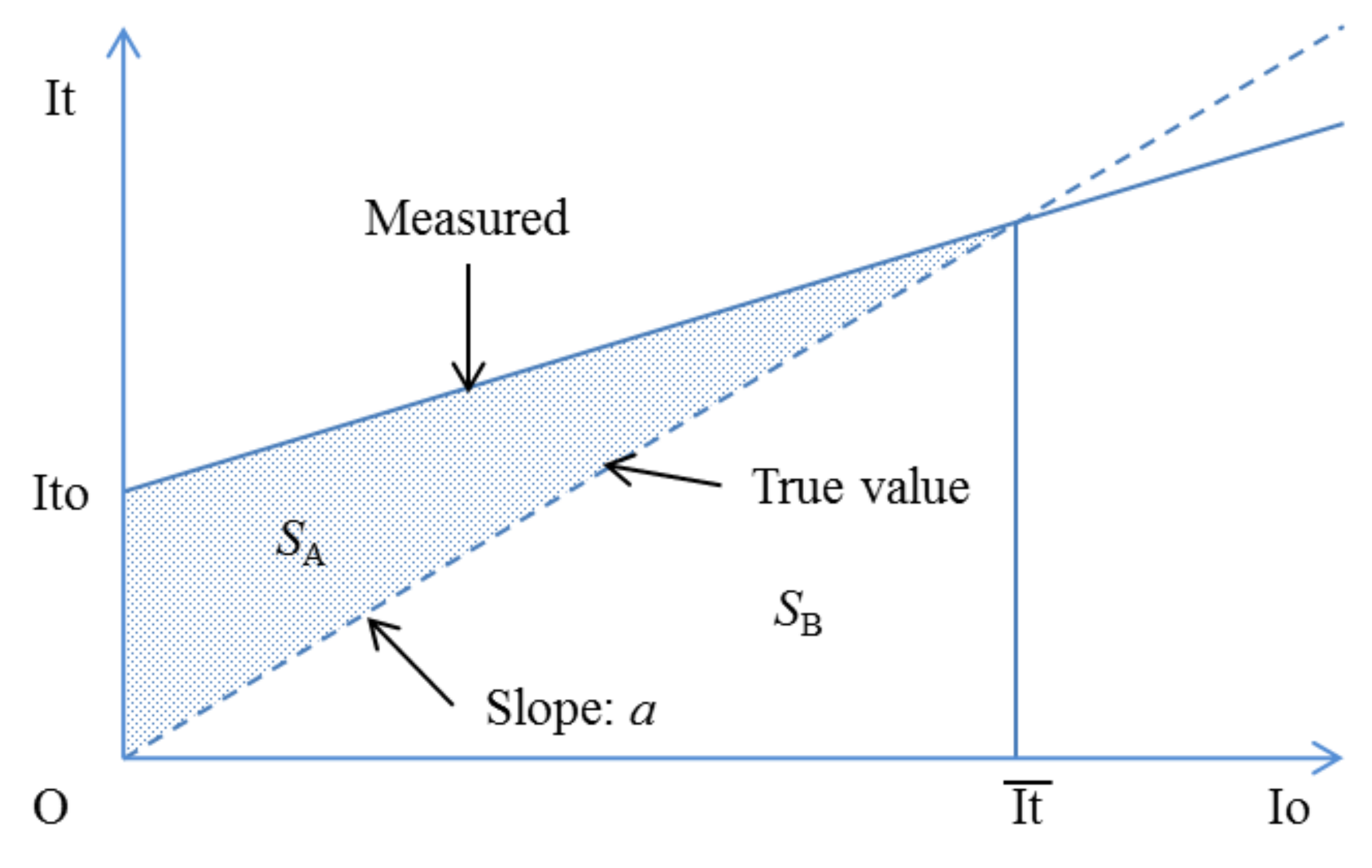}
      \caption{The schematic diagram to show the mathematics used to calculate the relative error $S_A$/$S_B$.}
         \label{fig:10}
   \end{figure}

In an unbiased and ideal system, $I_{ro}$, $I_{to}$ and $I_{tr}$ should be zero which means that no reflective light nor refractive light will be detected when the output light from the fibre is so weak that can be ignored. And it is always true when the light source is shut off. While in the experiments, the accuracy and the response of the devices or some other noise would bias the results and make $I_{ro}$, $I_{to}$ and $I_{tr}$ deviate from its true value. This deviation will bias the value of constant $C$. According to the error transfer formula, the relative error of constant $C$ is given by
\begin{equation}\label{eq24}
\begin{split}
& \frac{{\Delta C}}{C} = \frac{{\Delta {{{I_t}} \mathord{\left/
 {\vphantom {{{I_t}} {{I_r}}}} \right.
 \kern-\nulldelimiterspace} {{I_r}}}}}{{{{{I_t}} \mathord{\left/
 {\vphantom {{{I_t}} {{I_r}}}} \right.
 \kern-\nulldelimiterspace} {{I_r}}}}}\\
& \qquad = \left| {\frac{{\partial \ln {{{I_t}} \mathord{\left/
 {\vphantom {{{I_t}} {{I_r}}}} \right.
 \kern-\nulldelimiterspace} {{I_r}}}}}{{\partial {I_t}}}} \right| \cdot \Delta {I_t} + \left| {\frac{{\partial \ln {{{I_t}} \mathord{\left/
 {\vphantom {{{I_t}} {{I_r}}}} \right.
 \kern-\nulldelimiterspace} {{I_r}}}}}{{\partial {I_r}}}} \right| \cdot \Delta {I_r}\\
& \qquad = \left| {\frac{{\Delta {I_t}}}{{{I_t}}}} \right| + \left| {\frac{{\Delta {I_r}}}{{{I_r}}}} \right|
\end{split}
\end{equation}
In equation (\ref{eq24}),  $\left| {\frac{{\Delta {I_t}}}{{{I_t}}}} \right|$ and  $\left| {\frac{{\Delta {I_r}}}{{{I_r}}}} \right|$ represent the power error of $I_t$ and $I_r$ with respect to their true values, respectively. In Fig.\ref{fig:10}, we use the shaded area to present the relative difference. The dashed line is the true value and the solid line is the measured result. The ratio of the area $S_A$/$S_B$ is the relative error. Assuming that the slope of the dashed line is $s$. The ratio of $S_A/S_B$ can be written as
\begin{equation}\label{eq25}
\left\{ {\begin{array}{*{20}{c}}
{{S_A} = \frac{1}{2} \cdot \left| {{I_{to}}} \right| \cdot \overline {{I_t}} }\\
{{S_B} = \frac{1}{2} \cdot s \cdot \overline {{I_t}}  \cdot \overline {{I_t}} }
\end{array}} \right. \Rightarrow \left| {\frac{{\Delta {I_t}}}{{{I_t}}}} \right| \approx \frac{{{S_A}}}{{{S_B}}} = \frac{1}{s}\left| {\frac{{{I_{to}}}}{{\overline {{I_t}} }}} \right|
\end{equation}
Similarly, the ratio has the same form for the reflective light.
\begin{equation}\label{eq26}
\left| {\frac{{\Delta {I_r}}}{{{I_r}}}} \right| \approx \frac{{{S_A}}}{{{S_B}}} = \frac{1}{s}\left| {\frac{{{I_{ro}}}}{{\overline {{I_r}} }}} \right|
\end{equation}
So equation (\ref{eq24}) can be written in the form
\begin{equation}\label{eq27}
\frac{{\Delta C}}{C} = \left| {\frac{{\Delta {I_t}}}{{{I_t}}}} \right| + \left| {\frac{{\Delta {I_r}}}{{{I_r}}}} \right| = \frac{1}{{{s_t}}}\left| {\frac{{{I_{to}}}}{{\overline {{I_t}} }}} \right| + \frac{1}{{{s_r}}}\left| {\frac{{{I_{ro}}}}{{\overline {{I_r}} }}} \right|
\end{equation}
where $\overline {{I_t}} $  and $\overline {{I_r}} $  are the average values and $s_t$ and $s_r$ are the slopes of the fitting curves. The empiric value of $s_t$ and $s_r$ are around 0.3$\sim$0.5. Since the light source is not rigorously stable, the measurements of the power reading of each group of $I_o$, $I_r$ and $I_t$ repeat five times to acquire the average value. Table \ref{tab:sintercept} shows the intercepts of $I_{to}$, $I_{ro}$ and the slopes of $s_t$, $s_r$.

\begin{table}
\caption{The intercepts for the determination of $s$ (the empiric value is around 0.3$\sim$0.5).}             
\label{tab:sintercept}      
\centering                          
\begin{tabular}{c c c c c c}        
\hline                 
   Splitter & Wavelength (nm) & $I_{ro}$ & $s_r$ & $I_{to}$ & $s_t$ \\    
\hline
   SP1 & 632.8 & 0.058 & 0.369 & 0.067 & 0.444 \\
   & 532 & 0.057 & 0.357 & 0.075 & 0.449 \\
   & LED & 0.081 & 0.352 & 0.91 & 0.445 \\
   SP2 & 632.8 & 0.038 & 0.410 & 0.028 & 0.449 \\
\hline                                   
\end{tabular}
\end{table}

The same method was implemented in measuring $C$ of SP1 illuminated by a LED. Table \ref{tab:errorofc} shows the results of $s_r$ and $s_t$ and E$C$ represents the relative error of the constant $C$.

According to DEEM, the diameter is determined by $k$ which is a function of the constant $C$. The relative error of $C$ will contribute to the uncertainties in FRD. So improving the accuracy of $C$ is an effective way to reduce the error. And equation (\ref{eq27}) suggests that one should choose a proper range of $I_t$ and $I_r$ to reduce the relative error which requires $\overline {{I_t}} $ , $\overline {{I_r}}$  and the interval of $I_t$ and $I_r$ should not be too large. Generally, the power of $I_t$ and $I_r$ would be better to match the best response interval of the detectors to avoid underexposure or saturation.

\begin{table}
\caption{Relative error of the constant $C$.}             
\label{tab:errorofc}      
\centering                          
\begin{tabular}{c c c c c c}        
\hline                 
   Light source & Wavelength (nm) & \multicolumn{2}{c}{SP1} & \multicolumn{2}{c}{SP2} \\    
   & & $C$ & E$C$ (per cent) & $C$ & E$C$ (per cent) \\
\hline
   Laser & 632.8 & 1.20 & 1.4 & 1.10 & 0.9 \\
   & 532 & 1.26 & 1.0 & & \\
   LED & & 1.12 & 2.2 & & \\
\hline                                   
\end{tabular}
\end{table}

\subsubsection{Uncertainties from EE}
In the DEEM system, the ratio of $k$ is also a function of EE. EE determines the maximum of the energy usage. According equation (\ref{eq9}), the partial differential of the power $P$ is derived as:
\begin{equation}\label{eq:power}
\left\{ {\begin{array}{*{20}{c}}
{P \propto \exp \left( { - {r^2}} \right)}\\
{\frac{{\Delta P}}{P} \propto \left| {\frac{{d\left( {\ln \left( {\exp \left( { - {r^2}} \right)} \right)} \right)}}{{dr}}} \right| \cdot \Delta r = 2r \cdot \Delta r \propto \Delta EE}
\end{array}} \right.
\end{equation}
Equation (\ref{eq:power}) suggests that the uncertainty of the power reading is related to $\Delta EE$. And the changes of the diameter of the spot is proportional to $\Delta EE$. So an appropriate choice of EE is important to investigate FRD properties because different EE will directly affect the estimation of the diameters of output spots. If the EE ratio is too small, though the output focal ratio is large which seems to be better for the design of a spectrograph, the energy loss is so great that the spectra efficiency decreases too fast to acquire high-quality spectra with high SNR. Generally, EE85$\sim$95 is commonly used to encircle the effective energy.

\begin{figure}
   \centering
   \includegraphics[width=\hsize]{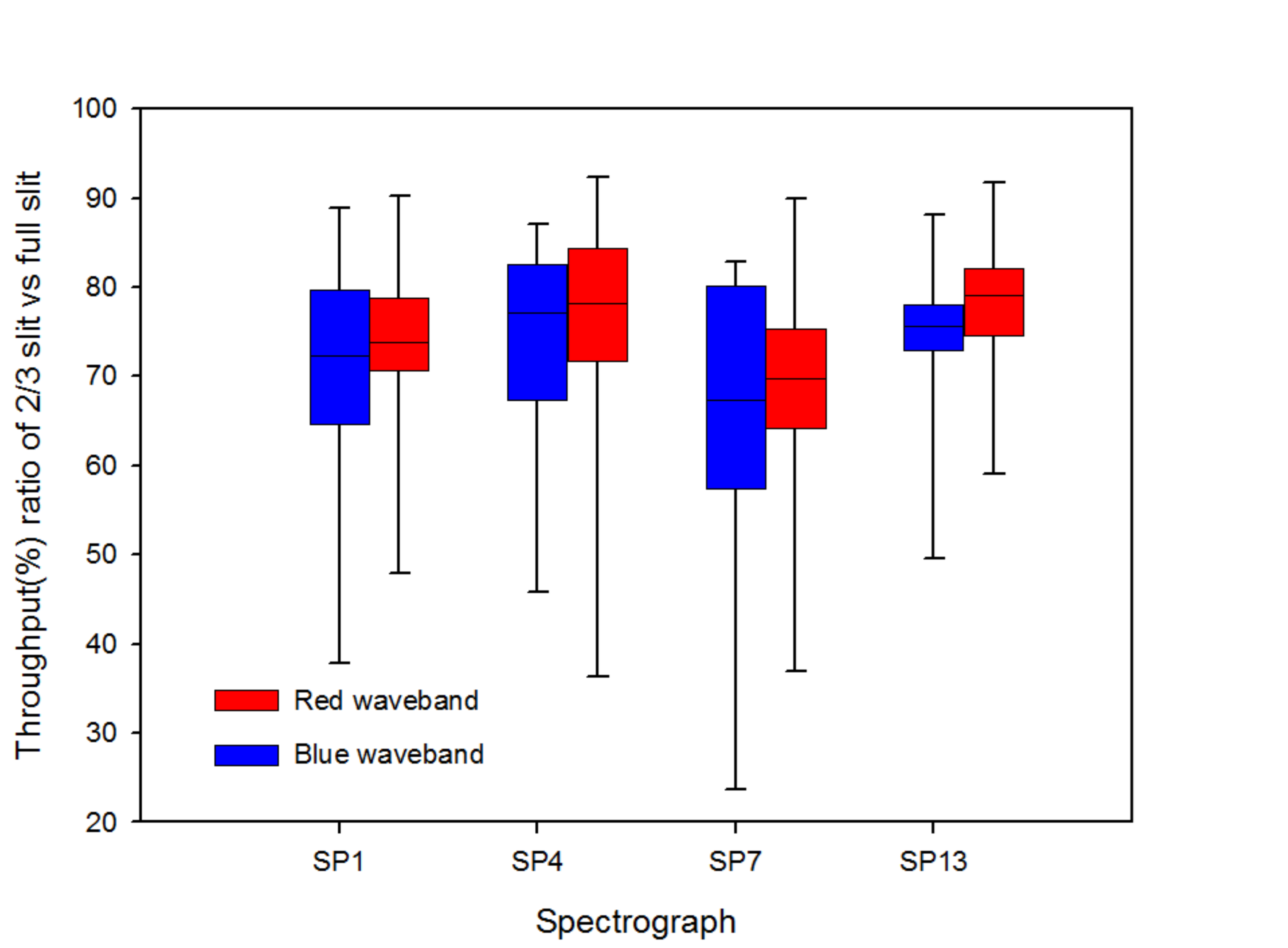}
      \caption{The throughput ratio of 2/3 slit mode vs full slit mode of 4 spectrographs in LAMOST. Blue and red bricks are the throughput in blue and red waveband, respectively. The throughput ratio mostly distributes in the range of 60$\sim$80 per cent due to the offset in the position and angles errors of the fibres in the V-grooves.}
         \label{fig:11}
   \end{figure}

Four out of sixteen spectrographs in LAMOST (No. 1,4,7,13) were chosen to investigate the influence of different EE on the throughput and SNR. A slit of 2/3 of the full width is placed in front of the fibre slit, the efficiency of 2/3 slit mode is supposed to be 78 per cent of the full slit theoretically. In practice, this proportion varies from 60 per cent to 80 per cent based on a series of experiments on four spectrographs as shown in Fig.\ref{fig:11}. Assuming that LAMOST has an effective aperture of 4.0m and the exposure time is 1800s, the theoretical SNR per pixel of a stellar object is predicted at the wavelength of 4770 angstrom and 7625 angstrom with magnitudes of 16, 17, 18 and 19, respectively. The SNR results of different slit modes with efficiency of 100 per cent, 78 per cent and 60 per cent are listed in Table \ref{tab:3}. It is clear to notice the decrease of SNR more than 10 per cent compared with the case of the full slit. And the efficiency of slit has almost the same influence on different bands.

\begin{table}
\caption{The SNR per pixel for stars with different magnitudes under specific slit and efficiency modes.}             
\label{tab:3}      
\centering                          
\begin{tabular}{c c c c c c}        
\hline                 
   Band & Efficiency mode & \multicolumn{4}{c}{Magnitude} \\    
   & & 16 & 17 & 18 & 19 \\
\hline
   $g$-band & Full slit & 51 & 29 & 15 & 7 \\
   & 2$/$3 slit (78 per cent) & 45 & 25 & 13 & 6 \\
   & 2$/$3 slit (60 per cent) & 39 & 22 & 11 & 5 \\
   $i$-band & Full slit & 51 & 29 & 15 & 7 \\
   & 2$/$3 slit (78 per cent) & 45 & 25 & 13 & 6 \\
   & 2$/$3 slit (60 per cent) & 39 & 22 & 11 & 5 \\
\hline                                   
\end{tabular}
\end{table}

\begin{figure}
   \centering
   \includegraphics[width=\hsize]{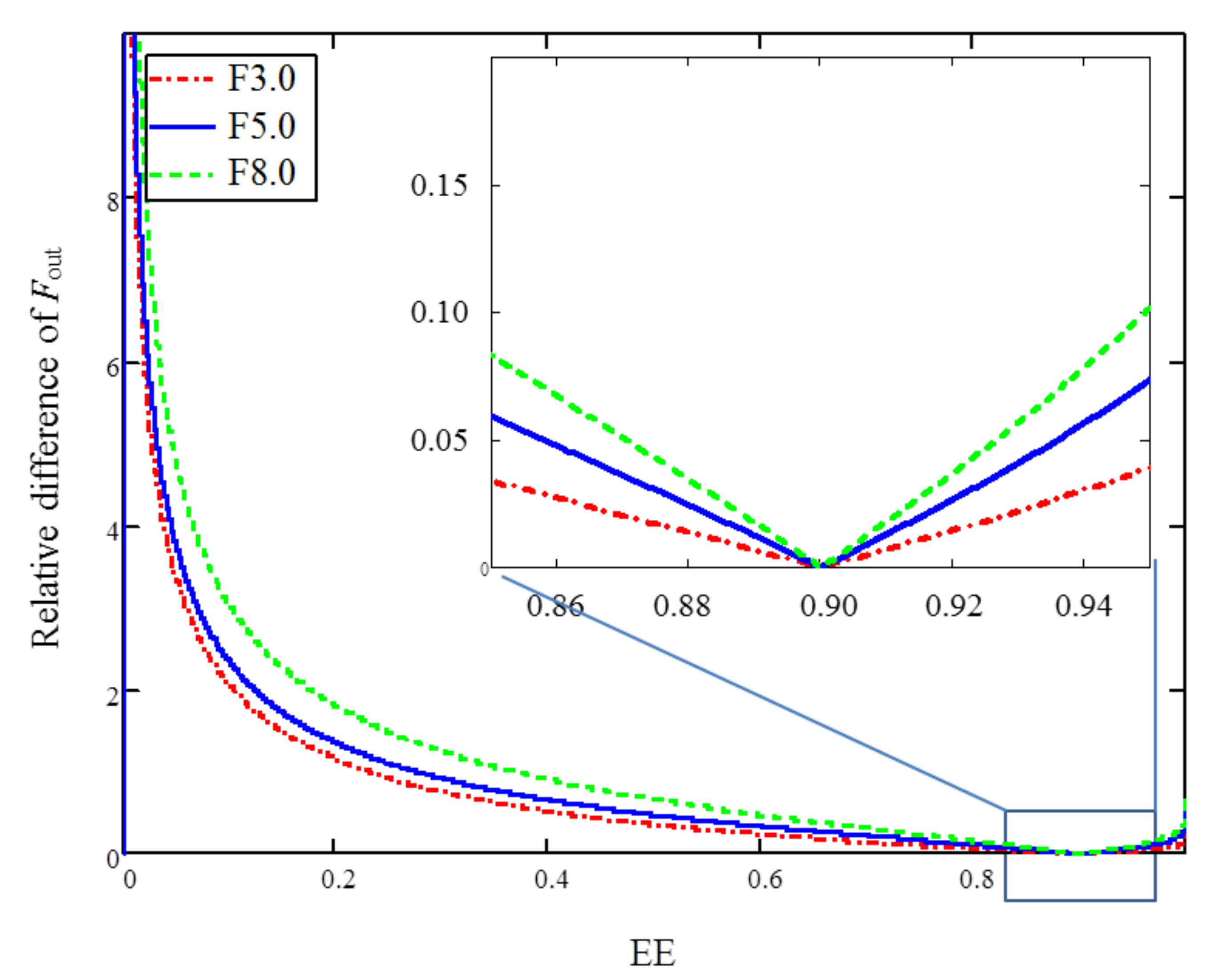}
      \caption{Simulation results by PDM of the relative difference of output focal ratio caused by different EE ratio. Choosing EE89$\sim$EE91 can suppress the relative difference down to 1 per cent.}
         \label{fig:12}
   \end{figure}

In order to acquire data with high SNR, the EE ratio cannot be too low. On the contrary, a large EE ratio will increase the energy usage and SNR but decrease the output focal ratio. In practice people usually take the intervals of 2$\sigma$ resulting in the 95 percent confidence intervals in a Gaussian function. The approximation is also fit for the Gaussian beam. But a little modification should be made that the EE ratio is the square of the confidence intervals as mentioned above in Fig.\ref{fig:2}. The simulation results in Fig.\ref{fig:12} shows the relative difference of output focal ratio caused by different EE ratio. In the intervals from EE85 to EE90, the relative variation is smaller than that of EE90 to EE95. The input focal ratio can also influence the distribution. The variation becomes smaller with decreasing input focal ratio. Compared with EE90, the relative difference can be up to 10.5 per cent for the input focal ratio of $F_{in}$=8.0 and it is only about 3.5 per cent much smaller for the input focal ratio of $F_{in}$=3.0. One can also notice that the difference between EE89 and EE91 is very small and less than 1 per cent. So control the EE ratio precisely can also contribute to the accuracy of measuring FRD.

\subsubsection{Summary of error analysis}
The major difference in the setup of DEEM and CCD-IM is the testbed in the output end. DEEM uses a diaphragm to directly encircle the output energy to measure the spot size. CCD-IM records a series of output spots by a CCD camera to estimate the diameters within a certain EE ratio. Two systems have different contributions of uncertainties in the error sources for FRD measurement. These uncertainties are combined in quadrature to give the upper limits of the measurement errors. Table \ref{tab:errorsummary} lists the maximum errors in the two systems.

\begin{figure}
   \centering
   \includegraphics[width=\hsize]{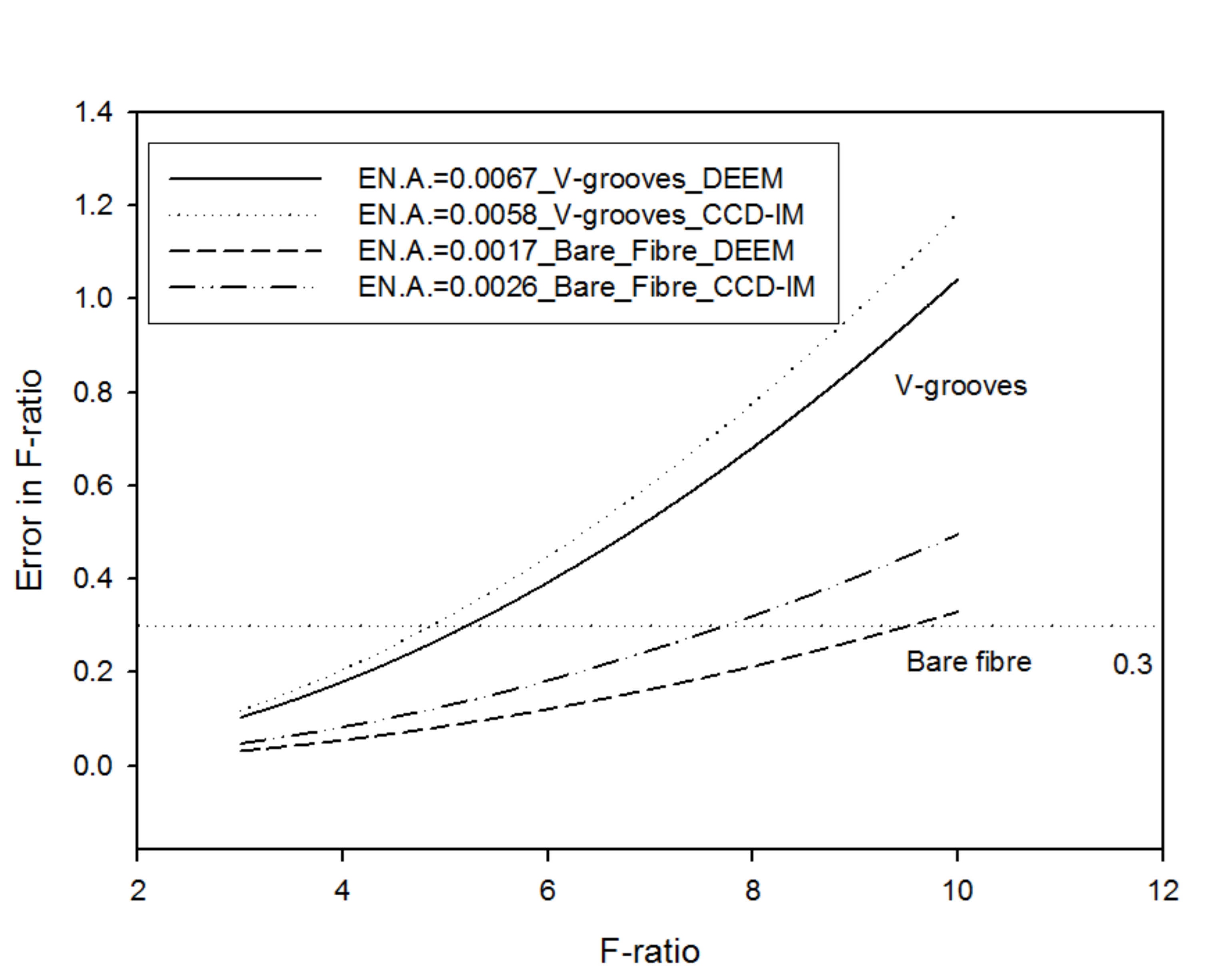}
      \caption{The maximum errors in F-ratios for fibre slit in V-grooves and a bare fibre. The errors in F-ratio decrease with decreasing input focal ratio. The errors for a bare fibre are suppressed and much smaller than fibre slit since no contributions come from the uncertainties of angles and positions in the V-grooves.}
         \label{fig:ena}
   \end{figure}

In the FRD measurement for the fibre slit, the errors of the angles and the positions of the fibres on the V-grooves are the major contributions. And the alignment uncertainties can be significantly reduced for a bare fibre by RIM. The error in F-ratio needs the conversion from E$N.A.$ to focal ratio according to equation (\ref{eq:sinf}) as shown in Fig. \ref{fig:ena}. The error value in F-ratio depends on the input condition and decreases with decreasing input focal ratio. In the whole range of the input focal ratio from 3.0 to 10.0, the output focal ratio is smaller than 7.5, so the maximum error for a bare fibre between DEEM and CCD-IM is less than 0.47 in F-ratio.

The error in F-ratio in Fig. \ref{fig:ena} is the maximum value of the two methods, which determines the uncertainty between the measured result and the theoretical value (or the pre-set value). And it is different from the error bar shown on the following plots like Fig. \ref{fig:16} (and similar). The error bar is evaluated by the standard deviation of multiple repeated measurements (no less than 5 times in our tests). So the error bar quantifies the dispersion of the measured results and describes the precision and the stability of the two methods which also depends on the input condition and has the similar trend as shown in Fig. \ref{fig:ena}. Therefore, the result is only valid if the difference between the two method is less than the maximum error in Fig. \ref{fig:ena}. For example, the E$N.A.$ of the input system is 0.001, so the pre-set input focal ratio of $F$/10.00 will lead the practical input beam to be $F_{in}$=10.00$\pm$0.02. The measured results of the output focal ratio in Fig. \ref{fig:21} are 7.32$\pm$0.09 and 6.98$\pm$0.16 from DEEM and CCD-IM, respectively. The error bars of 0.09 and 0.16 are the standard deviations of multiple measurements (5 times). And the difference of the results between the two methods is 0.34 (7.32-6.98=0.34) in F-ratio. And the maximum difference in Fig. \ref{fig:ena} is 0.43 (0.18+0.25=0.43, where 0.18 is the maximum error for the output focal ratio of 7.32 of DEEM and 0.25 for the output focal ratio of 6.98 of CCD-IM) larger than 0.34, which means the measured results of DEEM and CCD-IM are valid. If not, the data reduction should be checked and reprocessed, especially for the background subtraction. And this is one of the criteria to testify the feasibility of DEEM. Other criteria including the f-intercept, linearity of r-square, the increment of output focal ratio in Fig. \ref{fig:increment} and the shift of the focal point in equation (\ref{eq:fittingposition}) are also used to determine the confidence level of the measured results in the comparison of DEEM and CCD-IM.

\begin{table}
\caption{Error sources in DEEM and CCD-IM. The maximum errors are listed in the table and the uncertainties in F-ratios can be acquired according to equation (\ref{eq:sinf}). The E$N.A.$ of 0.0007 of V-grooves in the brackets is the maximum value among the 77 fibres (95 percent out of 81 fibres).}             
\label{tab:errorsummary}      
\centering                          
\begin{tabular}{c c c}        
\hline                 
   \multicolumn{1}{l}{Error source} & \multicolumn{2}{c}{Errors in E$N.A.$} \\
   & DEEM & CCD-IM \\
\hline
   \multicolumn{1}{l}{(1)Alignment of input end} & & \\
   \multicolumn{1}{l}{Travel stage} & \multicolumn{2}{c}{0.00033} \\
   \multicolumn{1}{l}{Aperture size of diaphragm} & \multicolumn{2}{c}{0.0003} \\
   \multicolumn{1}{l}{Centre position of diaphragm} & \multicolumn{2}{c}{0.0003} \\
   \multicolumn{1}{l}{Coupling point on the fibre end} & \multicolumn{2}{c}{0.00007} \\
   \\
   \multicolumn{1}{l}{(2)Alignment of output end} & & \\
   \multicolumn{1}{l}{Angle of V-grooves} & \multicolumn{2}{c}{<0.0041(0.0007)} \\
   \multicolumn{1}{l}{Centre of fibres in V-grooves} & \multicolumn{2}{c}{0.00011} \\
   \multicolumn{1}{l}{Focus point of fibre end} & 0.00045 & 0.0018 \\
   \multicolumn{1}{l}{Alignment of CCD} & - & 0.0064 \\
   & & (compensated) \\
   \\
   \multicolumn{1}{l}{(3) Other uncertainties} & & \\
   \multicolumn{1}{l}{Uncertainty of $C$, EE} & 0.0001 & - \\
   \multicolumn{1}{l}{Uncertainty of spot size} & - & 0.0007 \\
   \\
   \multicolumn{1}{l}{Total error of (1)$\sim$(3)} & <0.0058 & <0.0067 \\
   \multicolumn{1}{l}{Total error for a bare fibre} & <0.0017 & <0.0026 \\
   \\
   \multicolumn{1}{l}{Throughput} & & \\
   \multicolumn{1}{l}{Light source} & \multicolumn{2}{c}{<3.6 per cent} \\
   \multicolumn{1}{l}{Uncertainty of coupling efficiency} & \multicolumn{2}{c}{<2.8 per cent} \\
   \multicolumn{1}{l}{Power readings} & \multicolumn{2}{c}{<0.33 per cent} \\
\hline                                   
\end{tabular}
\end{table}

\subsection{Comparison of DEEM vs CCD-IM}
Before adopting DEEM for processing fibres in practice, two kinds of tests were conducted to make sure that the new method DEEM is at least no worse than CCD-IM. One kind of the tests is to check whether DEEM gives the same FRD measurements as the conventional method and the other one is to compare the accuracy and stability with the conventional method under the condition of a fixed output focal ratio.

Three types of fibres were prepared by cleaving and fixed into a fibre tube to easily handle the incident position and reduce the stress effect. The parameters of the fibres are shown in Table \ref{tab:4}. The stability of the light source indicates the status during the experiments. STB in the table stands for stable source and UNS for unstable source.

\begin{table*}
\caption{The parameters of three types of fibres. STB represents for stable light source and UNS for unstable light source.}             
\label{tab:4}      
\centering                          
\begin{tabular}{c c c c c}        
\hline                 
   Core & $N.A.$ & Length & Splitter & Light source (Stability) \\
\hline
   320$\mu$m & 0.22$\pm$0.02 & $\sim $20m & SP1,SP2 & Laser(STB$/$UNS),LED(STB$/$UNS) \\
   125$\mu$m & 0.22$\pm$0.02 & $\sim $20m & SP1 & Laser(STB),LED(STB) \\
   50$\mu$m & 0.22$\pm$0.02 & $\sim $20m & SP1 & Laser(STB) \\
\hline                                   
\end{tabular}
\end{table*}

According to the principle of DEEM, the corrected ratio of $k$ corresponding to a certain EE ratio in equation (\ref{eq23}) should preset in the console panel in order to measure the diameters of output light field. Table \ref{tab:5} shows the pre-set value of $k$ for specific EE ratios.

\begin{table}
\caption{The pre-set ratio of $k$ within a specific EE.}             
\label{tab:5}      
\centering                          
\begin{tabular}{c c c c c c}        
\hline                 
   Light source & Wavelength & \multicolumn{2}{c}{SP1} & \multicolumn{2}{c}{SP2} \\    
   & & EE90 & EE95 & EE90 & EE95 \\
\hline
   Laser & 632.8nm & 1.08 & 1.14 & 0.99 & 1.04 \\
   & 532nm & 1.13 & 1.19 & & \\
   LED & & 1.01 & 1.07 & & \\
\hline                                   
\end{tabular}
\end{table}

\subsubsection{Feasibility of DEEM}
\begin{figure}
   \centering
   \includegraphics[width=\hsize]{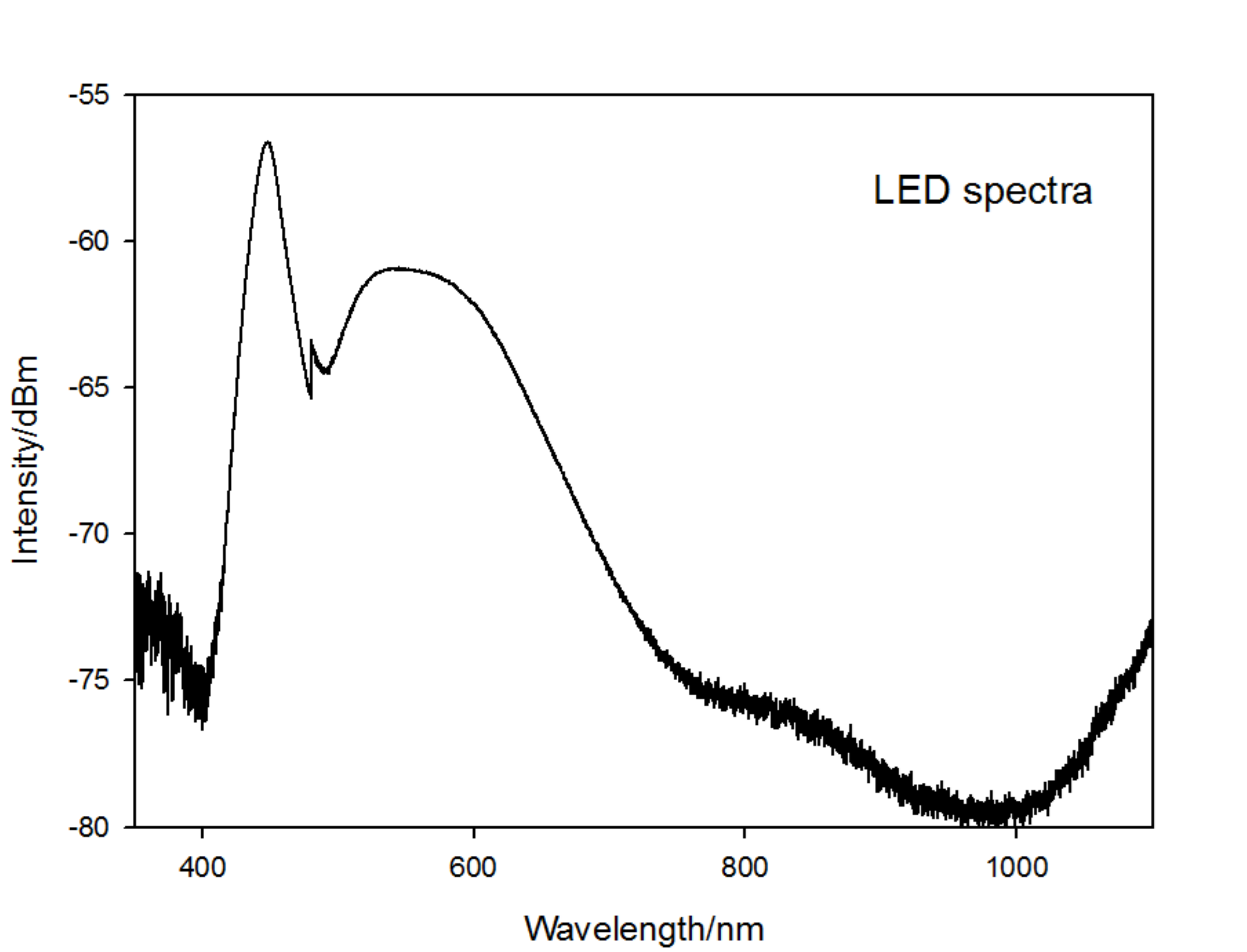}
      \caption{The spectra of LED. It covers the wavelength from 350nm to 1100nm. The peaks locate at 447.7nm and 548.2nm.}
         \label{fig:ledspectra}
   \end{figure}

\begin{figure}
   \centering
   \includegraphics[width=\hsize]{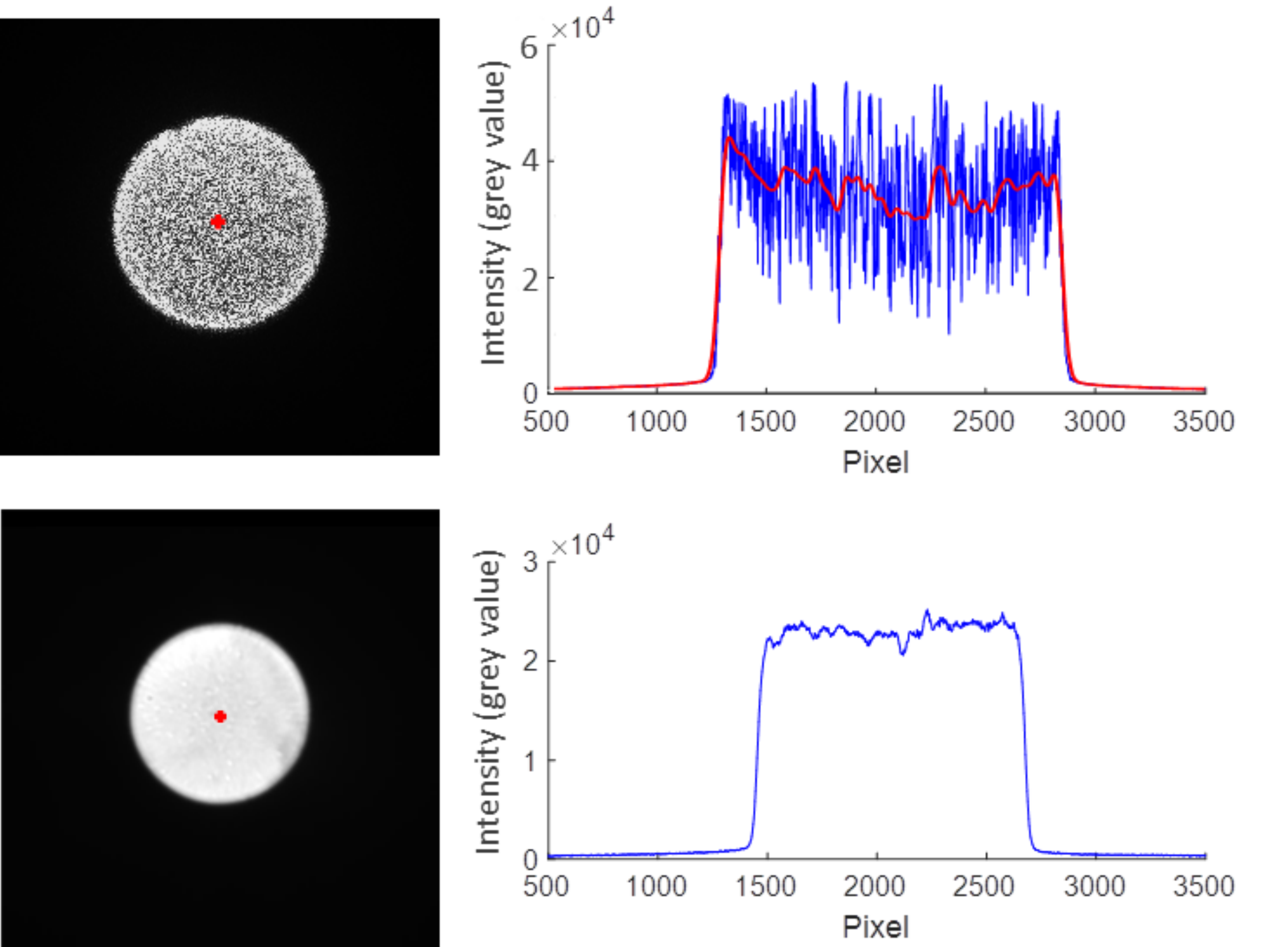}
      \caption{The input spot and the profile across the barycentre. The spot is recorded between the lens L2 and the fibre end. The speckle patterns bring huge fluctuations on the profile. The light of LED is a smooth flat function.}
         \label{fig:inputspot}
   \end{figure}

The tests were implemented on three types of fibres with two kinds of laser sources and a LED. Fig.\ref{fig:ledspectra} shows the LED spectrum and it covers the required wavelength range from 400nm to 900nm. The peaks of intensity occur at the wavelengths of 447.7nm and 548.2nm. The input focal ratio is controlled by the electric-driven diaphragm and the profile of the input spot is shown in Fig.\ref{fig:inputspot}. The input spot of laser has serious speckle patterns. On the contrary, the spot of LED is much smoother which can be treated as a flat function.

\begin{figure}
   \centering
   \includegraphics[width=\hsize]{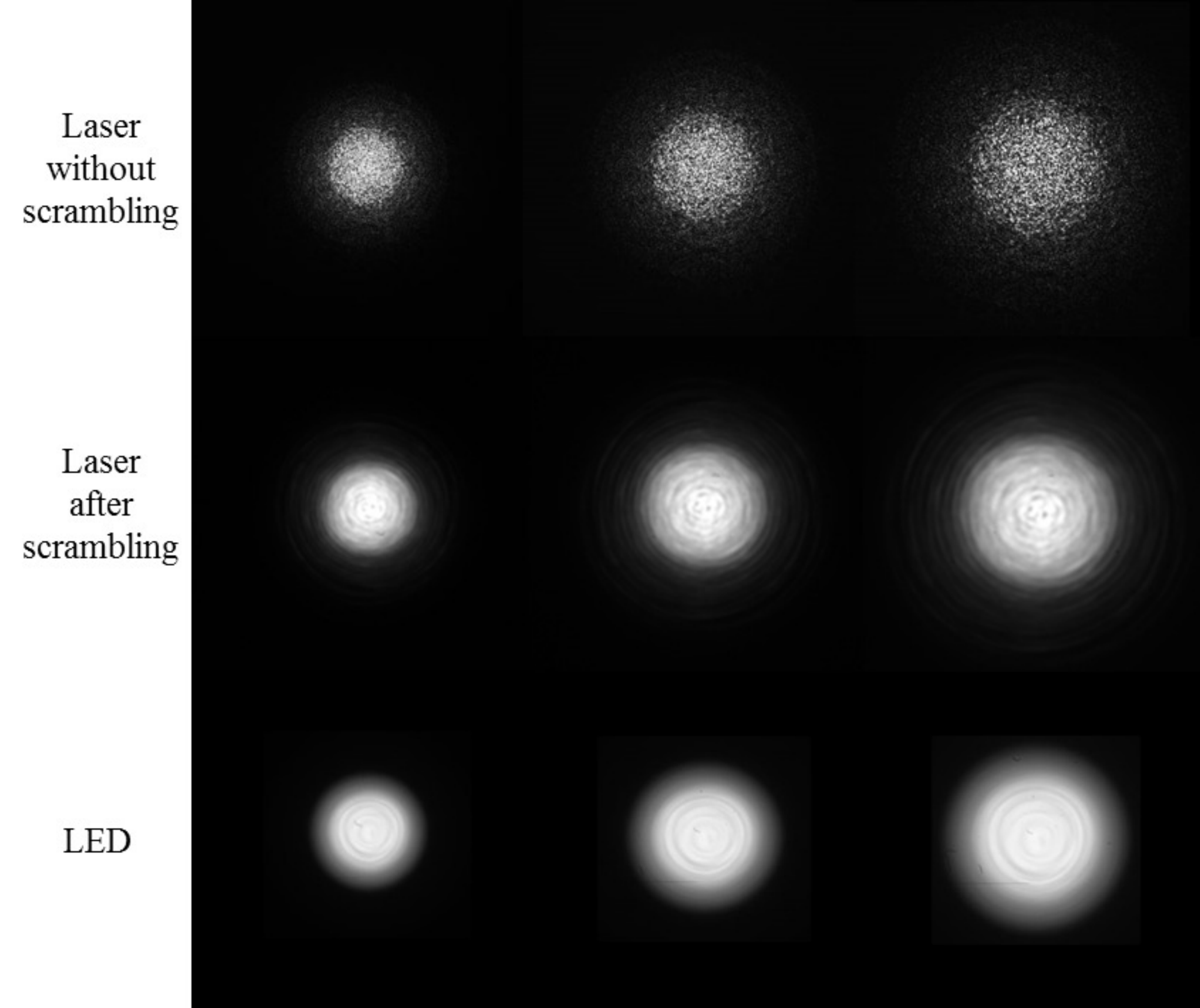}
      \caption{The output spots captured by CCD. The upper three images are spots without scrambling under the laser illumination. The middle three images are spots with scrambling at 65Hz. The bottom three images are spots under the LED illumination.}
         \label{fig:14}
   \end{figure}

\begin{figure*}
   \centering
   \includegraphics[width=\hsize]{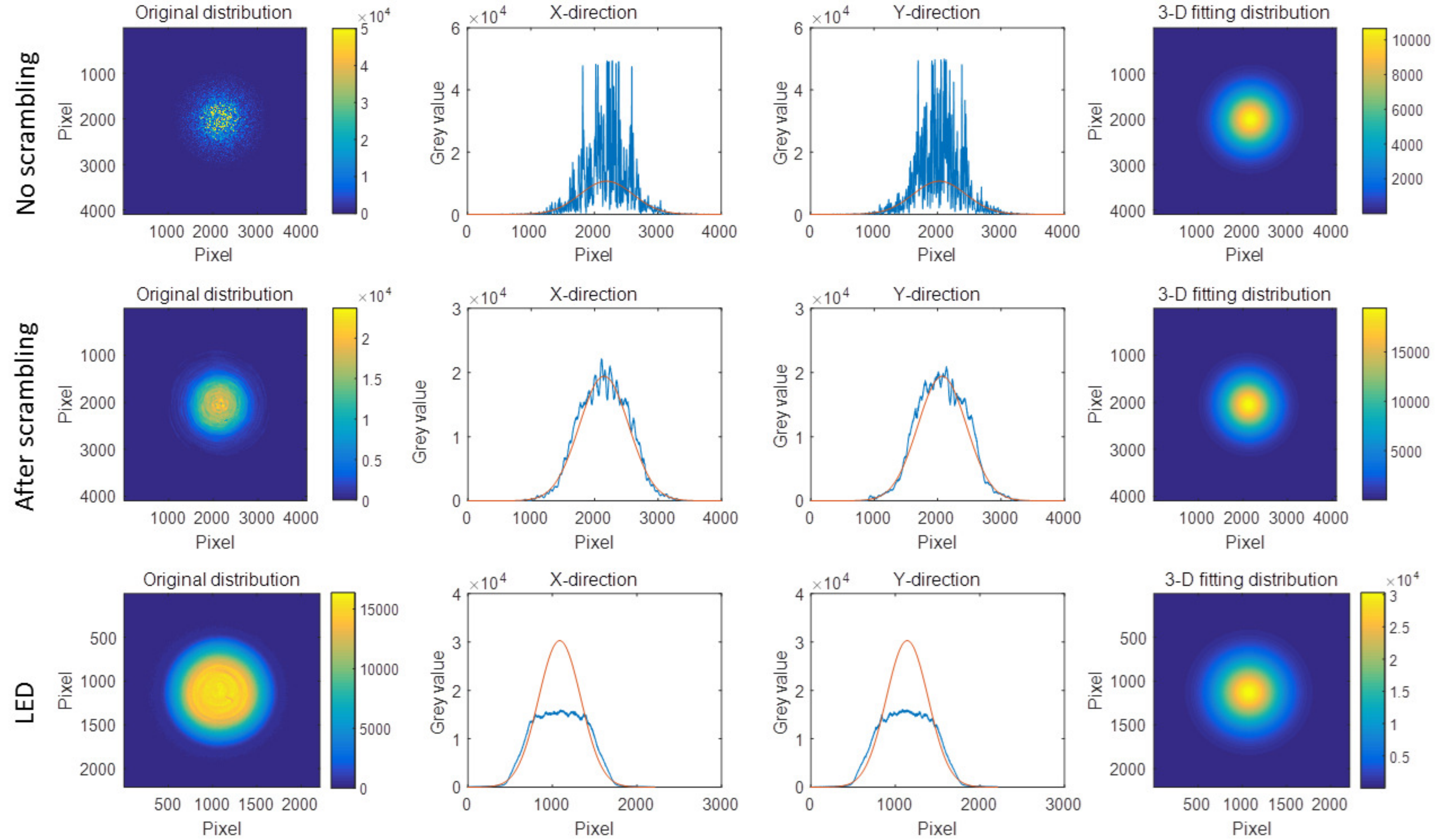}
      \caption{The profile of the output spots under the conditions of different input light sources. The colour bar shows the scale of the intensity (grey value in the image). Laser speckle can bring offset to the fitted centre point. After scrambling, the spot becomes smooth and the fitting surface is much closer to the original distribution. The profile of the output spot illuminated by LED is more like a broadened top-hat distribution. The 3D fitting method can still locate the centre position. Apparently, the EE ratio must be done with the original power distribution since not all the output spots are Gaussian-like.}
         \label{fig:15}
   \end{figure*}

The input focal ratio was set from $F_{in}$=3.0 to $F_{in}$=10.0. The output focal ratio was determined with EE90 and EE95 by both of the two methods. To suppress laser speckle effects, a vibrating scrambler is introduced into the system in order to smooth the output spots when the system is illuminated by a laser in CCD-IM. The output spots captured by CCD are partially illustrated in Fig.\ref{fig:14}. We use 3D fitting method to locate the barycentre. In the fitting curve of Fig.\ref{fig:15}(a), the noise caused by laser speckle makes very serious fluctuation on the profile. After scrambling, the spot is well smoothed and the profile across the barycentre is a Gaussian-like distribution. When illuminated by LED, the output spot appears to be a top-hat function blended with two Gaussian wings. But the three dimensional fitting surface can give a Gaussian function that matches well around the boundary. So the 3D fitting method can accurately determine the centre point to calculate the EE ratio. And this is why it must be pointed out that the fitting surface is to help determine the barycentre but not to measure the EE ratio and the diameter within a specific EE ratio is measured on the original power distribution as discussed in Section 2.4.

In DEEM system, the diameter is determined similarly by weighting averaging. When the diaphragm is moved to the pre-set position on the translation stage, the ratio $k$ increases with the expanding aperture size $D$ of the diaphragm. Six pairs of $k$ around the theoretical value and the corresponding diameter of $D$ are recorded.
\[\left[ {\begin{array}{*{20}{c}}
{{k_1}}&{{k_2}}&{{k_3}}&{{k_4}}&{{k_5}}&{{k_6}}\\
{{D_1}}&{{D_2}}&{{D_3}}&{{D_4}}&{{D_5}}&{{D_5}}
\end{array}} \right]\]
Then the diameter is averaged according to equation (\ref{eq:kave}):
\begin{equation}\label{eq:kave}
{k_{ave}} = \frac{1}{n} \cdot \frac{{\sum\limits_{j \ne i} {\left( {{k_j} - k} \right){k_i}} }}{{\sum {\left( {{k_i} - k} \right)} }}
\end{equation}
where $k$ is the pre-set value.

\begin{figure*}
   \centering
   \includegraphics[width=\hsize]{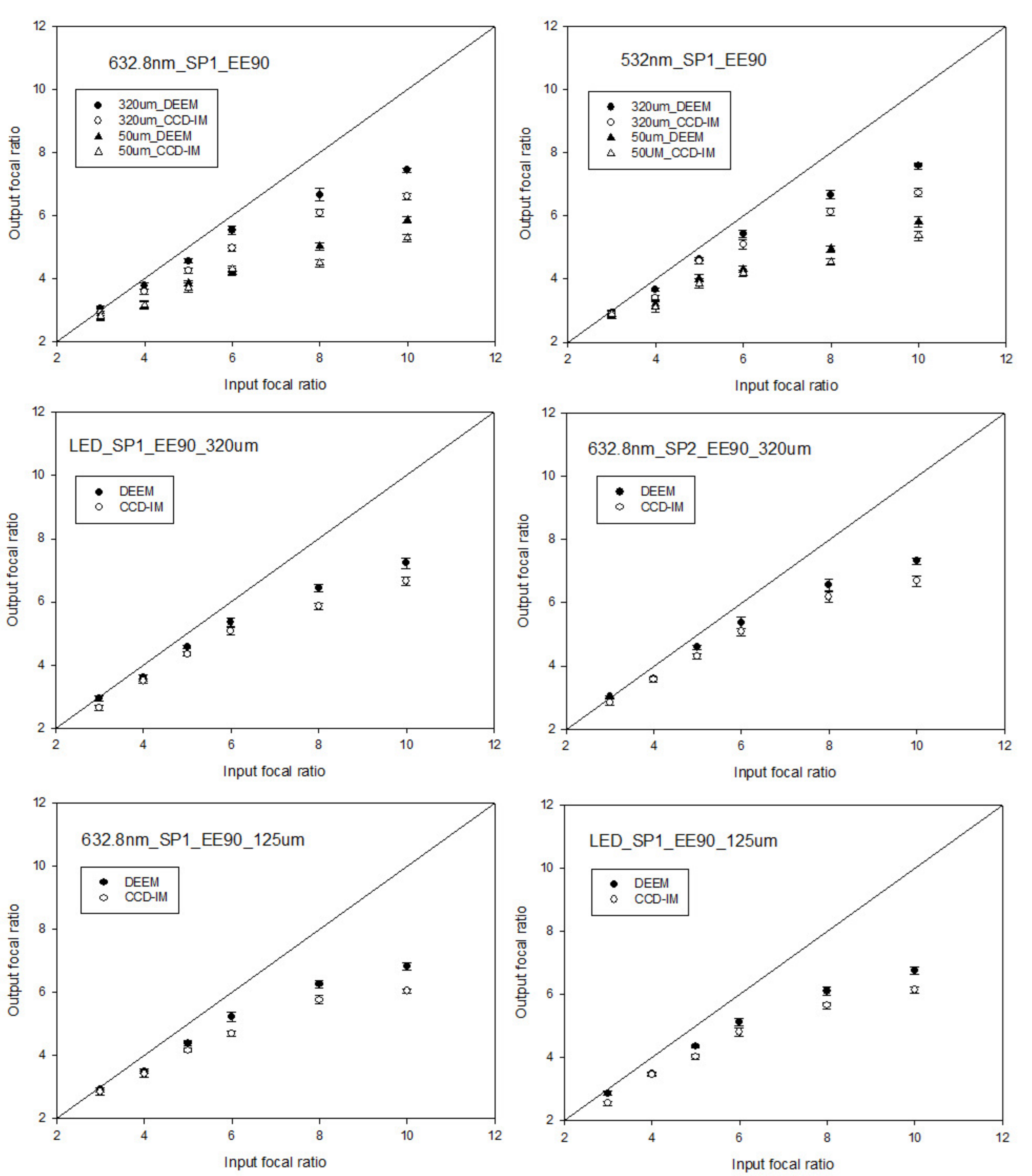}
      \caption{Comparison of the output focal ratio of DEEM and CCD-IM determined by EE90. When the input focal ratio is smaller than 5.0, the output focal ratios are almost the same within the error. But the difference between the two methods reaches to the maximum of 0.9 in F-ratio for input focal ratio larger than 5.0. In the first two images on the upper panel, the FRD dependence on core size can be seen that higher output focal ratio occurs for larger core fibre. But the core size dependence becomes much weaker between the results of 320$\mu$m and 125$\mu$m core fibres.}
         \label{fig:16}
   \end{figure*}

\begin{figure*}
   \centering
   \includegraphics[width=\hsize]{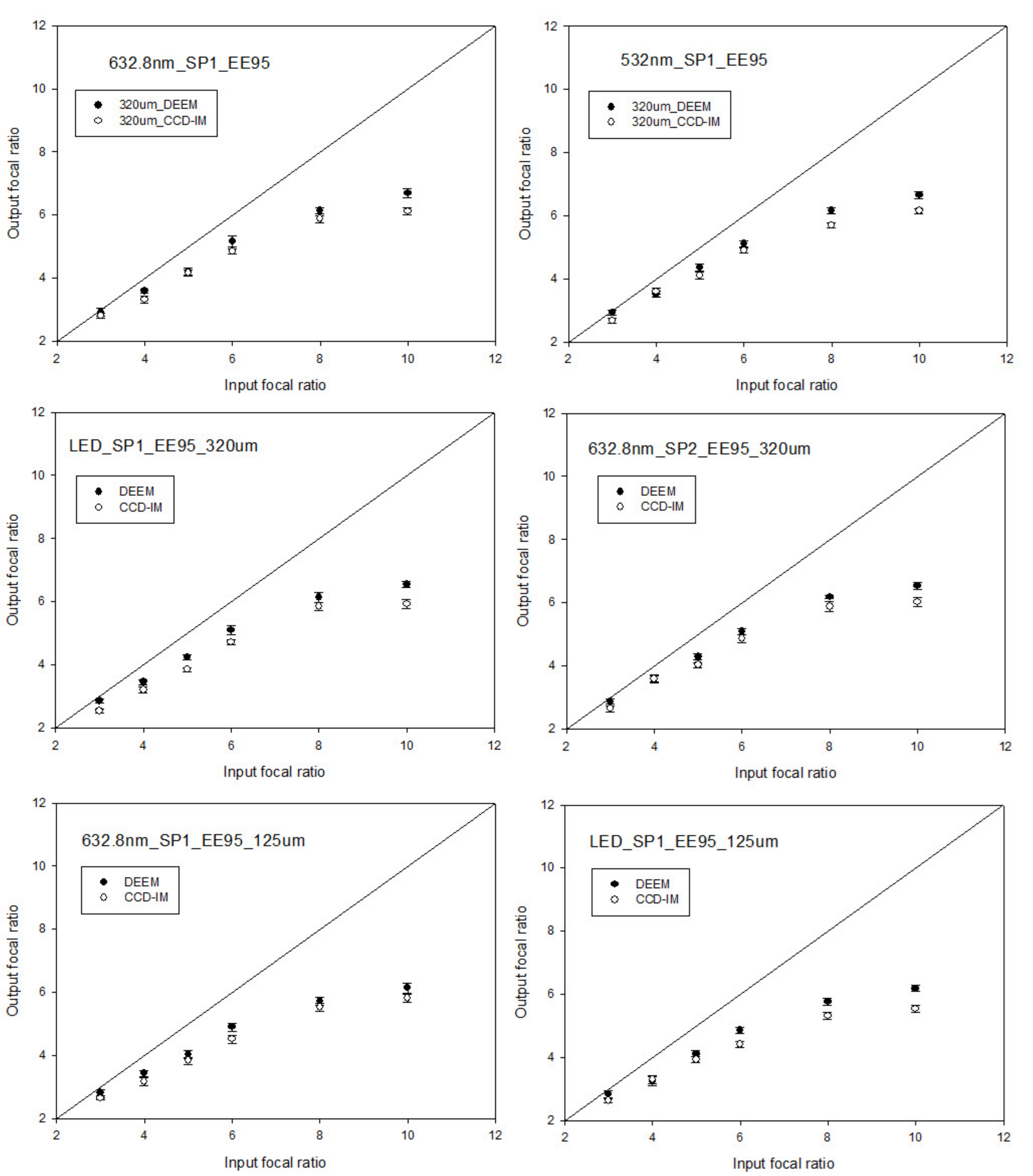}
      \caption{Comparison of the output focal ratio of DEEM and CCD-IM determined by EE95. The difference is also larger than the error when the input focal ratio is higher than 5.0.}
         \label{fig:17}
   \end{figure*}

The results derived from DEEM and CCD-IM in the cases of EE90 and EE95 are shown in Fig.\ref{fig:16} and Fig.\ref{fig:17}, respectively. When the input focal ratio is faster than 5.0, both methods give the same FRD measurements and the output focal ratios are well matched within the error bars. For slower input focal ratios of $F_{in}$=5.0$\sim$10.0, the differences between the two methods become larger and the output focal ratio of DEEM is larger than that of CCD-IM with the difference of 0.6$\sim$0.9 in F-ration.

\begin{figure}
   \centering
   \includegraphics[width=\hsize]{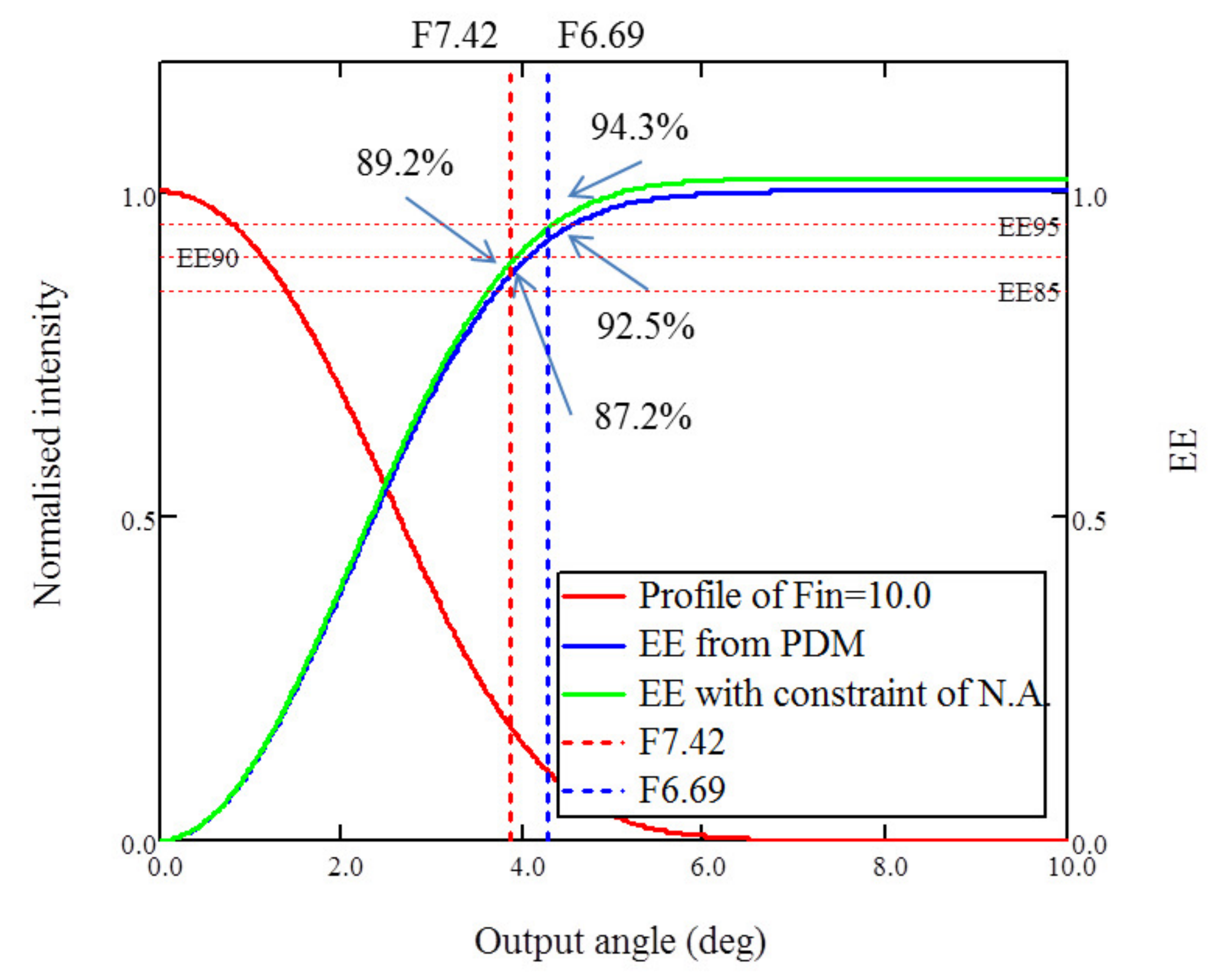}
      \caption{Comparison of the output focal ratio and EE ratio of DEEM and CCD-IM. The EE ratio of green line is corrected with the constraint of $N.A.$. F7.42 is measured by DEEM and F6.69 is from CCD-IM. The EE ratios after correction with $N.A.$ constraint are 89.2 per cent and 94.3 per cent for DEEM and CCD-IM, respectively. So the result of DEEM is closer to the pre-set value of EE=90 per cent.}
         \label{fig:eewithinna}
   \end{figure}

Take the results in Fig.\ref{fig:16} for instance, the output focal ratios derived from DEEM and CCD-IM are 7.42$\pm$0.09 and 6.69$\pm$0.12, respectively, for the input focal ratio of $F_{in}$=10.0 within EE90. The larger output focal ratio in the results indicates that the diameters of the output spots of DEEM are smaller than that of CCD-IM. According to the PDM, the theoretical EE ratios are 87.2 per cent for output focal ratio $F_{out}=7.42$ and 92.5 per cent for $F_{out}=6.69$. In the experiment, the total output power is integrated within the area limited by numerical aperture $N.A.$ of the fibre. So the simulation with the constraint of $N.A.$ is corrected and shown in green line in Fig.\ref{fig:eewithinna}. The results of simulation show that the EE ratios are corrected to 89.2 per cent and 94.3 per cent. Compared with the required EE ratio of EE90, DEEM is closer to the theoretical value than CCD-IM. On the other hand, the difference of the output focal ratio between DEEM and CCD-IM reaches to 0.73(7.42-6.69=0.73) which is larger than the maximum error of 0.41 (0.18+0.23=0.41, where 0.18 is the maximum error for the output focal ratio of 7.42 of DEEM and 0.23 for the output focal ratio of 6.69 of CCD-IM according to Fig. \ref{fig:ena}). Therefore the results are not valid and the measurements need to be modified for accurate background subtraction.

Although the difference of the output focal ratio is small within the error when the input focal ratio is <5.0, an apparent trend occurs that the average value of the output focal ratio of DEEM is larger than that of CCD-IM. An assumption is proposed that the inconsistence of DEEM and CCD-IM is mainly biased by the background noise. Since the setup and the data processing are different in the two methods, the background noise is divided into three types in order to separate the different kinds of noises in the two methods.

Type I. The dark current and Poisson noise. The detectors including the power meters and the CCD camera are working with cooling system and the temperature is set to 0$^\circ$C to reduce the noise.

\begin{figure}
   \centering
   \includegraphics[width=\hsize]{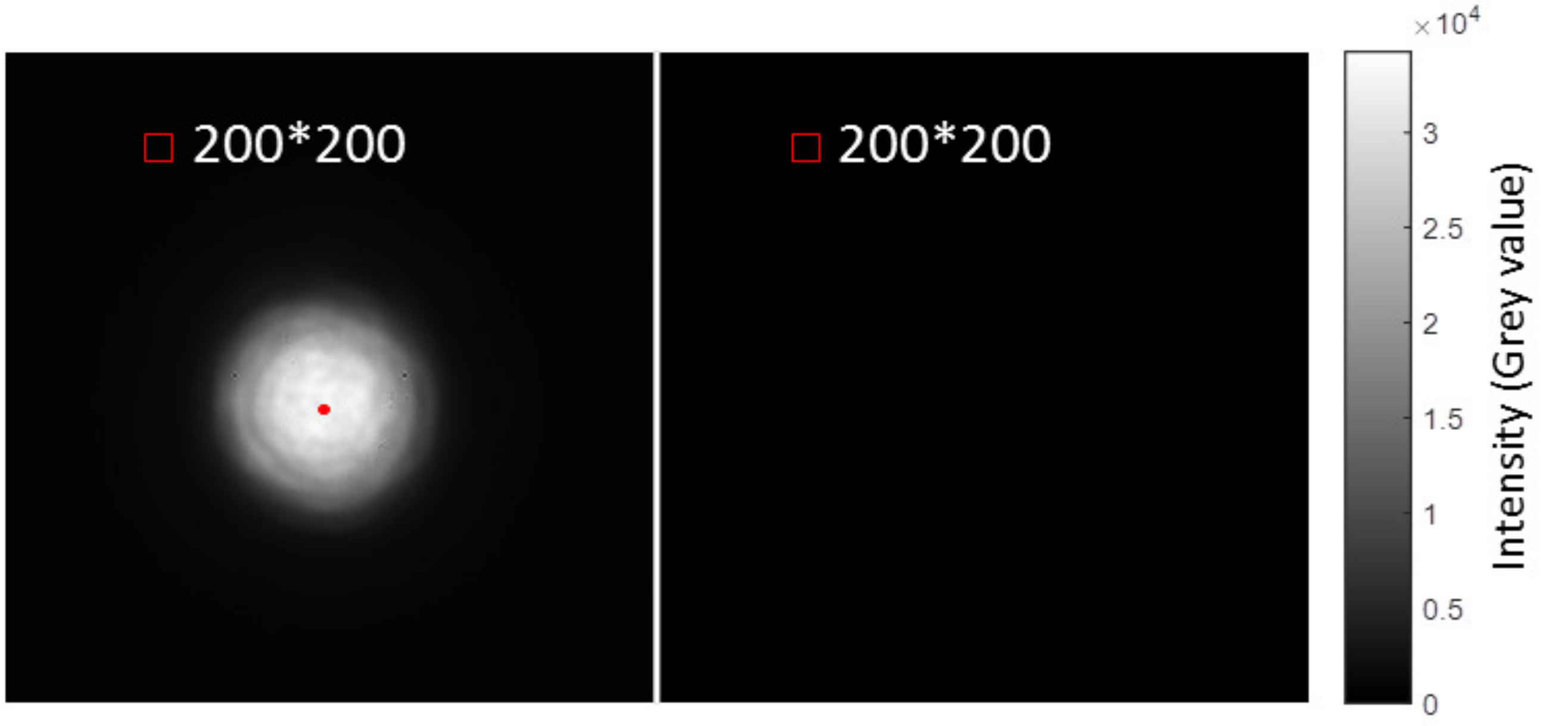}
      \caption{The output spot (left) and the corresponding dark image (right). Due to the large scale of the grey value, the emission in the dark image is hard to see. An area of 200$\times$200 pixels is magnified to reveal the noise of ambient light as shown in Fig. \ref{fig:residuallight}.}
         \label{fig:dk}
   \end{figure}

Type II. The ambient light. Though the detectors including the CCD camera and the power meters are covered with a box to reduce the influence, still the dark image with light source being obstructed by a black plate is taken in advance to be used as the background of ambient light as shown in Fig. \ref{fig:dk}. Then the background of each image of the output spots is subtracted using the corresponding dark image. For the power meter, the zero checking is to eliminate the ambient light.

\begin{figure*}
   \centering
   \includegraphics[width=\hsize]{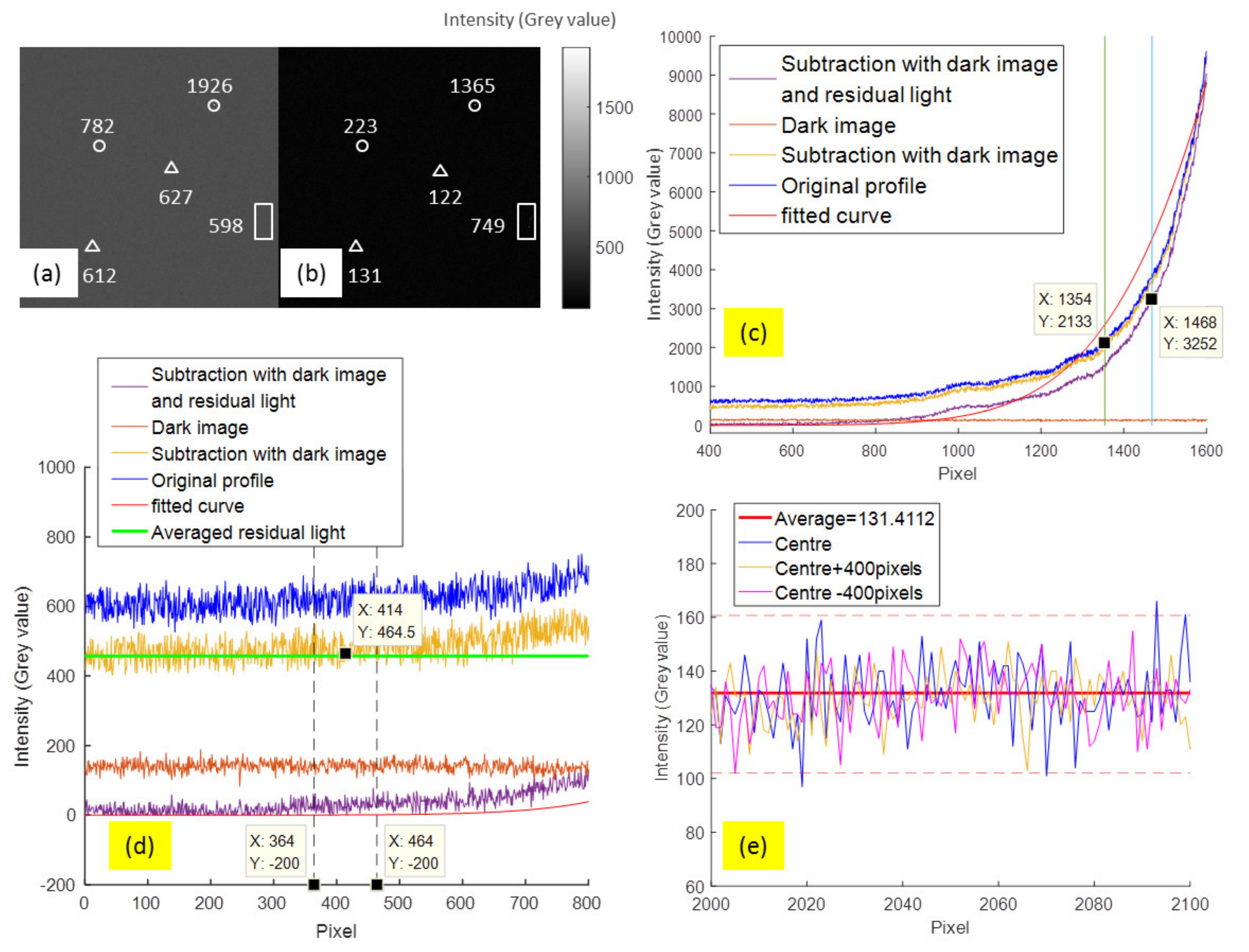}
      \caption{The residual light after the subtraction of the dark image and the magnified view of the red boxes of the output spot (a) and the dark image (b). Triangles in (b) are normal noise in dark image of type II. Some values of the circles in (b) are abnormal emission points and they exist in both the dark image and the output spot. These points can be subtracted. But the noise in the rectangle is different from the others. Several bright points of different grey values which are larger than the normal noise of type II exist in these rectangles in the dark image (b), but disappear in the output spot (a). And some of these bright points in other places behave in a opposite way that they show up in (a) and vanish in (b). These points are supposed to be the cosmic rays and spread randomly in CCD. For these points, the deduction of the dark image will be over or under subtracting the background. But it is negligible compared with the residual light in the 3D fitting method to determine the EE ratio. (c): The fitted curve derived from EE90 and EE95 in CCD-IM. (d): The determination of residual light: the average value of the integral grey value within the ring area of pixels from 364 to 464. Pixel position of 414 is calculated by equation (\ref{eq28}). (e): The distribution of the dark image (partially illustrated). Although the residual light value of each pixel is very small compared to the counts in the centre area, the integral sum cannot be ignored and it is not trivial to evaluate the EE ratio.}
         \label{fig:residuallight}
   \end{figure*}

Type III. The residual light. This part of background is the remnant in the data analysis of CCD-IM after the subtraction of the corresponding dark image. The residual light exists in the image as shown in Fig.\ref{fig:residuallight}. The subtraction of residual light is important for the determination of the diameter of the output spot. The existence of the residual light will enlarge the output spot size when the indicator of the diameter reaches to the pre-set EE ratio in the program code.

The noise of type I and type II exist in both of DEEM and CCD-IM and can be recorded and subtracted. For type III, the subtraction of dark image is not sufficient to eliminate the influence in CCD-IM. And in the experiments, we notice that the residual light became larger when the the fibre was coupled with higher input light power and longer exposure time. And the distribution in Fig.\ref{fig:residuallight} shows that the residual light is spreading uniformly. So it is supposed to be the diffuse reflection in the box. Since the dark image is recorded with the light being obstructed, the light of the diffuse reflection is not included in the type II.

In DEEM system, the two detectors can record both of the reflective light and the transmitted light simultaneously. Assuming that the proportion of the diffuse reflection is $\eta $ of the output power $I_o$, equations (\ref{eq14}) and (\ref{eq15}) can be written as:
\begin{equation}\label{eq:irh}
{I_r} = {I_o} \cdot f\left( {T,R,A} \right) \cdot {b_2} + {I_o} \cdot \eta
\end{equation}
\begin{equation}\label{eq:ith}
{I_t} = {I_o} \cdot g\left( {T,R,A} \right) \cdot EE \cdot {b_1} + {I_o} \cdot \eta
\end{equation}
Then the EE is set to 100 per cent to measure the value of constant $C$:
\begin{equation}\label{eq:cwithita}
C = \frac{{g\left( {T,R,A} \right) \cdot {b_1} + \eta }}{{f\left( {T,R,A} \right) \cdot {b_2} + \eta }}
\end{equation}
For a specific EE ratio in FRD measurement, the recorded ${I_r}^,$ and ${I_t}^,$ are as follows:
\begin{equation}\label{eq:irhr}
{{I_r}^,} = {I_o} \cdot f\left( {T,R,A} \right) \cdot {b_2} + {I_o} \cdot \eta
\end{equation}
\begin{equation}\label{eq:ithr}
{{I_t}^,} = {I_o} \cdot g\left( {T,R,A} \right) \cdot EE \cdot {b_1} + {I_o} \cdot \eta
\end{equation}
The ratio of $k^,$ with the diffuse reflection in equation (\ref{eq18}) is written as:
\begin{equation}\label{eq:khr}
\begin{split}
& {k^,} = \frac{{{I_t}^,}}{{{I_r}^,}}\\
& \quad  = \frac{{g\left( {T,R,A} \right) \cdot {b_1} \cdot EE + \eta }}{{f\left( {T,R,A} \right) \cdot {b_2} + \eta }}\\
& \quad  = C \cdot EE + \eta  \cdot \left( {1 - EE} \right)
\end{split}
\end{equation}
Generally, the ratio of $\eta$<0.1 and the encircled energy ratio of EE>90 per cent, and $\eta  \cdot (1-EE)\ll 1$, so that equation (\ref{eq:khr}) can be approximated to be
\begin{equation}\label{eq:kh}
{k^,} = C \cdot EE + \eta  \cdot \left( {1 - EE} \right) \approx C \cdot EE = k
\end{equation}
The ratio $k^,$ with the diffuse reflection is still the same within the relative error less than 1 per cent compared with the theoretical $k$. Then the difference in the output focal ratio is less than 1 per cent in FRD according to the simulation results in Fig.\ref{fig:12}. The influence of the noise of type III is suppressed by the bi-detector design.

In CCD-IM, the EE ratio is derived from the images of output spots. Assuming that EE is the pre-set value (90 per cent or 95 per cent), $EE_1$ is the measured value of the spot, and $EE_2$ is the corresponding ratio of the residual light. Then the EE ratio is written as follows:
\begin{equation}\label{eq:eeccdim}
EE = \frac{{{I_o} \cdot E{E_1} + {I_o} \cdot \eta  \cdot E{E_2}}}{{{I_o} + {I_o} \cdot \eta }}
\end{equation}
Then the difference $\Delta EE$ between the measured value and the pre-set ratio is
\begin{equation}\label{eq:eedifference}
\Delta EE = E{E_1} - EE = \left( {EE - E{E_2}} \right) \cdot \eta
\end{equation}
$EE_2$ is the percentage of the encircled residual light. As the residual light is uniform in the assumption, the corresponding fraction of the residual light $EE_2$ equals to the proportion of the encircled area with respect to the total area of $N.A.$. According to the geometric relationship and equation (\ref{eq7}), (\ref{eq8}) and (\ref{eq19}), the maximum ratio of $EE_2$ is given by
\begin{equation}\label{eq:ee2max}
E{E_{2\max }} = \frac{{\pi  \cdot {{\left( {{{{f \mathord{\left/
 {\vphantom {f {2F}}} \right.
 \kern-\nulldelimiterspace} {2F}}}_{out}}} \right)}^2}}}{{\pi  \cdot {{\left( {f \cdot N.A.} \right)}^2}}} = \frac{1}{{{{\left( {2{F_{out}} \cdot N.A.} \right)}^2}}}
\end{equation}
For fibres with $N.A.$=0.22, the input focal ratio is close to the up limits of 3.0, so the maximum of $EE_2$ is 57 per cent. Substituting into equation (\ref{eq:eedifference}), the difference $\Delta EE$ is
\begin{equation}\label{eq:ee2sim}
\Delta EE = \left( {EE - E{E_2}} \right) \cdot \eta  > 0.3 \cdot \eta
\end{equation}
Equation (\ref{eq:ee2sim}) can also explain why the difference of the output focal ratio in fast input beams is smaller than that of slow beams. When a beam with the fast focal ratio is injected into the fibre, $F_{in}$=3.0 for instance, the difference of EE ratio $\Delta EE$ is small (around 0.3$\times$0.1=3.0 per cent in FRD, that is, the difference of 3$\times$3.0 per cent=0.09 in F-ratio), so the induced error in FRD is small, which is close to the simulation results of 2.7 per cent according to PDM in Fig.\ref{fig:12}. While the value of $\Delta$EE increases to more than 8 per cent when the input focal ratio is $F_{in}$=10.0, and the error in FRD will be more than 12 per cent (10.0$\times$12 per cent=1.2 in F-ratio), which is similar to the difference of 0.9 in F-ratio in the measured data.

\begin{figure}
   \centering
   \includegraphics[width=\hsize]{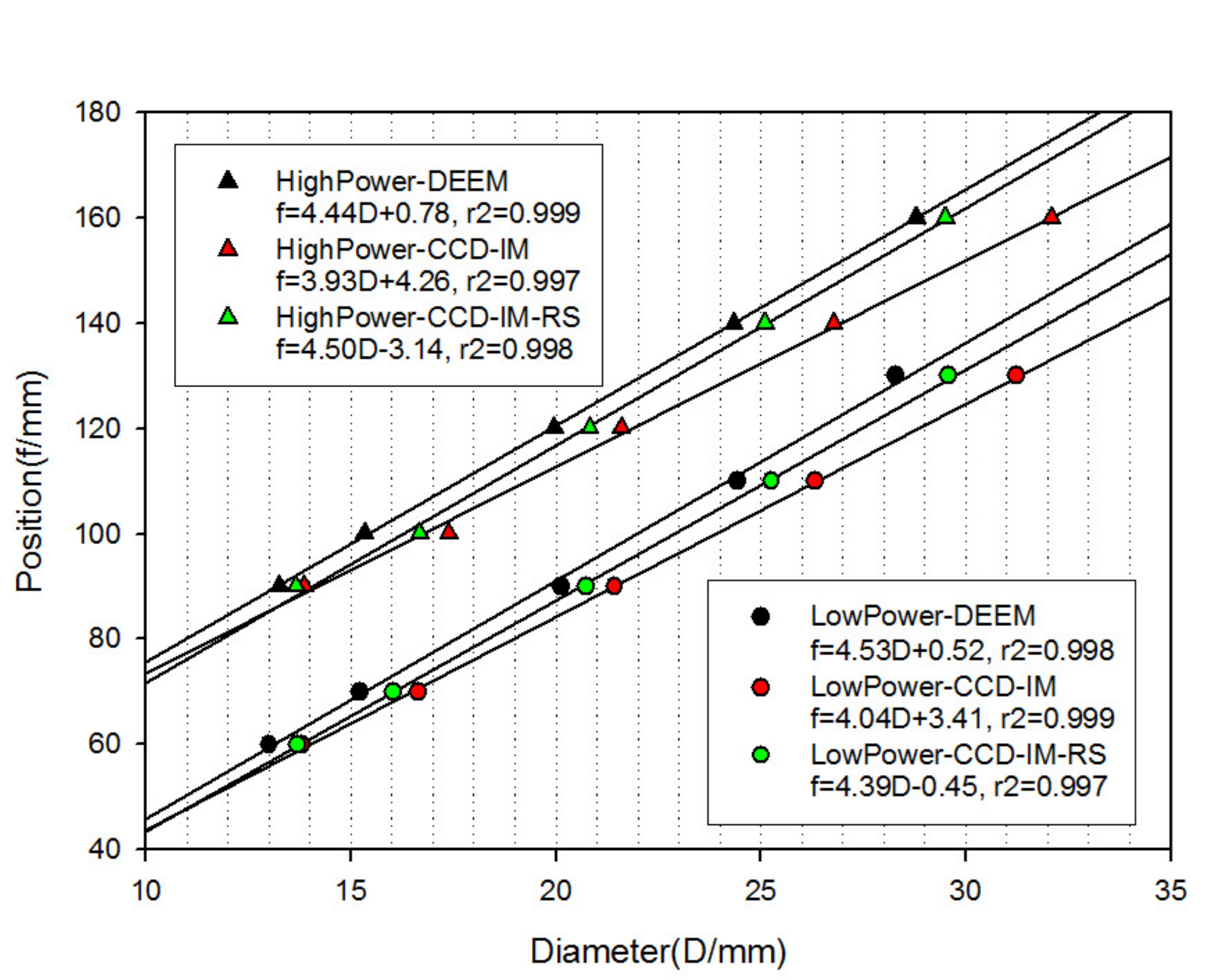}
      \caption{Comparison of the diameters of DEEM and CCD-IM in low and high input power. The two groups of diameters are sorted into high power and low power, and the results have the offset of 30mm on vertical axis for clarity. Whatever high or low input power, the diameter of DEEM is stable, and the diameter of CCD-IM is larger than that of DEEM. While the diameter of CCD-IM becomes smaller and is close to DEEM after the subtraction of residual light, which infers that the intensity of the input power is not the major factor that affect the spot size, otherwise we would find out some data that the diameter of CCD-IM should be smaller than that of DEEM. Thus the residual light which depends on the input intensity contributes to the changes of the diameters in CCD-IM.}
         \label{fig:diameteroflowandhighpower}
   \end{figure}

Technically, the spot size of the output spot should be the same no matter how much light is coupled into the fibre and the diameter should be independent of the input light power. However, the existence of the residual light changes the measured EE ratio and the diameter. Fig. \ref{fig:diameteroflowandhighpower} shows the measured diameters within EE90 of the output spots in different input power. The results show that the diameters of DEEM are stable and independent of the input power. While the spot sizes of CCD-IM increase with increasing input power. In other words, higher output power enlarges the diameter of the spot without the subtraction of the residual light, which is undesirable for measuring FRD. The slope of the fitting curve in Fig. \ref{fig:diameteroflowandhighpower} is the focal ratio. The intercept on $f$-axis ($f$-intercept, theoretical value is zero) indicates the offset between the fibre end and the fitting zero point, which is the smaller the better. Also we can see that the diameters of CCD-IM are larger than that of DEEM no matter the input power is high or low. With the correction of subtracting the residual light, the diameters of CCD-IM move close to DEEM, and so does the output focal ratio. This infers that the variation of the intensity of the input power is not the major cause that makes the spot sizes different (if so, there should be some diameters smaller than that of DEEM without the residual light subtraction, which is unseen from Fig.\ref{fig:diameteroflowandhighpower}), but the residual light has more contribution to the difference occurred in low and high power conditions.

Another problem that needs to pay attention is that the residual light varies in CCD-IM when the CCD camera is moved to different positions, while the power meters are fixed in the optical path in DEEM. To modify and correct the output focal ratio of CCD-IM, different residual light of every recorded image should be evaluated for subtraction to reduce the offset of the EE ratio and the diameter. Firstly, the boundary of a spot is evaluated according to the uncorrected focal ratio $F_{unc}$ and the uncorrected diameter $D_{unc}$ in CCD-IM. Secondly, the area of effective energy region limited by $N.A.$ is determined according to equation (\ref{eq28}):
\begin{equation}\label{eq28}
\left\{ {\begin{array}{*{20}{c}}
{{F_{unc}} = \frac{f}{{{D_{unc}}}}}\\
{{F_{\min }} = \frac{f}{{{D_{\max }}}}}\\
{{F_{\min }} = \frac{1}{{2 \cdot N.A.}}}
\end{array}} \right. \Rightarrow {D_{\max }} = 2 \cdot N.A. \cdot {F_{unc}} \cdot {D_{unc}}
\end{equation}
Finally, the total intensity of the output spot is integrated within the limited circle of $D_{max}$. For images of spots in different positions the residual light is averaged from the grey value integrated in the area of a ring limited by the boundary of $D_{max}\pm$50 pixels, and all the images were checked by eye one after another. However, even though the effective total energy in the limited region was reprocessed for a long and complicated procedure and checked by eye, the background is not for sure to be best fitted. The newly computed results of residual light, diameters, and output focal ratios are assembled in Table \ref{tab:6}. ORG represents the diameters before residual light subtraction and NEW for the diameters after residual light subtraction and RS for residual light subtraction.

\begin{table*}
\caption{The diameters and the output focal ratio after the residual light subtraction (RS, counts in grey value). The NEW diameters become smaller and the output focal ratio increases with 0.06$\sim$0.48 after the residual light subtraction. The difference between DEEM and CCD-IM become smaller and is less than 0.3 in F-ratio.}             
\label{tab:6}      
\centering                          
\begin{tabular}{c c c c c c c c c}        
\hline                 
   & EE90 & EE95 & \multicolumn{3}{c}{EE90} & \multicolumn{3}{c}{EE95} \\
   & & & ORG & NEW & RS & ORG & NEW & RS \\
\hline
   & \multicolumn{2}{c}{320$\mu$m$@$632.8nm$\_$DEEM} & \multicolumn{6}{c}{320$\mu$m$@$632.8nm$\_$CCD-IM} \\
   Diameter & 8.08mm & 8.68mm & 8.622mm & 8.082mm & 486.5 & 9.072mm & 8.604mm & 410.2 \\
   & 12.26mm & 13.19mm & 12.924mm & 12.330mm & 462.9 & 14.652mm & 13.230mm & 384.6 \\
   & 17.84mm & 19.71mm & 19.620mm & 18.666mm & 477.8 & 21.204mm & 19.692mm & 398.2 \\
   $F_{out}$ & 7.33 & 6.68 & 6.70 & 7.06 & & 6.21 & 6.69 & \\
   \\
   & \multicolumn{2}{c}{320$\mu$m$@$LED$\_$DEEM} & \multicolumn{6}{c}{320$\mu$m$@$LED$\_$CCD-IM} \\
   Diameter & 8.08mm & 9.34mm & 8.946mm & 8.874mm & 330.1 & 9.324mm & 8.946mm & 269.1 \\
   & 13.14mm & 13.92mm & 14.238mm & 13.266mm & 321.8 & 14.706mm & 13.536mm & 258.3 \\
   & 18.07mm & 19.55mm & 19.962mm & 18.522mm & 345.4 & 20.718mm & 20.790mm & 280.1 \\
   $F_{out}$ & 7.12 & 6.54 & 6.50 & 6.88 & & 6.27 & 6.33 & \\
\hline                                   
\end{tabular}
\end{table*}

   \begin{figure*}
   \centering
   \includegraphics[width=\hsize]{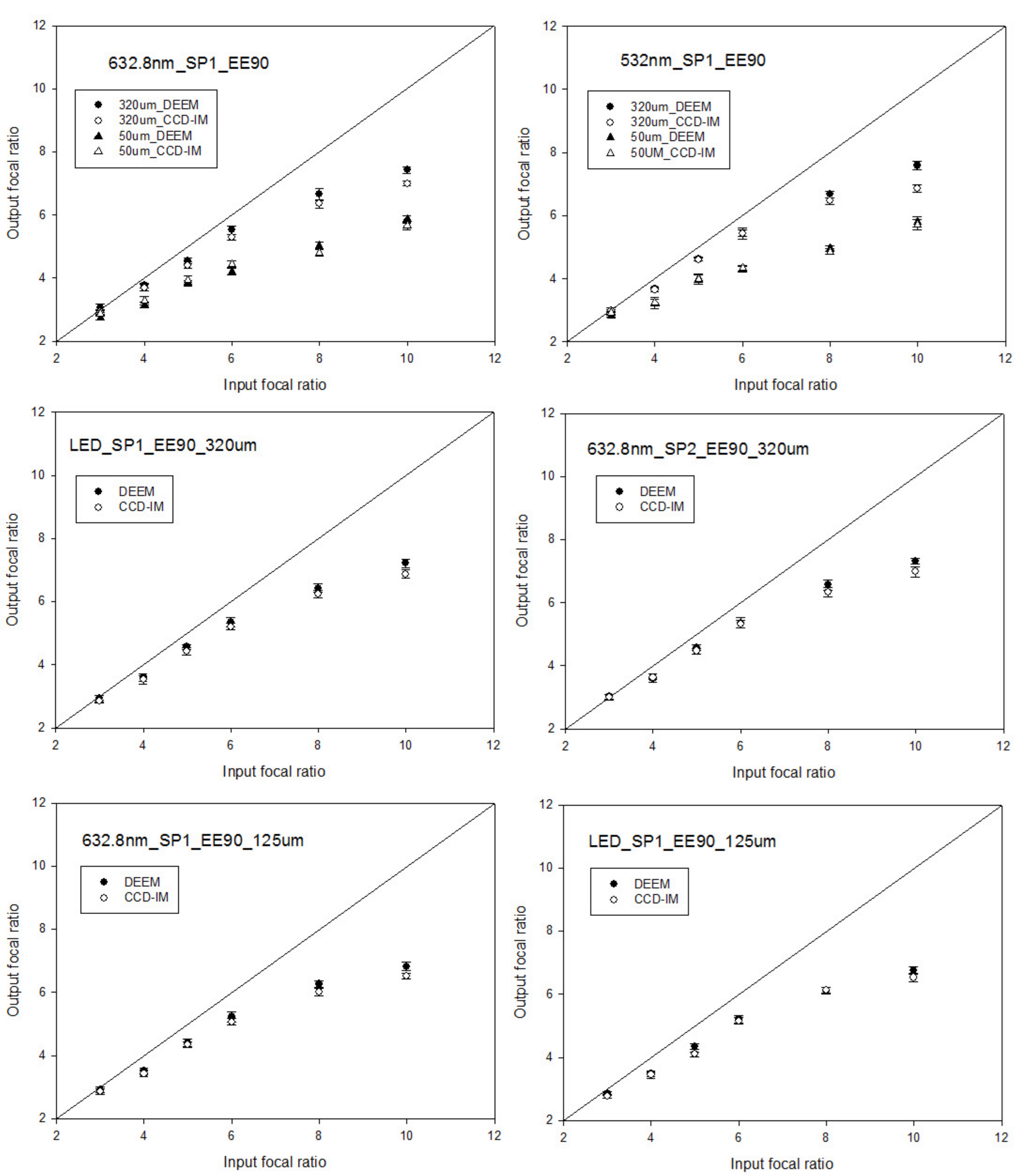}
      \caption{Comparison of the output focal ratio determined by EE90 of DEEM and the corrected results of CCD-IM after the residual light subtraction. When the input focal ratio is 3.0$\sim$6.0, the results are consistent with each other of the two methods within the error. And a small difference occurs that the output focal ratio of DEEM is larger than that of CCD-IM when the input focal ratio increases to 8.0$\sim$10.0. Still most of the output focal ratios locate within the acceptable range of 0.3 in F-ratio, which is less than the maximum error of E$N.A.$.}
         \label{fig:21}
   \end{figure*}

   \begin{figure*}
   \centering
   \includegraphics[width=\hsize]{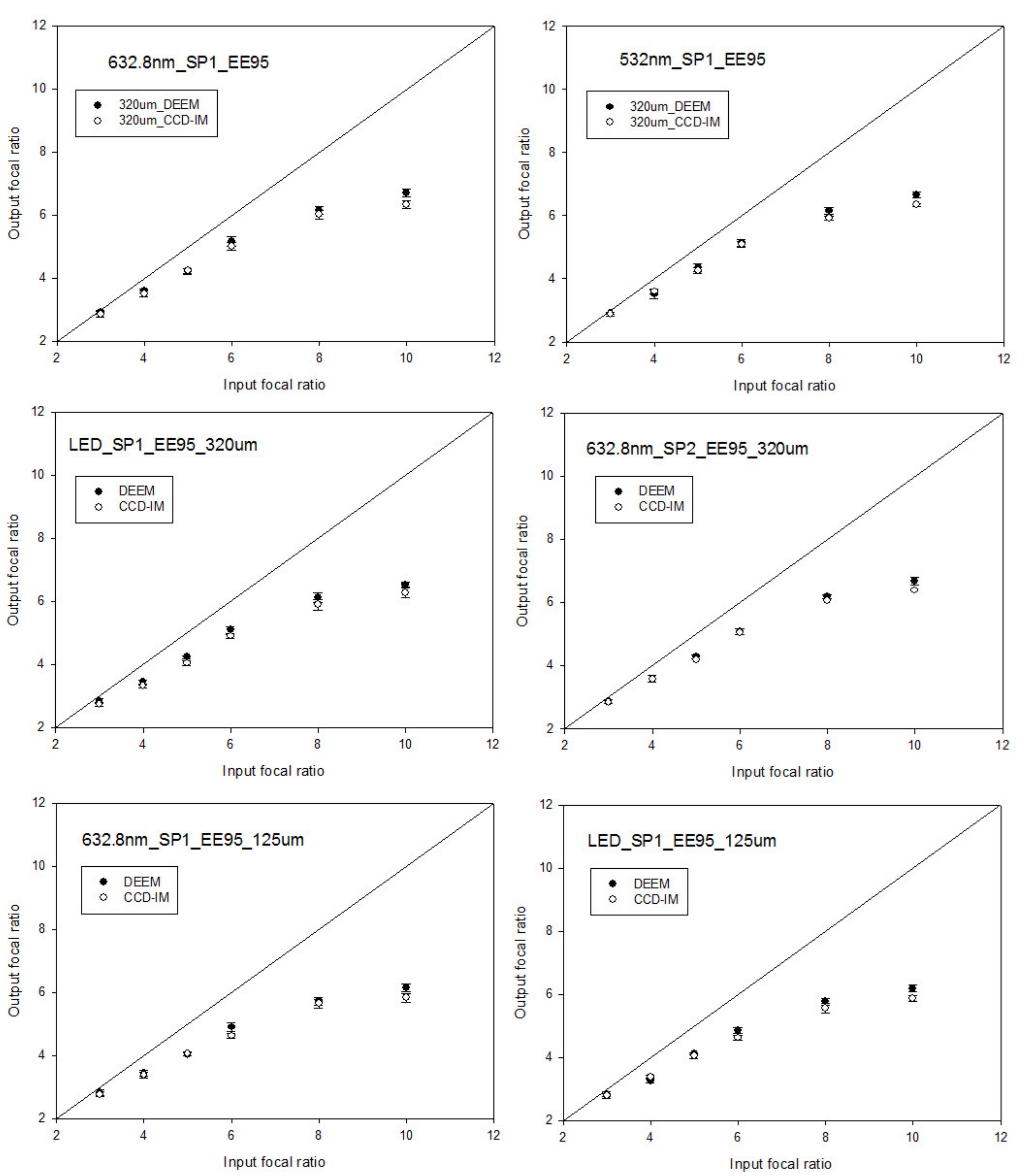}
      \caption{Comparison of the output focal ratio determined by EE95 of DEEM and the corrected results of CCD-IM after the residual light subtraction. And the consistence is also true within the error of 0.3 in F-ratio.}
         \label{fig:22}
   \end{figure*}

The differences of output focal ratios from DEEM and CCD-IM in Table \ref{tab:6} and Fig.\ref{fig:21} and Fig.\ref{fig:22} are significantly suppressed and this infers that the residual light and the noise indeed bias the diameters. But still the results of CCD-IM are smaller than that of DEEM and the difference is 0.1$\sim$0.3 in F-ratio, most of which are consistent within the errors. According to Table \ref{tab:6}, the background of the residual light cannot be simply subtracted using the same value in all the images, but a variable noise changing with the positions of CCD. In the residual light subtraction, it is treated as radially uniform noise, though the power distribution is not always smooth as shown in Fig. \ref{fig:residuallight}. After the correction, the output focal ratio is constrained in the acceptable range within the uncertainty of 0.3 in F-ratio. If a better model is constructed to characterise the residual light, the uncertainty can be further suppressed.

\begin{figure}
   \centering
   \includegraphics[width=\hsize]{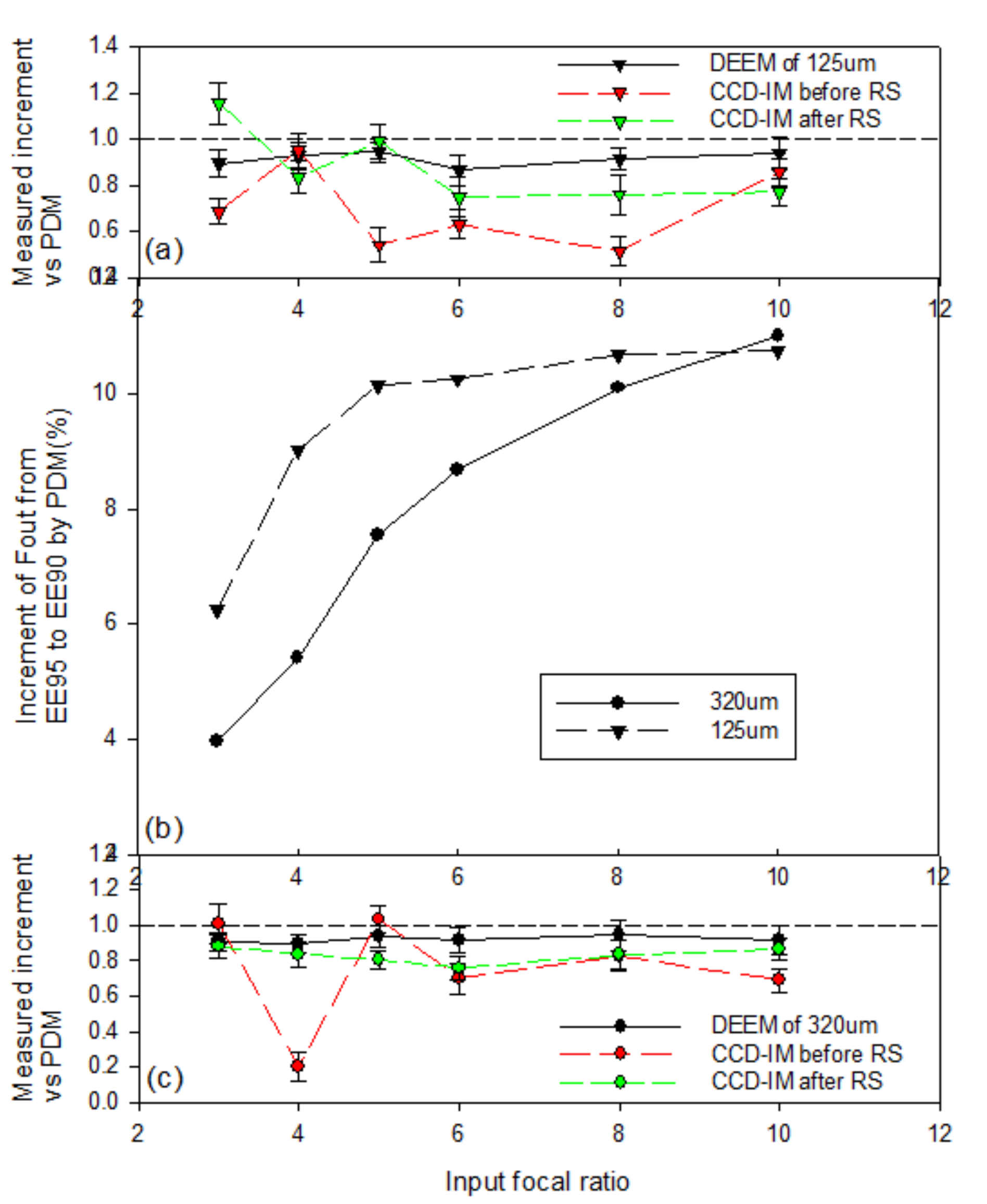}
      \caption{Comparison of the increment of output focal ratio from EE90 to EE95. Image (b) is the theoretical value predicted by PDM. Image (a) and (c) are the measured increments in the experiments, from which we can see that the increments of DEEM are closest to the PDM. The difference of the increment between PDM and the measured value of CCD-IM is reduced and smooth after the residual light subtraction.}
         \label{fig:increment}
   \end{figure}

Comparison of the increment of output focal ratio from EE90 to EE95 is shown in Fig. \ref{fig:increment}. The line with circles is the result of the fibre with 320$\mu$m core and triangles of the fibre with 125$\mu$m core. A larger EE ratio encloses more energy within a larger diameter of faster output focal ratio, so the focal ratio decreases with increasing EE ratio as shown in Fig.\ref{fig:increment}(b). And it is the theoretical value of the increment of output focal ratio from EE95 to EE90 simulated by PDM. Image (a) is the increment of the measured output focal ratio in the experiments of the fibre with 125$\mu$m core size versus the theoretical value of PDM and image (c) of the fibre with 320$\mu$m core size. RS stands for residual light subtraction for CCD-IM. The ratios of the measured increment vs PDM of two kinds of fibres are improved after the residual subtraction. Comparing the ratios of the two methods, DEEM has a better consistence with the predicted values than CCD-IM. It is also more compatible for a large range of input focal ratio from 3.0 to 10.0 with higher stability.

\subsubsection{Accuracy}
To investigate the accuracy performance of DEEM, we build up an optical system with a fixed output focal ratio of $F_{out}$=5.0 and then compare the measured values of the two methods with the pre-set value. The output spots with and without the limitation of the diaphragm EAD and the EE ratio will be recorded at the same time during the experiments to compare with the theoretical values.

In the experiment, the light source was red laser at 632.8nm and the beam splitter was SP1 of $C$=1.20. The input focal ratio was $F_{in}$=5.0. Four positions of 40.0mm, 60.0mm, 90.0mm, 110.0mm away from the output fibre end were chosen to place the diaphragm EAD to build the optical system with the output focal ratio of $F_{out}$=5.0. For instance, the aperture size of the diaphragm EAD was set to 8.0$\pm$0.05mm in the distance of 40.0$\pm$0.01mm to make the output focal ratio of $F_{out}$=5.00$\pm$0.04, and during the same time, the CCD camera was placed in the distance of 60mm to record the output spot. And then the CCD camera recorded another spot after the diaphragm EAD fully expanded to let the whole spot pass freely through the aperture.

Since the accuracy experiment is a fixed output focal ratio test, the value of EE or $k$ is unknown. The first step is to estimate the ratio of $k$ corresponding to the output focal ratio of $F_{out}$=5.0. The steps of measuring the ratio of $k$ is considered to be the inverse process of measuring the output focal ratio by DEEM.

Step 1. Build up the optical system of $F_{out}$=5.0. Move the electric-driven diaphragm to the pre-set positions (precision of $\pm$0.01mm) of 40.0mm, 60.0mm, 90.0mm, 110.0mm and 130mm. The corresponding aperture sizes (precision of <$\pm$0.05mm) of the diaphragm are 8.0mm, 12.0mm, 18.0mm, 22.0mm and 26.0mm, respectively.

Step 2. Record the ratio of $k$ in the five positions. Then the average value is used as the theoretical ratio of the pre-set $k$.

Step 3. Determination of the actual EE ratio according to equation (\ref{eq18}). Then the diameter of the output spot with the diaphragm fully expanded is determined by the actual EE ratio in CCD-IM.

\begin{figure}
   \centering
   \includegraphics[width=\hsize]{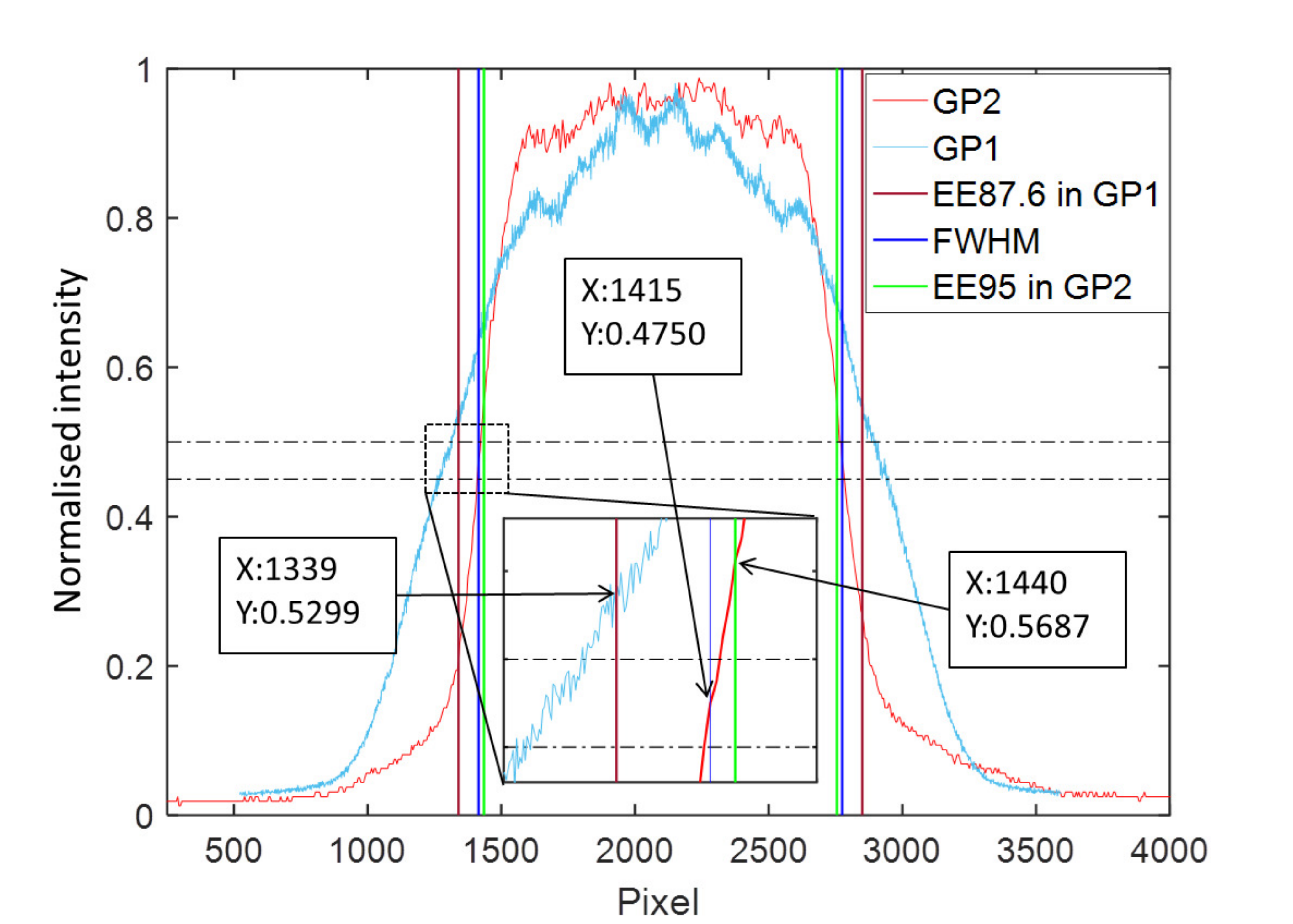}
      \caption{Profiles of the spots in GP1 (diaphragm fully expanded) and GP2 (with the obstruction of diaphragm to build $F$=5.0). The barycentre is shifted to the same point and the intensity is normalised.}
         \label{fig:23}
   \end{figure}

In this case, every time the diaphragm was moved to a specific position, the ratio of $k$ was recorded and the average value was 1.06. According to equation (\ref{eq18}), the actual EE ratio was 88.3 per cent from the aperture of $F$=5.0 which was very close to the theoretical result 88.0 per cent in the simulation of Fig.\ref{fig:eewithinna} with the constraint of $N.A.$. We recorded the output spots and sorted them into two groups: GP1 included the spots with the diaphragm fully expanded and GP2 assembled the images with the limitation of the diaphragm. We found that the power distribution was different in the two groups as shown in Fig.\ref{fig:23}. For GP1, the power distribution is Gaussian-like and they were processed by using EE88 to estimate the diameters. For GP2, the profile decreases rapidly near the edge and the diameter must not be simply determined by EE88, because the spot is only part of the whole spot, so we calculated the diameters by EE95 and FWHM. Table \ref{tab:7} shows the diameter and output focal ratio of the five measurements. The result of DEEM is consistent with the pre-set value, which confirms the accuracy of DEEM. The output focal ratios of CCD-IM differ with each other. The result from FWHM is the closest to the pre-set value. But the output focal from EE88.0 of CCD-IM is smaller than DEEM and the theoretical value. The difference is around 0.3 in F-ratio which also remains within the maximum error of E$N.A.$.

\begin{table*}
\caption{Comparison of the results from a specific output focal ratio of $F_{out}$ = 5.0 with SP1 ($C$=1.20) at the wavelength of 632.8nm. Since the actual EE ratio of $F_{out}$=5.0 corresponding to $F_{in}$=5.0 is unknown for DEEM, the pre-set value of $k$ is estimated from the average value of $k_{ave}$=1.06, which means the actual EE ratio is 88.3 per cent according to equation \ref{eq23}. EE88.0 is the theoretical value from PDM applied in CCD-IM.}             
\label{tab:7}      
\centering                          
\begin{tabular}{c c c c c c c}        
\hline                 
   Focal length & Theoretical & $k_0$ & DEEM & \multicolumn{3}{c}{CCD-IM} \\
   $\pm$0.01mm & Diameter & & & \multicolumn{2}{c}{GP2} & GP1 \\
   & <$\pm$0.05mm & & $k$=$k_{ave}$ & EE95 & FWHM & EE88.0 \\
\hline
   40.0mm & 8.0mm & 1.059 & 8.03mm & & & \\
   60.0mm & 12.0mm & 1.054 & 12.08mm & 11.718mm & 12.052mm & 13.536mm \\
   90.0mm & 18.0mm & 1.065 & 17.91mm & 17.532mm & 18.252mm & 19.980mm \\
   110.0mm & 22.0mm & 1.062 & 22.05mm & 21.456mm & 22.338mm & 24.588mm \\
   130.0mm & 26.0mm & 1.052 & 25.93mm & 25.290mm & 26.172mm & 28.854mm \\
   \\
   $F_{out}$ & 5.0$\pm$0.04 & $k_{ave}$=1.06 & 5.02 & 5.15 & 4.94 & 4.55 \\
   & & & 4.95, 4.93 & 5.05, 5.19 & 5.18, 4.94 & 4.65, 4.78 \\
   & & & 4.88, 5.02 & 5.34, 5.24 & 5.25, 5.08 & 4.69, 4.78 \\
   Average of $F_{out}$& 5.0$\pm$0.04 & & 4.96$\pm$0.06 & 5.19$\pm$0.10 & 5.08$\pm$0.12 & 4.69$\pm$0.08 \\
\hline                                   
\end{tabular}
\end{table*}

\subsubsection{Stability}
As is mentioned above, the noise of the background and the determination of the diameters of the output spots are not independent of the input light power, especially in CCD-IM as shown in Fig. \ref{fig:diameteroflowandhighpower}. The stability test of the sensitivity to the input power variation is needed to verify the performance of the two methods. The artificial unstable light source was controlled by rotating the neutral density filter (NDF) or regulating the laser power supply to change the input light intensity. Different input light power will affect the noise of the residual light of type III, which will bring uncertain changes in the diameters in CCD-IM because of the precision for the residual light subtraction.

\begin{figure}
   \centering
   \includegraphics[width=\hsize]{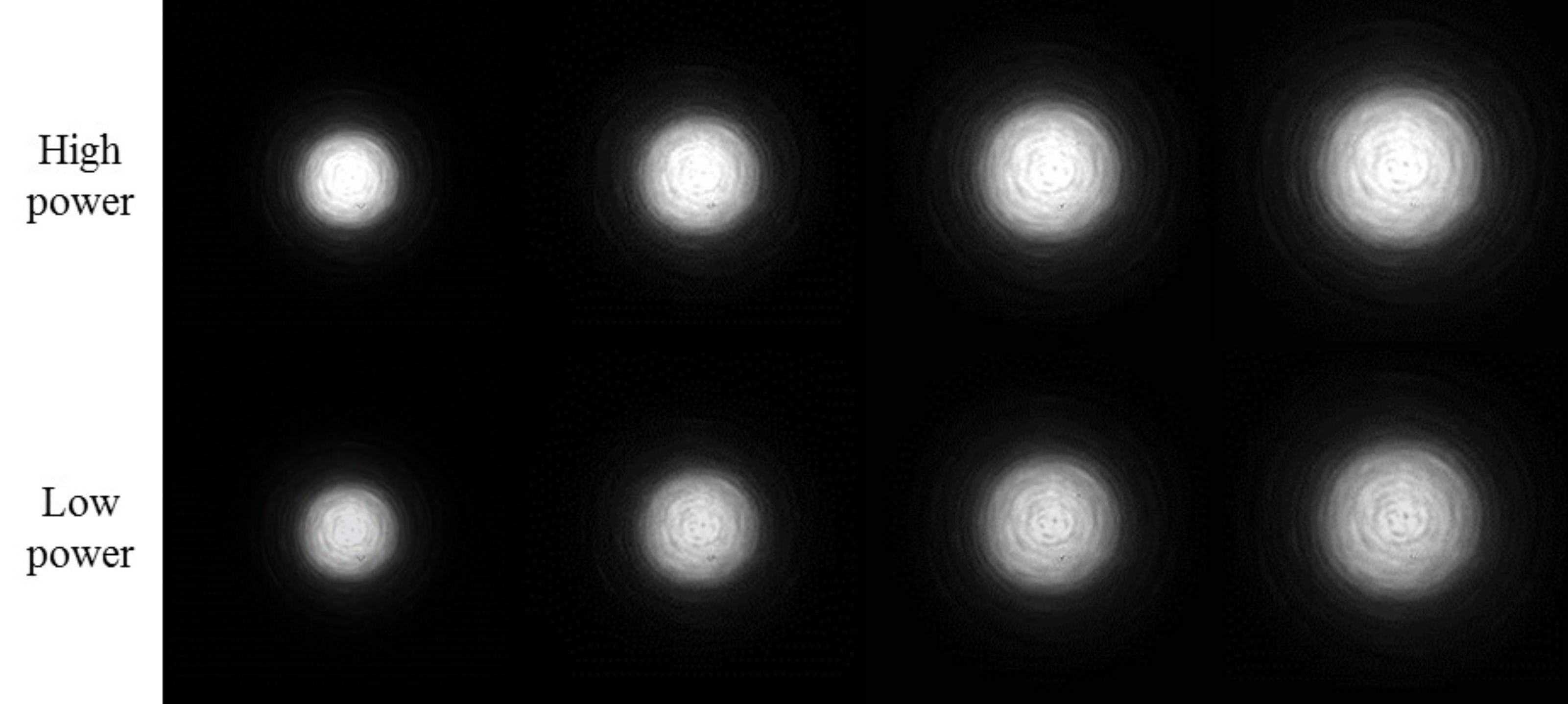}
      \caption{Output spots under different input powers.}
         \label{fig:24}
   \end{figure}

According to equations (\ref{eq14})$\sim$(\ref{eq18}) and (\ref{eq28}), (\ref{eq29}), the bi-detector design in the two-arm measurement system of DEEM has the advantage of compensating the bias caused by the residual light. So in the stability test, the residual light subtraction was implemented using the same method as described in the feasibility test. And the 2-D profile of the output spot was checked by eye to conserve the best fitted one after the subtraction of the dark image and the residual light. For laser illumination system, the input laser power can be easily controlled by changing the input current of the laser source adapter. For LED illumination system, we rotated the neutral density filter NDF to change the intensity, and the input light power was distributed randomly. Parts of the spots are shown in Fig.\ref{fig:24} for demonstration and reference. The detailed parameters are listed in Table \ref{tab:8}.

\begin{table*}
\caption{The diameters and output focal ratio for stability test determined by EE90. The relative error of output focal ratio reaches to 6.5 per cent for CCD-IM which is larger than 2.2 per cent for DEEM in unstable light source. The linear regression coefficient $r^2$ of CCD-IM is lower than DEEM because of the variation of the spot sizes as shown in Fig.\ref{fig:25}(b).}             
\label{tab:8}      
\centering                          
\begin{tabular}{c c c c c c c c c c c c}        
\hline                 
   LS & Stability & $F_{in}$ & Method & \multicolumn{5}{c}{Diameter (mm)} & $F_{out}$ & ave$F_{out}$ & $r^2$ \\
   & & & Focal length & 60.0 & 70.0 & 90.0 & 110.0 & 130.0 & & & \\
\hline
   Laser & STB & $F_{in}=$5.0 & DEEM & 12.82 & 15.70 & 19.87 & 24.40 & 28.06 & 4.60 & 4.53$\pm$0.09 & $\ge$0.998 \\
   & & & CCD-IM & 13.392 & 15.246 & 20.106 & 24.804 & 28.908 & 4.42 & 4.34$\pm$0.08 & $\ge$0.995 \\
   & & $F_{in}=$8.0 & DEEM & 8.84 & 10.67 & 13.36 & 16.71 & 19.35 & 6.66 & 6.68$\pm$0.08 & $\ge$0.994 \\
   & & & CCD-IM & 8.856 & 10.386 & 14.112 & 17.262 & 19.800 & 6.24 & 6.11$\pm$0.11 & $\ge$0.997 \\
   & UNS & $F_{in}=$5.0 & DEEM & 12.83 & 15.58 & 19.32 & 24.19 & 28.24 & 4.57 & 4.59$\pm$0.09 & $\ge$0.996 \\
   & & & CCD-IM & 12.564 & 15.984 & 18.558 & 24.552 & 27.324 & 4.67 & 4.38$\pm$0.27 & $\ge$0.929 \\
   & & $F_{in}=$8.0 & DEEM & 8.89 & 10.17 & 13.41 & 16.2 & 19.62 & 6.52 & 6.61$\pm$0.10 & $\ge$0.989 \\
   & & & CCD-IM & 9.450 & 11.844 & 15.012 & 17.082 & 21.348 & 6.16 & 6.19$\pm$0.29 & $\ge$0.933 \\
   LED & STB & $F_{in}=$5.0 & DEEM & 12.94 & 15.84 & 19.73 & 24.21 & 28.55 & 4.56 & 4.52$\pm$0.10 & $\ge$0.997 \\
   & & & CCD-IM & 12.798 & 15.300 & 20.106 & 23.994 & 29.178 & 4.34 & 4.27$\pm$0.09 & $\ge$0.997 \\
   & & $F_{in}=$8.0 & DEEM & 8.98 & 10.42 & 13.74 & 16.7 & 19.62 & 6.53 & 6.65$\pm$0.08 & $\ge$0.998 \\
   & & & CCD-IM & 9.828 & 11.412 & 14.292 & 18.126 & 20.664 & 6.32 & 6.17$\pm$0.12 & $\ge$0.992 \\
   & UNS & $F_{in}=$5.0 & DEEM & 13.00 & 15.23 & 20.12 & 24.44 & 28.28 & 4.52 & 4.47$\pm$0.08 & $\ge$0.992 \\
   & & & CCD-IM & 12.186 & 16.002 & 20.970 & 23.580 & 29.736 & 4.12 & 4.03$\pm$0.26 & $\ge$0.911 \\
   & & $F_{in}=$8.0 & DEEM & 8.95 & 10.42 & 13.36 & 16.56 & 19.64 & 6.53 & 6.48$\pm$0.09 & $\ge$0.994 \\
   & & & CCD-IM & 10.674 & 10.260 & 13.446 & 17.262 & 20.340 & 6.49 & 6.53$\pm$0.27 & $\ge$0.924 \\
\hline                                   
\end{tabular}
\end{table*}

   \begin{figure}
   \centering
   \includegraphics[width=\hsize]{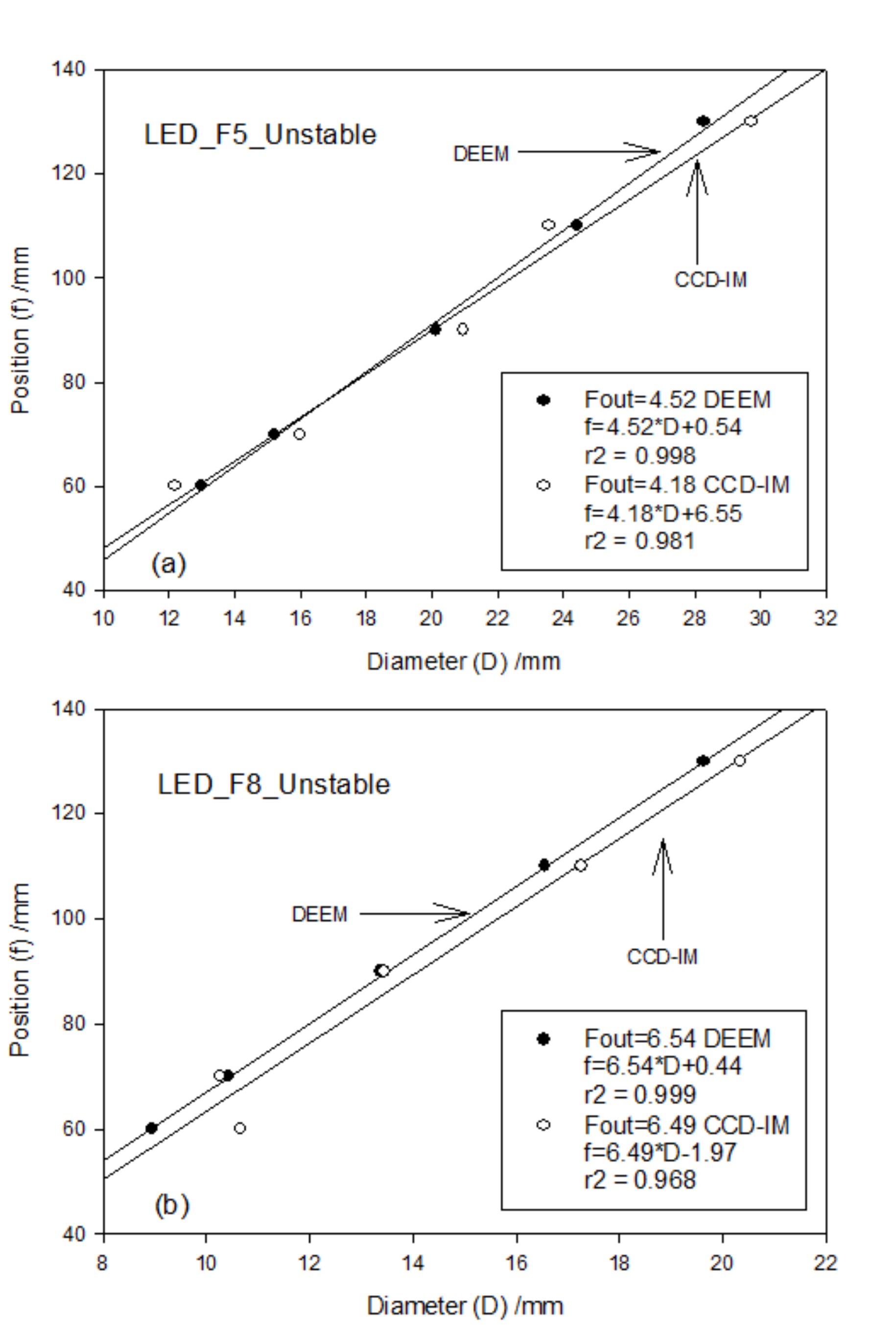}
      \caption{The fitted curve of output spots illuminated by unstable light source. The variation of the diameter in CCD-IM is unpredictable and the linear regression coefficient is lower. Though occasionally the output focal ratio of CCD-IM is the same as DEEM as shown in image (b), the diameters in the first two points of CCD-IM are abnormal. The $f-$intercept on vertical axis of DEEM is much smaller which matches the fitted zero position of the fibre end according to equation \ref{eq:fittingposition}.}
         \label{fig:25}
   \end{figure}

In the experiments, most of the results of the output focal ratio derived from the two methods are consistent with each other within the relative difference of 0.3 in F-ratio after the subtraction of the dark image and the residual light. Considering the results of the feasibility test in the condition of a stable light source and the stability test with an unstable light source, it is effective that the subtraction of the residual light can improve the accuracy of CCD-IM and reduce the difference between the two methods.

When the light source is unstable, the diameter of the output spot derived from CCD-IM becomes unpredictable and the output focal ratio is not for sure either. As is shown in Fig.\ref{fig:25}, although the output focal ratio of CCD-IM rises up in some cases of the unstable light source and becomes larger than the result of DEEM, the linearity of the fitting curve is very low. Another important parameter is the intercept on the $f$-axis which indicates the fitted zero point of the fibre end. In the Fig. \ref{fig:25}(a), the $f$-intercepts of DEEM and CCD-IM are 0.54 and 6.55, respectively, which means the distance of the fitted zero point of the fibre end is offset with 0.54mm and 6.55mm from its actual position. The regulation precision of the aperture size of the electric-driven diaphragm is $\Delta r$=0.05mm, and the corresponding uncertainty of the fitted position of the fibre end is derived from
\begin{equation}\label{eq:fittingposition}
\Delta f = 2 \cdot {F_{in}} \cdot \Delta r
\end{equation}
And the estimated distance is $\Delta f$=0.50mm, which is the same level of 0.54mm of the fitting curve from DEEM, but much smaller than 6.55mm from CCD-IM. So the confidence level of the output focal ratio of CCD-IM in Fig.\ref{fig:25}(a) is too low compared with DEEM. On the contrary, the regression linearity is much higher and the $f$-intercept is smaller so that the robustness of DEEM is acceptable to measure the output focal ratio in the unstable environment.

\section{Applications on FRD measurements}
Multi-fibre telescopes (e.g. SDSS, AAT and LAMOST) can efficiently amass the catalogues of galaxies and stars to investigate their universal properties. When carry out multi-object survey with these telescopes, the fibre units are discretely distributed in the independent area to observe the specific target. So these surveys have fundamental limitations that they are unable to resolve the spatial and morphological relations. And the spectra acquired by these surveys suffer from positioning problem caused by alignment, especially when the fibre unit is misaimed to the object, it will record a different fraction and part of the galaxy depending on the scale and distance of the object. On the contrary, the light projected on the IFU head is an area light and the IFU combined with a micro lens array is used to sample the observe field and avoid the blind spot. It can spatially sample the spectra to resolve each galaxy spatially, giving the morphological and dynamical information at multiple positions across the galaxy and it is more robust to positioning error. Moreover, IFU can be made of a huge number of fibres and the telescope can assemble an array of many IFUs to improve the information density, such as the VIMOS-IFU with 6400 spatial elements \citep{Bonneville2003The} and TEIFU with 1000 elements \citep{Murray2000TEIFU}. With the fast increasing number of closely-packed fibres, the demand of efficient and accurate measurement of parameters and properties of IFU is urgently needed to enable rapid processing of massive fibres compared to a general multi-fibre telescope.

An IFU of high quality should have good performance on homogeneity and high throughput. Generally, the core diameter of fibres used in IFU is smaller than 200$\mu$m to acquire higher spatial resolving power as well as the spectra resolution. In terms of high resolving power, an IFU with 50$\mu$m core fibers for FASOT requires that the FRD should be suppressed to satisfy the requirement of the output focal ratio of $F_{out}$>4.5 for each fibre, which can help efficiently maintain the good performance of PSF and throughput.

\subsection{Performance of a bare fibre}
To achieve high spectral resolution, small core fibres of 50$\mu$m are chosen to build the first generation prototype IFU. In the FRD measurement, the input focal ratio was set to $F_{in}$=5.0 and the output focal ratio was determined by EE90. The splitter was SP1 and the constant $C$ was the same as in Table \ref{tab:5}. In the test of CCD-IM, the background was subtracted using the same method as in Section 3.3.

   \begin{figure}
   \centering
   \includegraphics[width=\hsize]{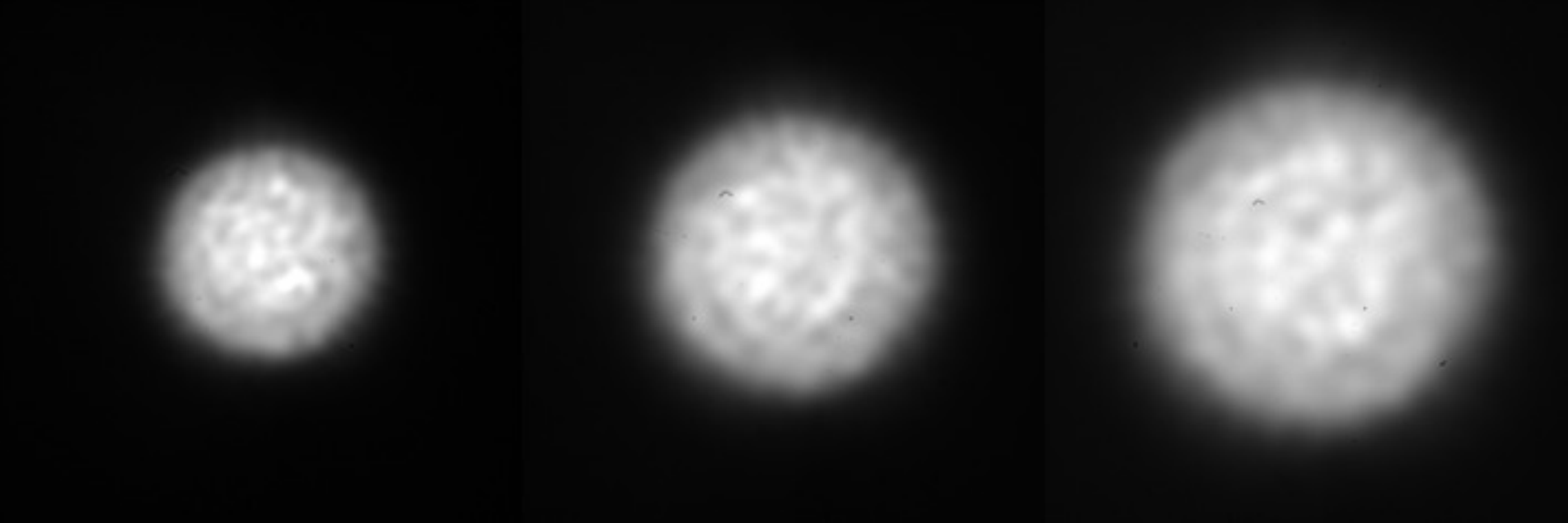}
      \caption{The output spots captured by CCD with input light source of laser at wavelength of 532nm.}
         \label{fig:26}
   \end{figure}

\begin{table*}
\caption{The performance of a bare fibre of 50$\mu$m core with $N.A.$=0.22.}             
\label{tab:9}      
\centering                          
\begin{tabular}{c c c c c c c}        
\hline                 
   LS & \multicolumn{2}{c}{Laser$@$532nm} & \multicolumn{2}{c}{Laser$@$632.8nm} & \multicolumn{2}{c}{LED} \\
   Method & DEEM & CCD-IM & DEEM & CCD-IM & DEEM & CCD-IM \\
\hline
   $F_{out}$ & 4.18, 4.25 & 3.96, 3.85 & 4.15, 4.13 & 3.99, 4.09 & 4.21, 3.95 & 4.00, 4.11 \\
   & 4.15, 4.06 & 3.76, 4.10 & 4.22, 4.05 & 3.95, 3.82 & 4.05, 4.02 & 4.15, 3.91 \\
   & 4.19 & 4.12 & 4.00 & 4.09 & 4.15 & 3.89 \\
   Average of $F_{out}$ & 4.17$\pm$0.06 & 3.96$\pm$0.14 & 4.11$\pm$0.08 & 3.98$\pm$0.10 & 4.08$\pm$0.09 & 4.01$\pm$0.10 \\
   $r^2$ & $\ge$0.996 & $\ge$0.988 & $\ge$0.995 & $\ge$0.989 & $\ge$0.986 & $\ge$0.987 \\
   Throughput & 86.9$\pm$1.1 per cent & & 89.9$\pm$0.9 per cent & & 88.5$\pm$1.2 per cent & \\
\hline                                   
\end{tabular}
\end{table*}

Fig.\ref{fig:26} shows part of the output spots. The results in Table \ref{tab:9} shows that the output focal ratio remains around 4.1, less than the required value of 4.5, but the throughput is high enough to meet the designed requirements and if the microlens array is equipped, the throughput should be much higher. Compared to the results of large core fibres of 320$\mu$m and 125$\mu$m, the FRD is more serious and this is a classical phenomenon which is observed in many research groups. Though the FRD performance is not satisfying, a prototype of 9$\times$9 IFU was built to investigate other characteristics like homogeneity and the validation of fabricating technology, especially the quality of the quartz plate, fibre arrangement and the break rate of fibres.

\subsection{The first generation prototype of 9$\times$9 IFU}
\begin{figure*}
   \centering
   \includegraphics[width=\hsize]{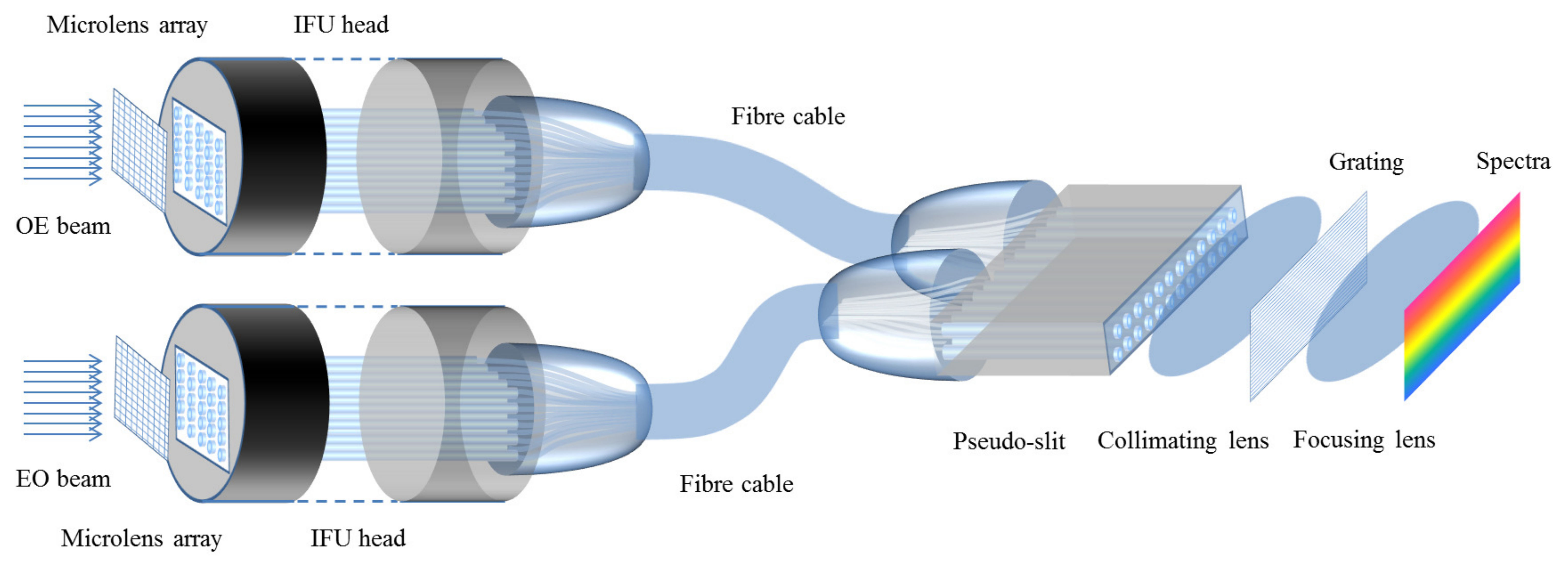}
      \caption{The structure of the prototype IFU.}
         \label{fig:28}
   \end{figure*}

The FASOT proposed by \citeauthor{Qu2011A} in 2011 is designed to be an efficient 3D spectropolarimeter with 8192 fibres working at the wavelength range from 400nm to 900 nm with high resolution of about 110,000. The light from the sun is split into two orthogonally polarized beams, ordinary-extraordinary beam (OE) and extraordinary-ordinary beam (EO), after the Savart polarizing system. And the two beams will transmit separately through two optical arms into a pair of IFUs. The basic structure of the IFU is shown schematically in Fig.\ref{fig:28}. A mocrolens array is placed before the IFU head to sample the OE or EO beam spatially to couple into the fibres. The fibres at the output ends are rearranged to be a two-row pseudo-slit for the spectrograph.

\begin{figure}
   \centering
   \includegraphics[width=\hsize]{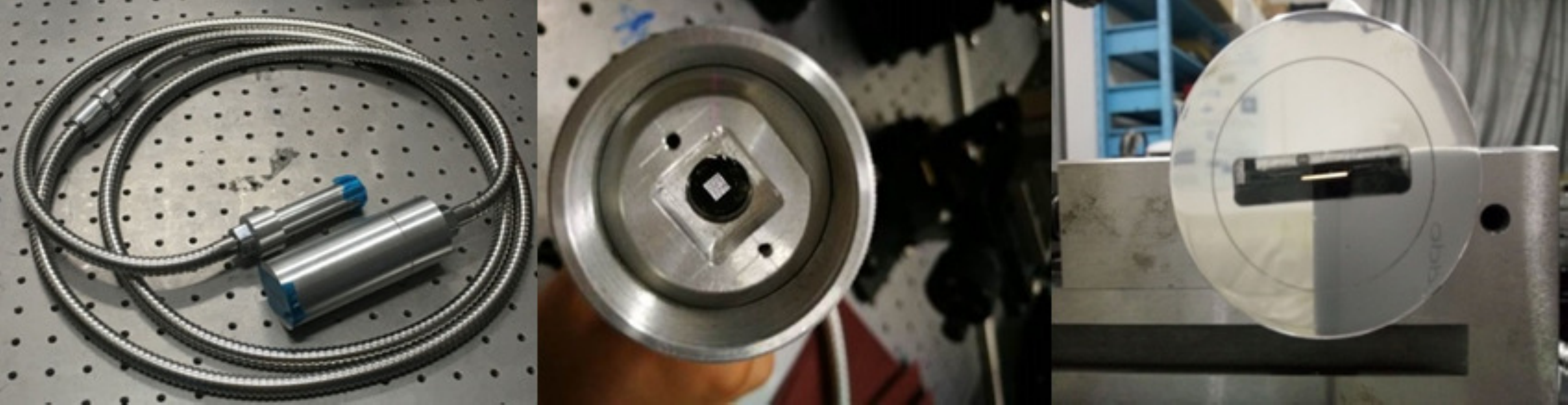}
      \caption{The first generation IFU with one-row pseudo-slit.}
         \label{fig:29}
   \end{figure}

In the tests, a small prototype of 9$\times$9 IFU with one-row pseudo-slit consists of 81 fibres was fabricated in Harbin Engineering University as shown in Fig.\ref{fig:29}. Those fibres with small cores of 50$\mu$m and $N.A.$=0.22 were required to improve the performance of high spectral resolution. Due to the small core and $N.A.$ of the fibre, high accurate positions of the both ends of IFU are crucial for high and uniform transmission efficiency. The platform of DEEM can easily conduct a 3D imaging system to monitor the relative position of the fibre array in the near-field. Two microscopes are equipped to image the fibre ends in both x- and y- directions and monitor the position shift in fibre array in Z-direction. The beam splitter was particularly placed in the optical path inside the collimating system to reduce the offset in the input point caused by the reflective and refractive light back and forth between the two surfaces of the splitter. Then the position distribution on the IFU head and the pseudo-slit end can be reconstructed.

\subsection{Results of the prototype IFU}
According to the design requirements, the input focal ratio should be slower than $F_{in}$=5.0 and the final output focal ratio should be controlled slower than $F_{out}$=4.5. As is known that the result in the test of a bare fibre was negative, but it is still necessary to study the feasibility of the process technique for reference to upgrade the next generation IFU. The input focal ratio was chosen $F_{in}$=5.0 at wavelength of 632.8nm.

\begin{figure}
   \centering
   \includegraphics[width=\hsize]{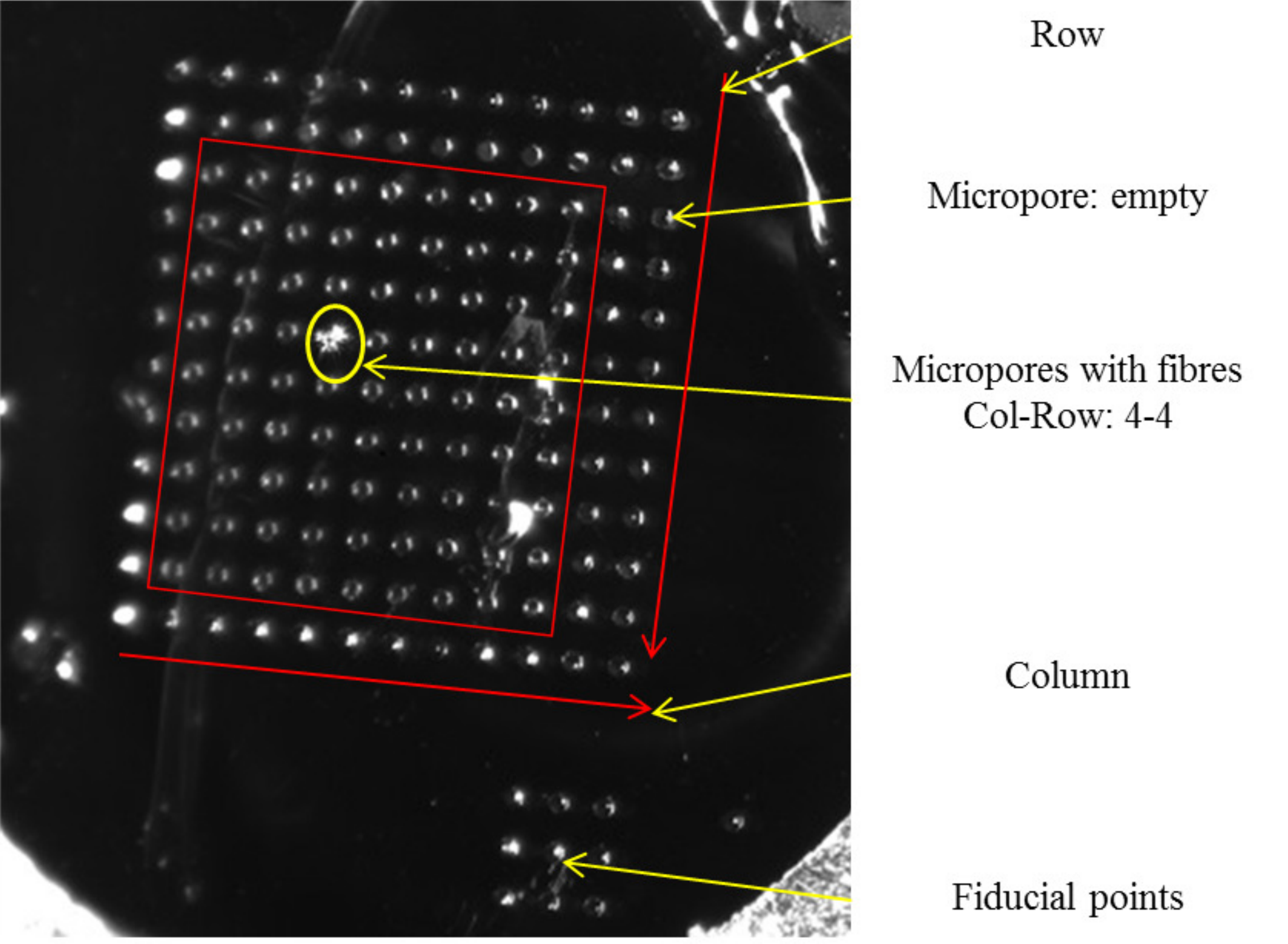}
      \caption{The near-field image of the fibre array. The quartz plate was designed with 11$\times $ 11 micropores. But only 9$\times $9 fibres were assembled in the centre area.}
         \label{fig:30}
   \end{figure}

The fibre array was fixed at suitable positions in a quartz plate with micropores made by laser. The near-field image of the fibre array is shown in Fig.\ref{fig:30}. Choosing quartz plate of the same thermal expansion coefficient with the fibres can reduce the different rates of material removal and stress effects. The first batch of the IFU was only assembled with 81 fibres limited by the length of the pseudo-slit in the output end. The identification number of each fibre and the effective area with fibres are marked on the figure. And we selected ten out of 81 fibres to test the homogeneity. All the fibres were polished and fixed in the holes by heating curing process.

\subsubsection{Focal ratio degradation}
\begin{figure}
   \centering
   \includegraphics[width=\hsize]{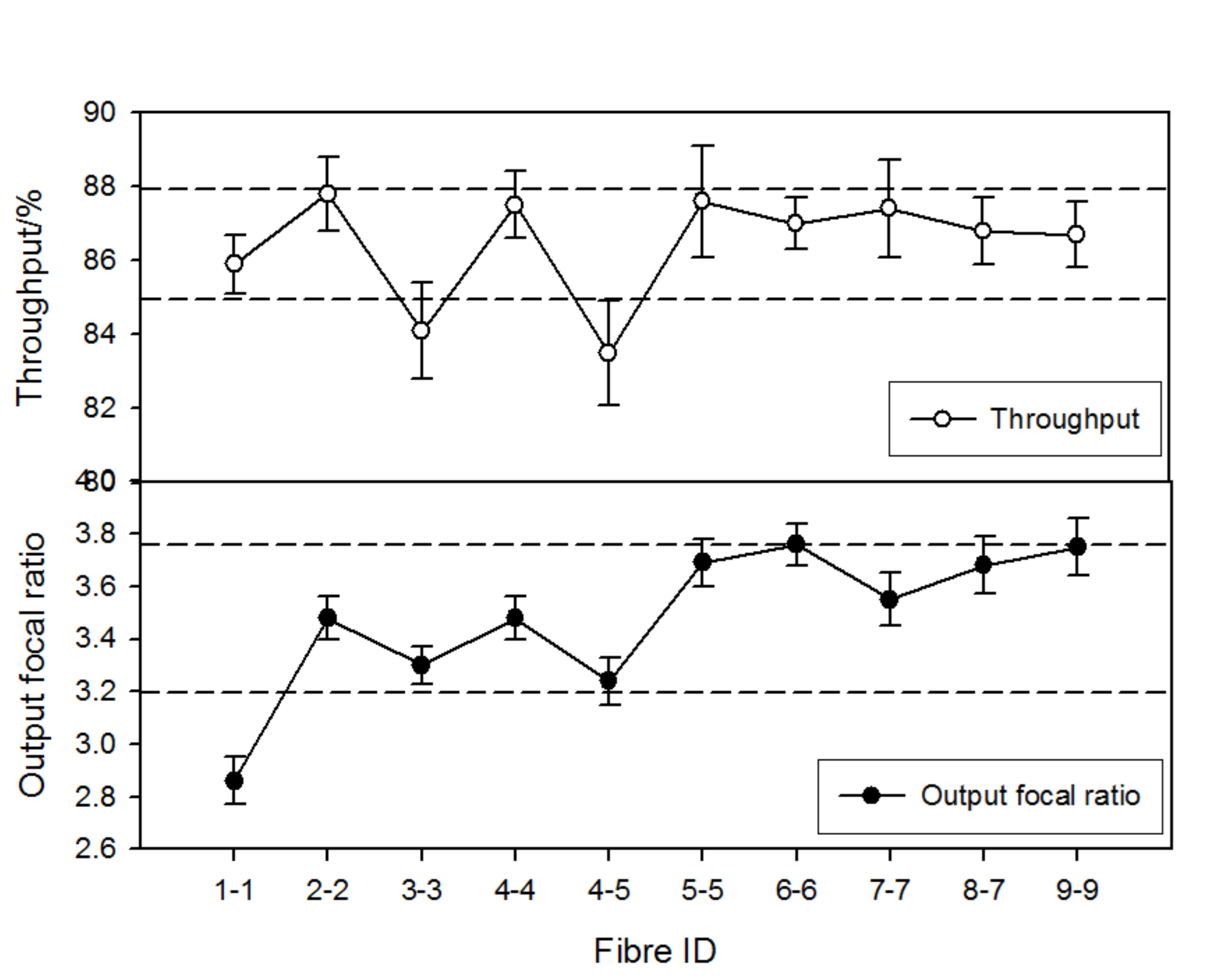}
      \caption{The performance of output focal ratio and throughput of the first generation IFU. The output focal ratio spread from 2.8 to 3.8, lower than the designed requirement. The throughput is relatively uniform and most of them are larger than 85 per cent.}
         \label{fig:31}
   \end{figure}

The homogeneity of output focal ratio and throughput of IFU are shown in Table \ref{tab:10} and Fig.\ref{fig:31}. Much more serious FRD occurs in IFU compared with the result of a bare fibre and the difference between the fibres can be up to 0.9 in F-ratio (the relative error of 30 per cent in FRD).
\begin{table}
\caption{The performance of homogeneity of the prototype IFU.}             
\label{tab:10}      
\centering                          
\begin{tabular}{c c c c}        
\hline                 
   Fibre ID & Output focal ratio & $r^2$ & Throughput (per cent) \\
\hline
   1-1 & 2.86$\pm$0.08 & $\ge$0.996 & 85.9$\pm$0.8 \\
   2-2 & 3.48$\pm$0.07 & $\ge$0.988 & 87.8$\pm$1.0 \\
   3-3 & 3.30$\pm$0.06 & $\ge$0.993 & 84.1$\pm$1.3 \\
   4-4 & 3.48$\pm$0.07 & $\ge$0.993 & 87.5$\pm$0.9 \\
   4-5 & 3.24$\pm$0.08 & $\ge$0.994 & 83.5$\pm$1.4 \\
   5-5 & 3.69$\pm$0.07 & $\ge$0.994 & 87.6$\pm$1.5 \\
   6-6 & 3.76$\pm$0.06 & $\ge$0.997 & 87.0$\pm$0.7 \\
   7-7 & 3.55$\pm$0.09 & $\ge$0.996 & 87.4$\pm$1.3 \\
   8-7 & 3.68$\pm$0.09 & $\ge$0.991 & 86.8$\pm$0.9 \\
   9-9 & 3.75$\pm$0.07 & $\ge$0.994 & 86.7$\pm$0.9 \\
\hline                                   
\end{tabular}
\end{table}

\begin{figure}
   \centering
   \includegraphics[width=\hsize]{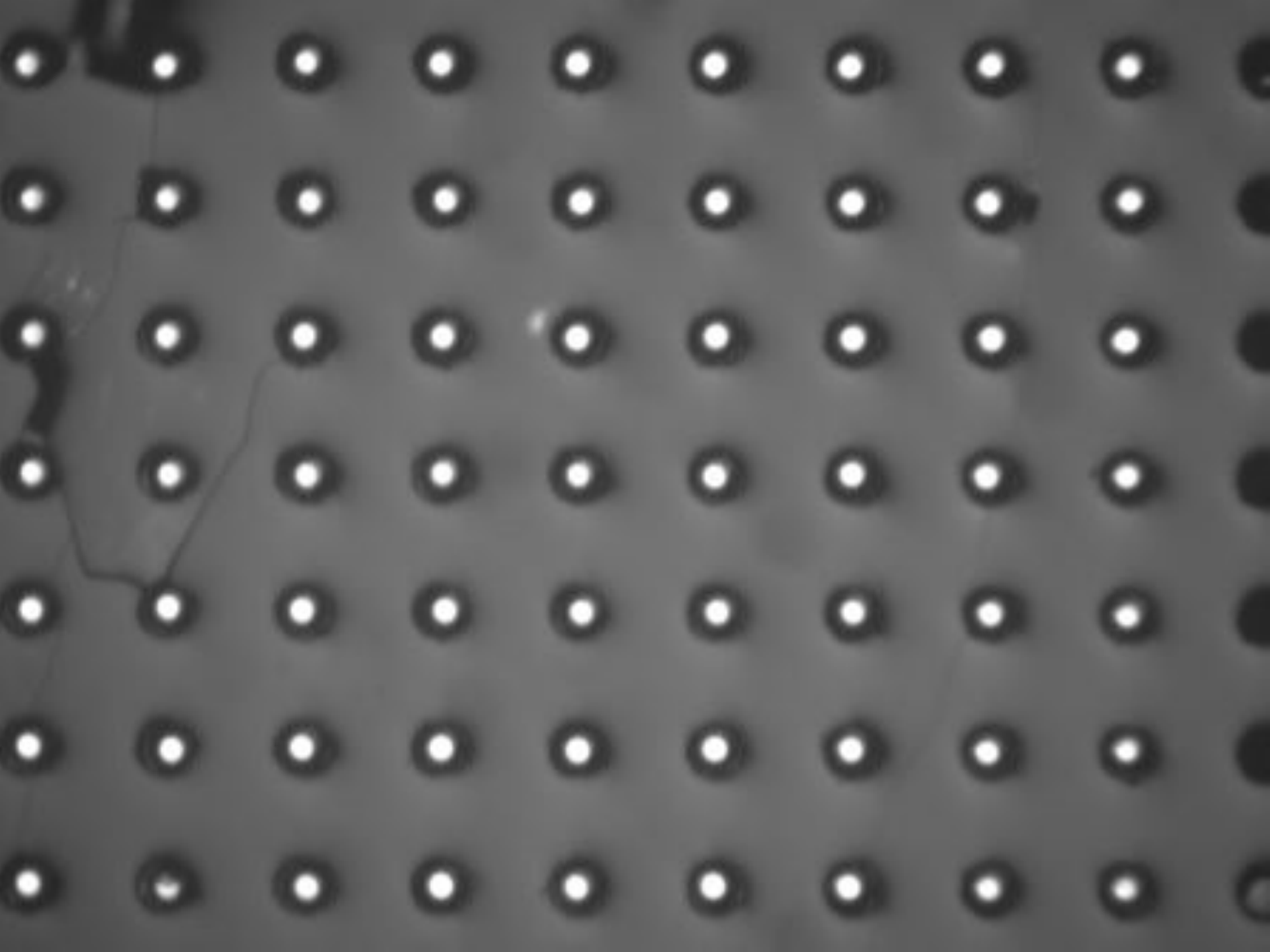}
      \caption{The image of the quartz plate. The bright points are fibres. Some defects of breaks occurs on the top left corner. And there are also some cracks and fissures spreading on the surface of the quartz plate. These imperfections mainly formed during the manufacturing process caused by stress. So they can also affect the distribution of fibres and burden the fibre with side stress to induce micro bending.}
         \label{fig:32}
   \end{figure}

Then we used the collimated beam from LED to illuminate the both ends of IFU head and the pseudo-slit separately to monitor the distribution of fibres on the opposite end. In the near-field image of the quartz plate in Fig.\ref{fig:32}, some defects are found out like the fissure and the break in micropores. These imperfections make the fibre stray away from the position where it should have been to be and they can also bring stress on fibres to form micro bending which contributes to FRD. From the results in Table \ref{tab:10}, the output focal ratio of fibres on top left in Fig. \ref{fig:32} where exists some cracks of different sizes is much smaller, which is probably caused by the stress on fibres from the squeeze by the cracks. However, no distinct defects of fissures exist in the centre area and the output focal ratio begins to increase.

\subsubsection{Throughput}
The transmission efficiency is relatively stable from the sample. The throughput remains around 86 per cent and is about 3 per cent lower on average compared with that of a bare fibre. The power used in calculating the throughput was integrated under the condition of EE100 which is limited by the numerical aperture $N.A.$. If we only collect the power within the limited critical focal ratio of $F_{critical}$=4.5, the throughput would decrease severely and differ far away from each other.

The lower throughput of IFU compared with a bare fibre is due to the stress effect and the coupling efficiency in the input end. Some individual fibres with lower throughput than others can be ascribed to the scattering of higher order modes inner the fibre core because of the imperfections like the mechanical damage and non-uniform core size in the drawing process. The cross-talk caused by scattering is negligible for throughput since the fibres are separately fixed in the quartz plate of the IFU head and the fibre coating is remained on the fibre in the cable. In Fig.\ref{fig:31}, though the output focal ratio of fibre 1-1 is much smaller, the throughput is consistent with others within errors. So the throughput losses from FRD is not susceptible in the tested range of input focal ratio from 3.0 to 10.0 because the FRD is not serious enough in this range to expand the output light to reach or exceed the asymptote value of $N.A.$. In the current tests, the input light was directly injected into the fibre without the microlens array and the fibre end was uncoated. It is a common method in the practical application of a telescope that the end surface is coated with multi-layer of antireflection materials or index-matching gel to reduce the Fresnel reflection. In the design of FASOT, the microlens array provides a precision incident position of less than 5$\mu$m which can improve the coupling efficiency in the input end. So the variation in the throughput is dominated by absorption in the fibre and it is wavelength dependent that the throughput is lower at shorter wavelengths. Other causes such as the fluctuation in the light source, precision of repeatability of the diaphragm, also affect the variations in throughput. These errors are discussed in Table.\ref{tab:errorsummary} and the throughput of the fibres is still in agreement within errors.

\subsubsection{Position distribution}
\begin{figure}
   \centering
   \includegraphics[width=\hsize]{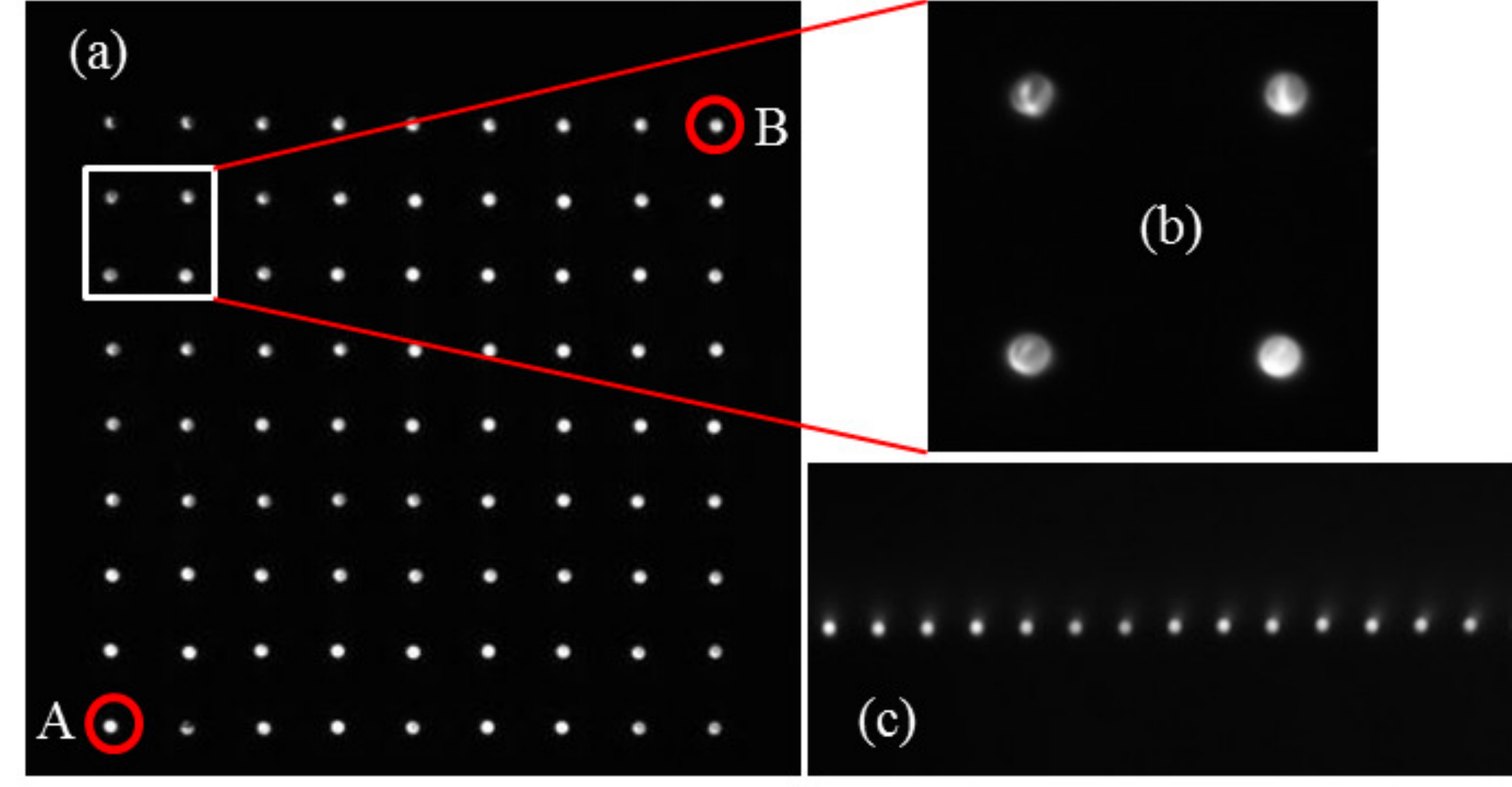}
      \caption{The image of IFU head and pseudo-slit in the near-field. Image (a) is the full view on the IFU head. The positive direction of Column and Row is from A to B in the red circles. Image (b) is to show the area in white box in magnified view. One can clearly see the asymmetric spots with some dark gaps and these phenomenon are very common in an imperfect  fibre with roughness on the end face. Image (c) is the partial view of the pseudo-slit end.}
         \label{fig:33}
   \end{figure}

\begin{figure}
   \centering
   \includegraphics[width=\hsize]{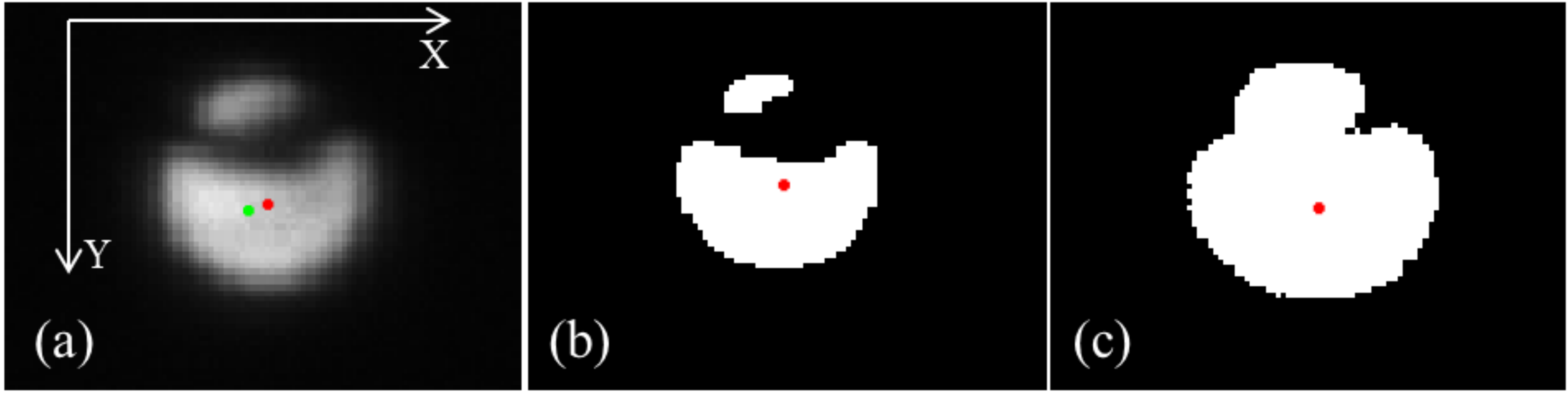}
      \caption{The zoomed view of the spot on the right hand of spot A in Fig.\ref{fig:33}(a) on the IFU head. Image (a) is the original picture of the spot. Image (b) is a spot after binarization with a BLC derived from Otsu algorithm (BLC=0.3984 for this spot). Image (c) is processed with BLC=0.0091 which is manually chosen by evaluating the ratio of the peak intensity and the background.}
         \label{fig:34}
   \end{figure}

\begin{figure}
   \centering
   \includegraphics[width=\hsize]{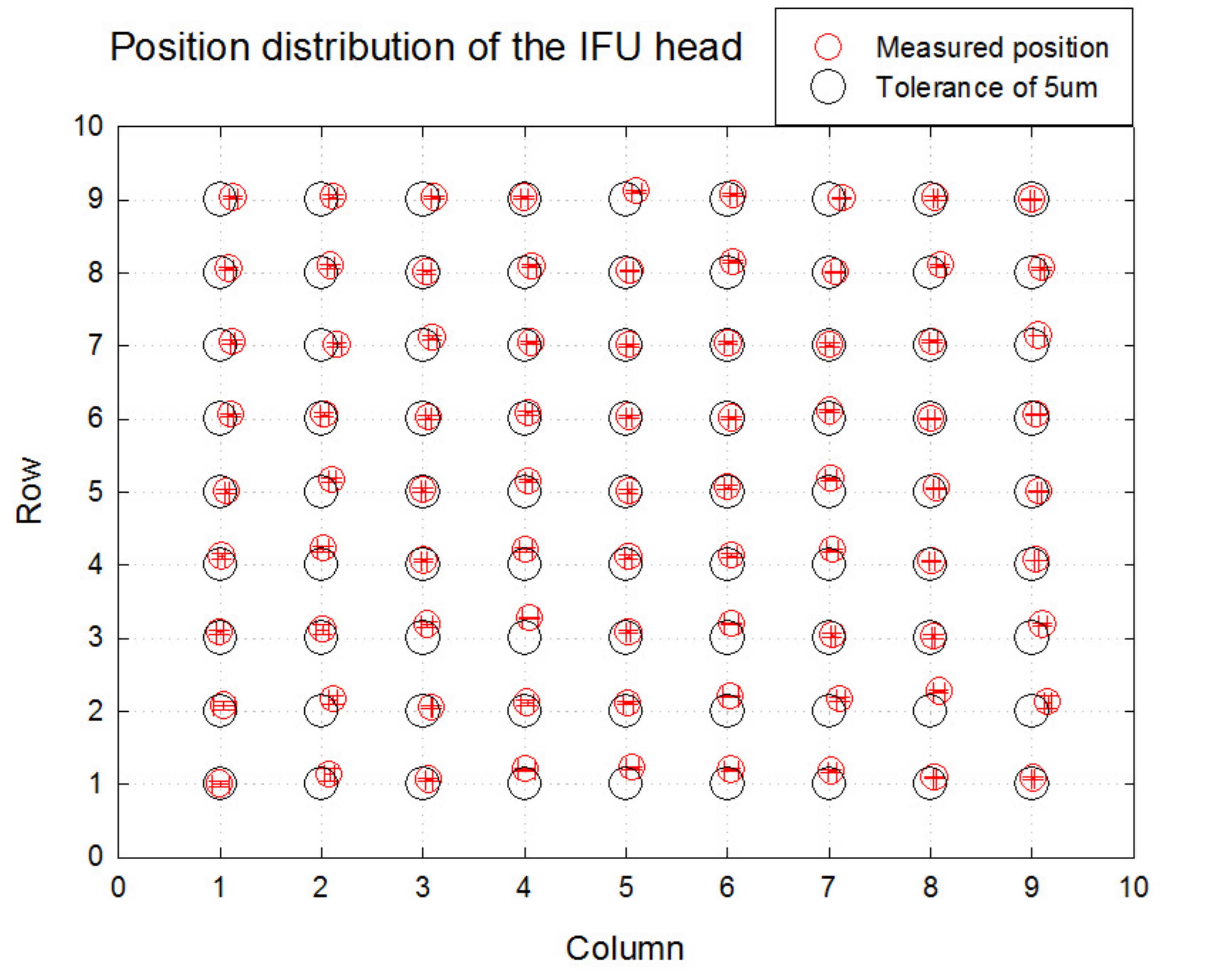}
      \caption{The position distribution of the IFU head. The red circle covers the maximum error bar of each spot in both Column- and Row- directions. The black circle with radius of 5$\mu$m limits the tolerance of the position error that can be accepted according to the design requirement.}
         \label{fig:35}
   \end{figure}

\begin{figure}
   \centering
   \includegraphics[width=\hsize]{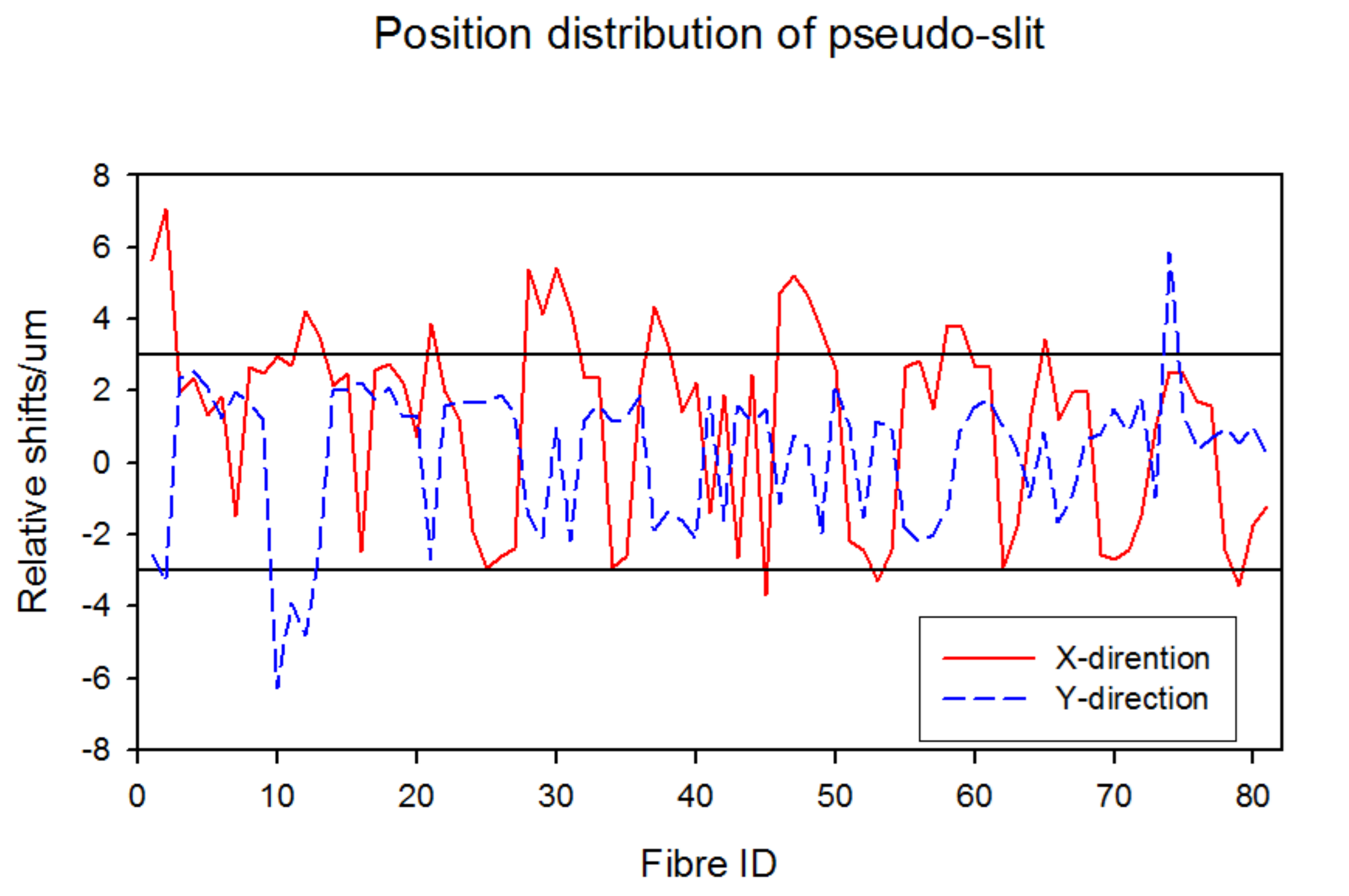}
      \caption{The position distribution of pseudo-slit. Most of the fibres are fixed in the central area within the shifts less than 4$\mu$m in both X- and Y- directions. The standard position deviation is about 3$\mu$m, which is fit for the situation of the designed system.}
         \label{fig:36}
   \end{figure}

The fibre position distribution in the quartz plate is shown in Fig.\ref{fig:33}. The fiducial points are used as the standard position to measure the registration between each fibre. Two methods were implemented to determine the centre position of the fibre. One is to find the barycentre by the integral method according to equation (\ref{eq:barycentre}). The other is to locate the centre of each output spot after binarization which converts the greyscale image to a binary image. The binary image replaces all pixels in the original image with luminance greater than the binarization level coefficient (BLC) with the value 1 (white) and replaces all other pixels with the value 0 (black). Two types of BLCs were preset to acquire the edge profile and the surface profile of the fibre end. To reveal the surface characteristics, Otsu algorithm \citep{Otsu1979Threshold} was used to constraint the BLC to uncover the power distribution and the surface profile on the fibre end face as shown in Fig.\ref{fig:34}. Usually the BLC derived from Otsu algorithm is so large that the edge profile is neglected but the detailed texture is shown up. Another BLC was chosen manually by evaluating an appropriate ratio of the peak value of the intensity and the background, around BLC=0.01 in the tests, with which the full edge profile appeared and the centre position could be determined. The position distribution is shown in Fig.\ref{fig:35} and Fig.\ref{fig:36}. The standard position deviations in the IFU head and slit end are about 7$\mu$m and 3$\mu$m, respectively. So the position precision should be improved to meet the designed requirements within 5$\mu$m.

It is a convenient way to use Otsu algorithm to reveal the internal and the external profiles. The missing gaps between the two bright parts of the spot in Fig.\ref{fig:34}(b) shows the dark illumination of the power distribution in the near-field. And it also indicates that the roughness would be more serious than other parts. In order to reconstruct the whole view of the spot, the BLC should be reduced to a smaller value to reserve more bright pixels to make up a full circle as in Fig.\ref{fig:34}(c). It can maximally restore the original spot in a binary image to show the edge profile. The barycentre coordinates are marked in the images. The red circle is the centre position of a binary image and the green circle is derived from equation (\ref{eq:barycentre}): (a): (525,1829) is the green circle, and (528,1828) is the average value of the binary centre position in the red circle, (b): (528,1827), (c): (528,1829). For some spots, the difference of barycentre positions among the three types are not sensitive to BLC like the spot in Fig.\ref{fig:34}, and for some spots, the difference can be up to 6 pixels, that is about 8$\mu$m considering the magnification factor of the imaging system.

The position distribution of the total 81 fibres in the IFU head is shown in Fig.\ref{fig:35}. For clarity, the size of the circles and the distance of the gap between the adjacent fibres are demonstrated in two different scale spaces. The standard positions of all the fibres are located on the crossing point on the dashed lines. The red points and circles are the measured values within the maximum error bar. Comparing the results with the near-field image of Fig.\ref{fig:32}, one can notice the large misalignments on the top left corner, and a possible speculation is that the defects of cracks and stress on the quartz plate worsen the position accuracy. In the meanwhile, the offsets of the positions in the centre area are much smaller than that in the outer region. This may be relevant to the stress distribution inner the quartz plate and the heating curing process.

\subsection{Modified design with smaller core fibres}
To better perform the scientific research for the telescope, the throughput and FRD properties are strictly designed to meet the requirements. Yet the first generation of the prototype cannot fully satisfy the scientific goal, especially the serious FRD and the position precision of the fibre array. A modified design of the IFU is in progress that smaller core fibres of 35$\mu$m with $N.A.$=0.12 are chosen to be assembled in a new type of quartz plate with staggered micropores. This construction can increase the number density of the fibre array and improve the spatial resolution at the same time. Also the staggered arrangement can perfectly match the microlens array before the IFU head. The filling factor and sample efficiency of the honeycomb-shaped microlens array is nearly 100 per cent, which significantly improve the coupling efficiency \citep{Corbett2009Sampling}.

In a fibre, $N.A.$ limits the maximum input and exit emission angle and the critical focal ratio. The critical focal ratio limited by $N.A.$ is the minimum value and is determined by the approximation:
\begin{equation}\label{eq29}
{F_{critical}} = \frac{l}{{{D_{\max }}}} = \frac{f}{{2\tan {{\theta _{\max }}} }} \approx \frac{1}{{2\sin {\theta _{\max }}}} = \frac{1}{{2N.A.}}
\end{equation}
where $\theta _{max}$ is the critical angle, $D_{max}$ is the maximum diameter of the input or output spot and $f$ is the distance between the fibre end and the observe plane. For a fibre with $N.A.$ = 0.12, the $F_{critical}$ is about 4.2 and in this way we can control the threshold of the final output focal ratio from the pseudo-slit. We tested the FRD performance of a bare fibre under different input conditions of the input focal ratio varying from 5.0 to 8.0, which covers the whole range of required design specifications.

\begin{figure}
   \centering
   \includegraphics[width=\hsize]{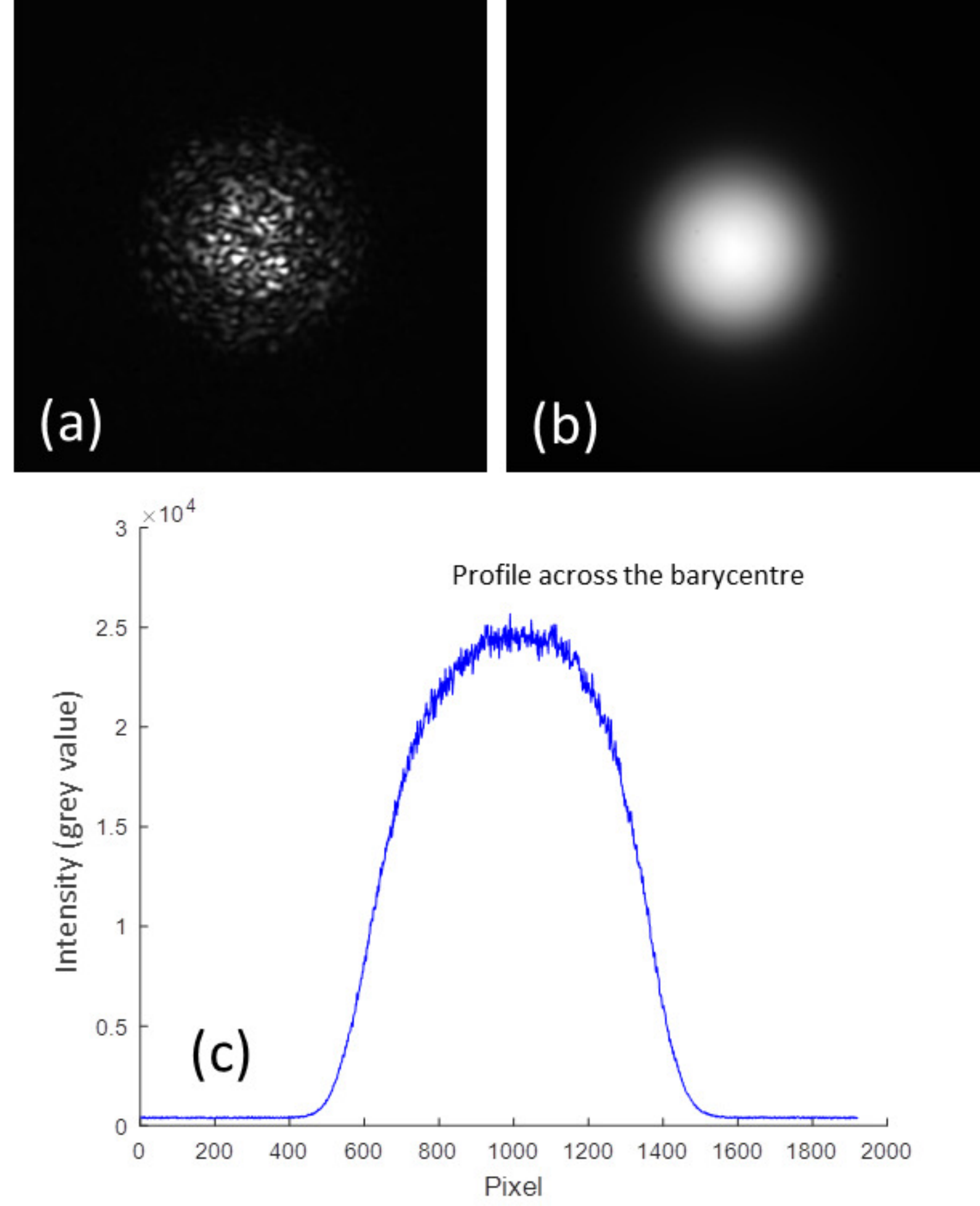}
      \caption{Serious laser speckle patterns (a) on the power distribution of a small core fibre of 35$\mu$m with $N.A.$=0.12. Broadband light source will significant suppress the speckle as shown in the image (b) of a LED. The image (c) is the profile cut across the barycentre, which is not a perfect Gaussian function.}
         \label{specklepattern}
   \end{figure}

Since the core of the fibre is smaller, the input light source is replaced with a single-mode fibre, which is multimode in the tested wavelength range, to control the incident position. As the focal ratio of the output light from the fibre is limited by $N.A.$, an expander is used to enlarge the solid angle to let the light in the centre area pass through the lens L1 so that the light coupled into the fibre is approximate to a uniform flat function. In addition, the excited modes in the fibre of $N.A.$=0.12 will decrease significantly compared with the fibre of $N.A.$=0.22. The guided mode volume $M$ can be calculated from the radius of the core $a$, wave number $k=2\pi /\lambda$ and $N.A.$ as follow:
\begin{equation}\label{eq:guidedmodes}
M = \frac{1}{2} \cdot {\left( {k \cdot a \cdot N.A.} \right)^2}
\end{equation}
Equation (\ref{eq:guidedmodes}) is the upper limit of the excited modes. The population of $M$ depends on the incident angle $\theta _{in}$ and the fibre-launch position as the offset on the axis. With the limit of small $N.A.$, it applies the relation of ${\theta _{in}} \approx \sin {\theta _{in}} \approx \tan {\theta _{in}} = {1 \mathord{\left/
 {\vphantom {1 {2{F_{in}}}}} \right.
 \kern-\nulldelimiterspace} {2{F_{in}}}}$, so that equation (\ref{eq:guidedmodes}) can be approximated to
\begin{equation}\label{eq:guidiedmodesf}
M = \frac{1}{8}{\left( {{{ka} \mathord{\left/
 {\vphantom {{ka} F}} \right.
 \kern-\nulldelimiterspace} F}} \right)^2}
\end{equation}
The excited mode number is about 200 for the input focal ratio of $F_{in}$=5.0 and it will be less for a slower beam. The decrease in excited modes will produce serious speckle patterns as shown in Fig.\ref{specklepattern}, which may introduce large offset into the determination of the energy barycentre and the spot size. Another important phenomenon is that the coherence in the output power distribution of the small core fibre makes some pixels so bright that the intensity reaches to the saturation range of the detector, while other black pixels are so dark. So a smooth and uniform distribution of the input light was required to acquire the stable FRD measurement, and the LED was used in the measurement of FRD and the throughput.

   \begin{figure}
   \centering
   \includegraphics[width=\hsize]{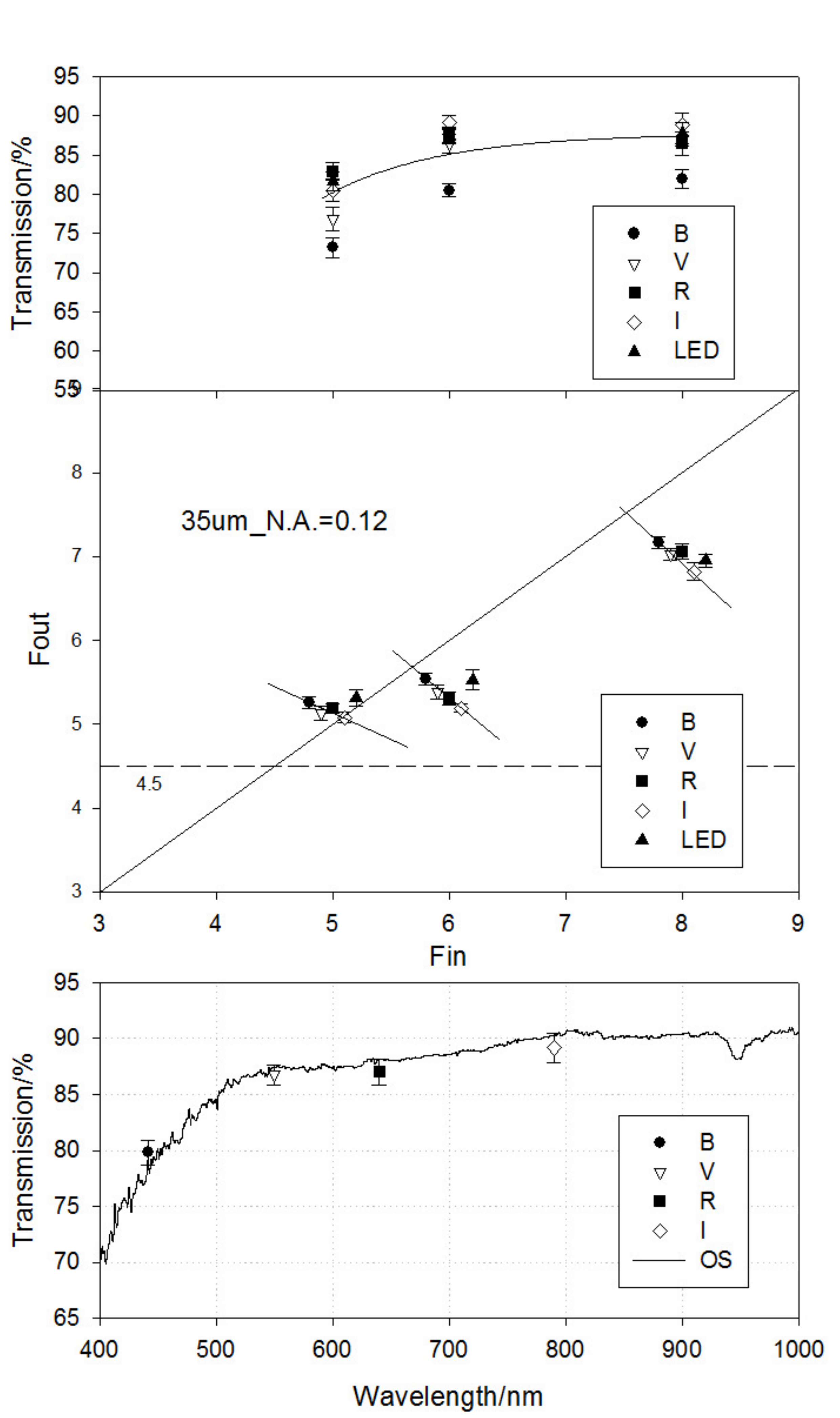}
      \caption{The FRD performance of a small core fibre of 35$\mu$m with $N.A.$=0.12. The small $N.A.$ can effectively lift the output focal ratio. Except for the output focal ratio of LED, the wavelength dependence can be seen when the input focal ratio is larger than 5.0 in other wavebands that the output focal ratio decreases with increasing wavelength which is the same trend as predicted by PDM. OS is the transmission efficiency acquired from the optical spectrometer. The throughput is larger than 85 per cent at the wavelength of larger than 500nm.}
         \label{fig:39}
   \end{figure}

The results of FRD measurements are shown in Fig.\ref{fig:39}. The output focal ratio maintains larger than 5.0 in the range of input focal ratio from 5.0 to 8.0. Notice that the input light is Gaussian-like which is different from the flat function in the previous experiments, so the power is more constrained in the centre area. The light in the fibre excites more power into the lower-order modes and that will make the output light a more constrained Gaussian spot. Then the diameter of the spot will be smaller within EE90 and a larger output focal ratio occurs compared with the results of the input light of a flat function.

A slight wavelength dependence of FRD is observed when the input focal ratio is larger than 5.0. The symbols are separated with 0.1 in x-direction for clarity. The negative slope of the trend line indicates that the output focal ratio decreases with increasing wavelength.

Changing the fibre into the small $N.A.$, the output focal ratio is slower than 4.5 in the range of input focal ratio from 5.0 to 8.0. The transmission of $F_{in}$=5.0$\sim$8.0 is larger than 75 per cent and the throughput is more than 85 per cent at the wavelength of larger than 500nm. The small $N.A.$ and the small core size also increase the sensitivity to the fibre-launch geometry including the incident angle and the axis-offset on the fibre end. In the future work, the stress effect and the micro and macro bending influence should be tested and the scrambling property is also under study.

\section{Discussion}
With the development of the technique in astronomy, more and more large-scale telescopes with large aperture, large field of view and high resolution are proposed. It costs plenty of time, even more than ten years, to build a large telescope until it is accomplished. During the time, several generations of prototypes will be assembled and tested, which needs the support of sufficient human and material resources. The calibration and measurement of the designed parameters is an essential work. For different scientific measurements, the experiment platforms are specifically designed to server the projects. A reliable platform is always of the first priority to be considered, of which a platform with high efficiency and multifunction is a better choice.

\subsection{Summary of DEEM}
The optimised platform of DEEM proposed in this paper is to meet the scientific needs of measuring the performance of highly multiplexed fibre systems in astronomy. We use the technique of controlling variables method to compare DEEM with the conventional method to testify its validation. In general, DEEM can support the required measurements with better operability compared with the conventional method of CCD-IM and it gives the same FRD measurement as CCD-IM within the errors which confirms the feasibility and accuracy. Compared with CCD-IM, DEEM mainly has the advantages on three aspects:

I. Closed loop system. First, a stable light source is crucial for FRD measurements and the throughput tests in the open loop system of conventional methods, because no feedback of the stability of the light source is acquired to compensate the variation in the output power. A closed loop system consists of two power detectors enhances the robustness of DEEM to a unstable light source. Second, when we carry out the throughput test, the variation in the output power not only comes from the light source, but also is affected by the noise including dark current, ambient light and residual light. If the light source and noise differ from each other in the twice measurements of input and output power, extra errors in throughput will be introduced into conventional method, but the closed loop of DEEM can suppress the influence. Finally, FRD and throughput can be measured simultaneously in DEEM, which can accelerate the progress of measurements.

\begin{figure}
   \centering
   \includegraphics[width=\hsize]{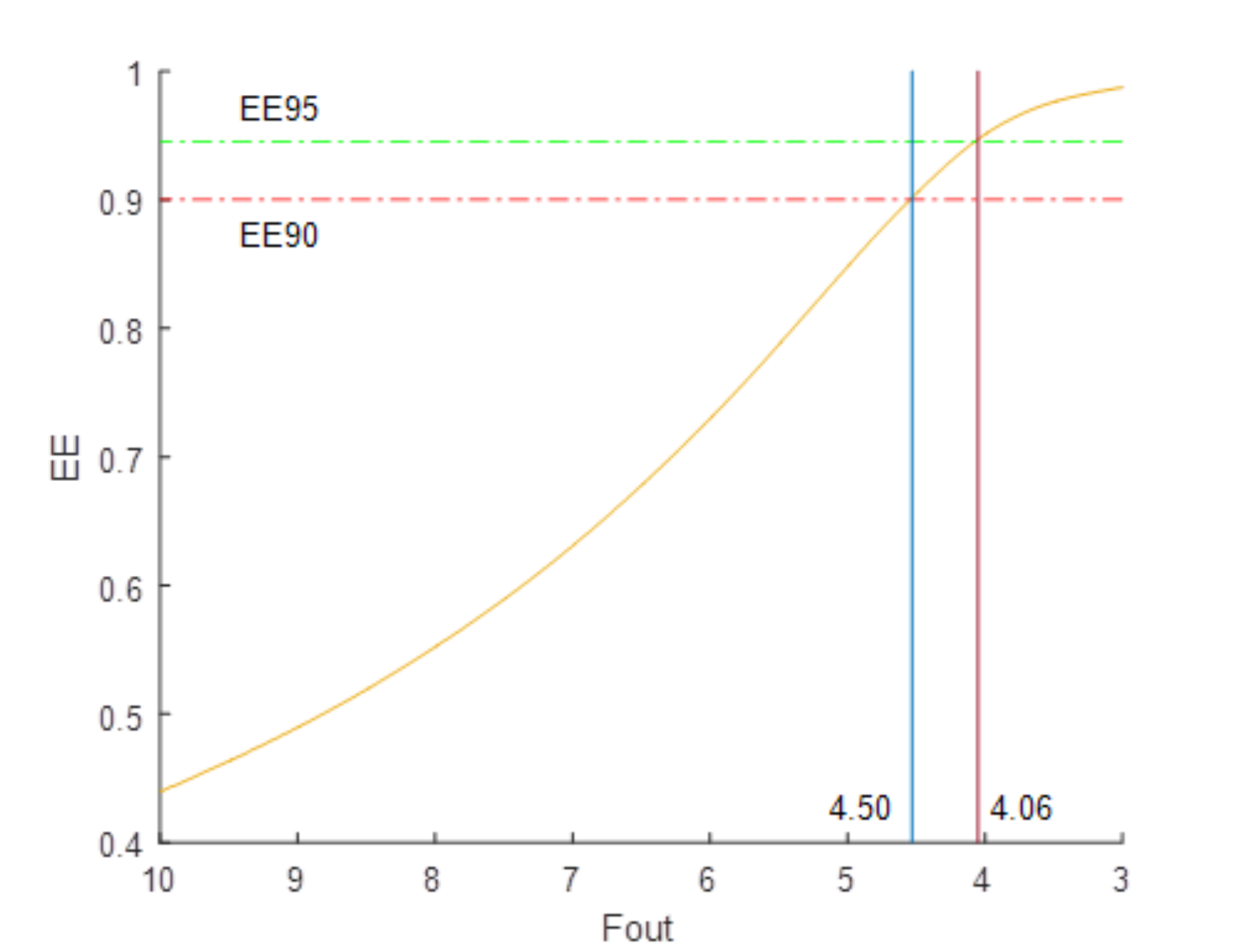}
      \caption{The results of EE ratio versus output focal ratio of 320$\mu$m core fibre. The easy operability of the two-arm design makes it possible to simultaneously record the reflective and the transmitted light, so one can easily acquire the relation between EE and $F_{out}$ by enlarging or shrinking the diaphragm in one derection.}
         \label{fig:eefrd}
   \end{figure}

II. Easy operability. DEEM acquires the spot size directly from the encircled energy to save time. It improves the accuracy by avoiding the potential errors in the intermediate steps such as the precision of the pixels, the approximation approach in the program code and so on. The reverse incidence method applied in DEEM ensures the alignment of input and output end. The monitor system in x-, y- and z- directions provides a visual way to control the input condition. The automation panel of DEEM supports automated regulation of travel stage and diaphragm to enable rapid measurements. Since the electric-driven diaphragm is enlarging or shrinking in one way, it is convenient to acquire the impact of FRD on the encircled energy curves as shown in Fig. \ref{fig:eefrd}.

III. Divided error sources. Different types of error sources based on the particular setup are analysed to help optimise the approach of DEEM. Some errors are suppressed in DEEM like the stability of light source and the uncertainties of ambient light. Considering the different setup of DEEM and CCD-IM, we analysed the error sources in the background of three types of noise, which explains the contributions of different error sources. And at the same time, the analysis of the error sources can help to optimise the optical design and the data reduction of DEEM and CCD-IM.

Although DEEM has good robustness in a complex environment, the precision still connects with the residual light of $( {1 - EE} ) \cdot \eta $. And this also relates to the precision of the power detector. In our system, the value of the power intensity can only accurate to two places decimal. The number in the third place is always changing with time because there are always some tiny fluctuations on the light source and the ambient light. So to better perform the experiment, a well-controlled darkroom will be perfect.

In the meantime, with the setup of DEEM in our experiments, the image of output spot is not captured simultaneously with the energy encircling process and one should switch the CCD and power meter to acquire the 2-D images. This would be a disadvantage of DEEM that needs to improve. A possible way to optimise the design of DEEM is to split the output light from lens L4 to record the output spot with another CCD camera to achieve the real-time imaging and energy encircling. Nevertheless, this kind of design will no doubt complicate the construction of optics, enlarge the capability of the experiment platform and bring more error sources from the alignment issue.

\subsection{FRD dependence}
It is well known that common sources of FRD include the poor fibre end termination (stress and roughness), alignment (offset of the input position and angle, repositioning in the input end) and the bending effect (macro and micro bending). And the simulation of PDM predicts a wavelength and length dependence that FRD increases with longer wavelength and fibre length and the core size dependence that FRD increases with smaller core size.

It is shown by \citet{Craig1988Measurement} that the difference in FRD between a properly cleaved fibre and a polished fibre is negligible for the various fibres they tested. While a poor polished fibre has more serious FRD as shown in \citep{Poppett2010A,Allington-Smith2013End,Haynes2008Focal}, which indicates that a faster constant polish produces less stress on the fibre end. The alignment is a common issue that exists in the telescope like LAMOST during the multi-object survey with discrete fibres. The offset of the input position on the fibre end will produce a scatter in arrival positions of the light rays over the entire output end face, of which many rays are confined to the outer parts of the fibre causing a near-field distribution that is edge-brightened \citep{Allington-Smith2012Simulation,Avila1998Results}. The incident angle is well controlled and calibrated within 1.0$^\circ$ for LAMOST, while the positioning precision is 100$\mu$m which is caused by several aspects including the accuracy of the fibre probe on the focal plane, the seeing and so forth. Bending of fibre is inevitable and many results \citep{Allington-Smith2012Simulation,Pazder2014The,Murphy2013The} show that the macro-bending (radius>5.0cm) has little effect on FRD but a bending radius less than 5.0cm can affect both throughput and far-field illumination. For length property, \citet{Poppett2010The} found that the FRD was not sensitive to the fibre length and no significant increase occurred in FRD for a longer fibre. \citet{Ramsey1988Focal} showed the decrease in FRD as core size increases.

The trends of FRD dependence mentioned above are basically consistent with different researchers. While the wavelength dependence of FRD differs from different scientific groups and it is somewhat controversial. The prediction of PDM by \citet{Gloge1972Derivation} has been observed by some researchers, the relative difference in FRD reached to 13.9 per cent in \citet{Carrasco1994A} and 20.0 per cent in \citet{Poppett2007Fibre}, but some other groups have tested for this effect but found no marked changes in FRD and the difference was less than 2 per cent \citep{Crause2008Investigation,Pazder2014The,Matthew2004SparsePak}.

\begin{figure*}
   \centering
   \includegraphics[width=\hsize]{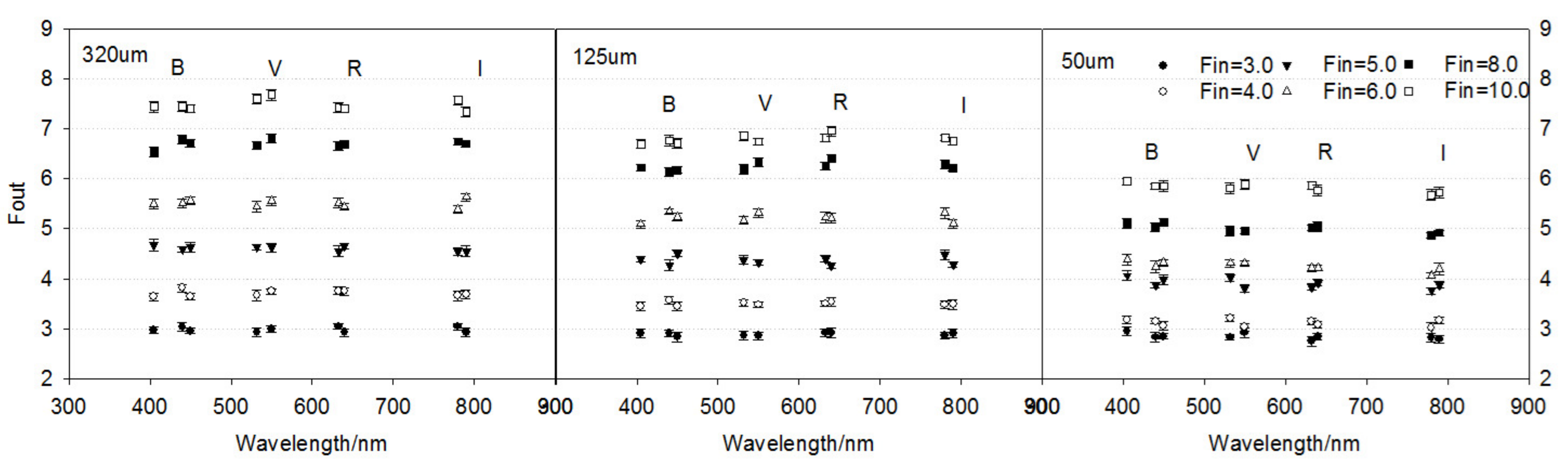}
      \caption{The FRD performance at different wavelengths. No evident wavelength dependence can be seen in the fibres of 320$\mu$m and 125$\mu$m core sizes. For 50$\mu$m core fibre, a slight trend occurs but it is very weak when the results mix in the output focal ratios of the broadband light. The FRD dependence on core size can be seen in horizontal direction that the output focal ratio decreases with decreasing fibre core size.}
         \label{fig:wavelengthdependence}
   \end{figure*}

In our experiments, the wavelength dependence in Fig.\ref{fig:16}, \ref{fig:17}, \ref{fig:21} and \ref{fig:22} is implicit between the wavelength of 532nm and 632.8nm for the fibre with the core size of 320$\mu$m and 125$\mu$m. But a weak dependence on wavelength is revealed in the results of 35$\mu$m core fibre. In order to confirm the wavelength dependence, a wider range of wavelength should be tested. Then we tested the FRD performance at other wavelengths from 400nm to 800nm. The results are shown in Fig.\ref{fig:wavelengthdependence}.

\begin{figure*}
   \centering
   \includegraphics[width=\hsize]{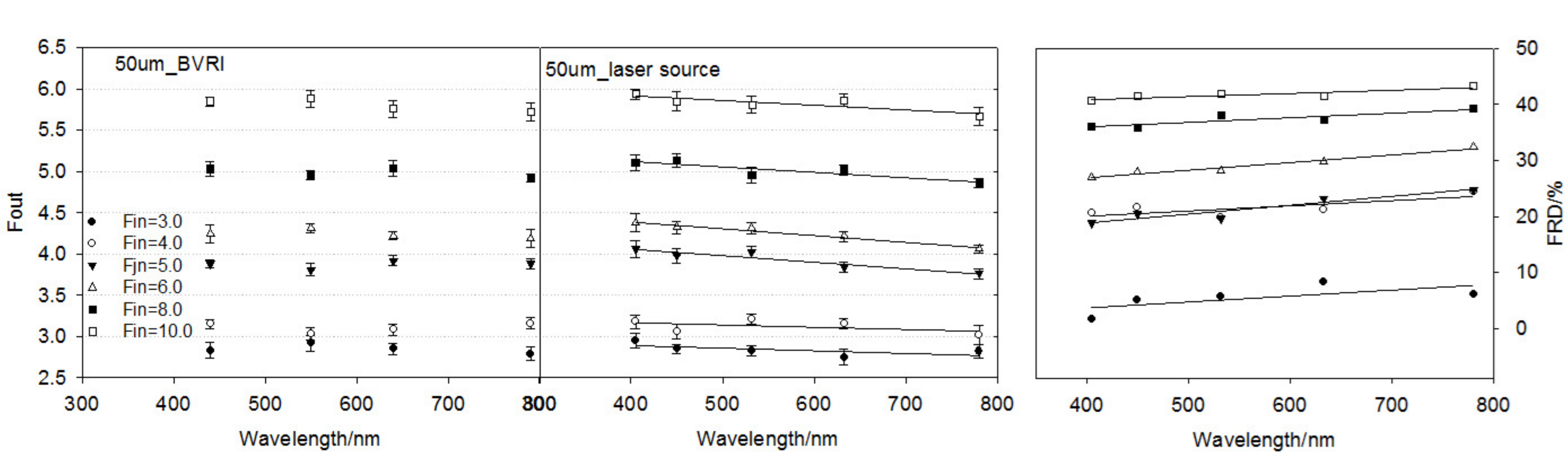}
      \caption{The FRD performance of 50$\mu$m core fibre in laser sources and broadband light sources. The wavelength dependence is distinct when the input focal ratio is larger than 5.0 in laser sources. But it disappears in the condition of broadband light sources. When the input focal ratio is small (3.0 or 4.0), the FRD is unapparent compared to slower beams where the output beam approaches the asymptotic value. So the wavelength dependence is also very weak in a fast input focal ratio.}
         \label{fig:50umfibre}
   \end{figure*}

\begin{figure*}
   \centering
   \includegraphics[width=\hsize]{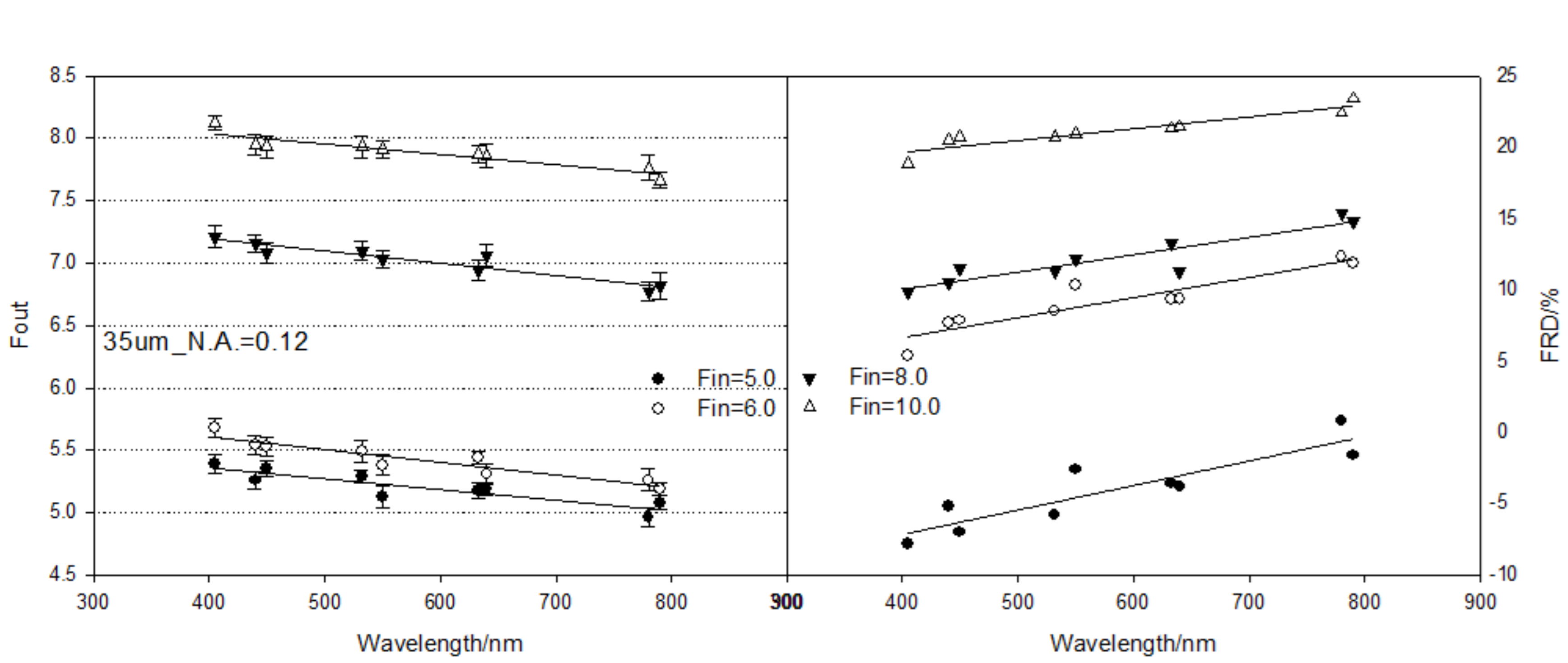}
      \caption{The FRD performance of a small core fibre of 35$\mu$m with $N.A.$=0.12. It differs from the situation in 50$\mu$m core fibre that the wavelength dependence can be seen in both laser sources and broadband sources when the input focal ratio is larger than 5.0. The relative difference between the maximum and the minimum value of output focal ratio is around 0.5 in F-ratio.}
         \label{fig:35umfibre}
   \end{figure*}

The wavelength dependence for fibres with 320$\mu$m and 125$\mu$m core is negligible that the difference of the output focal ratio is consistent within the size of error bar. While for fibre with smaller core of 50$\mu$m, the results show a slight trend to higher output focal ratio for a given input focal ratio at the shorter wavelength and it is true for almost all values when the input focal ratio is larger than 5.0, especially when illuminated by laser sources as shown in Fig.\ref{fig:50umfibre}. But the wavelength dependence disappears in the results of the broadband filters. The trend of increasing FRD with increasing wavelength is also true for 35$\mu$m core fibre with $N.A.$=0.12 as shown in Fig.\ref{fig:35umfibre}. The relative difference of FRD for input focal ratio of 5.0$\sim$10.0 is 2.7 per cent$\sim$6.0 per cent for 50$\mu$m core fibre and 3.6 per cent$\sim$8.6 per cent for 35$\mu$m core fibre from the wavelength of 405nm to 780nm. The wavelength dependence becomes weak in both fibres of 35$\mu$m and 50$\mu$m core sizes with broadband filters, which indicates that the coherence of light can enhance the FRD dependence on wavelength.

\begin{figure}
   \centering
   \includegraphics[width=\hsize]{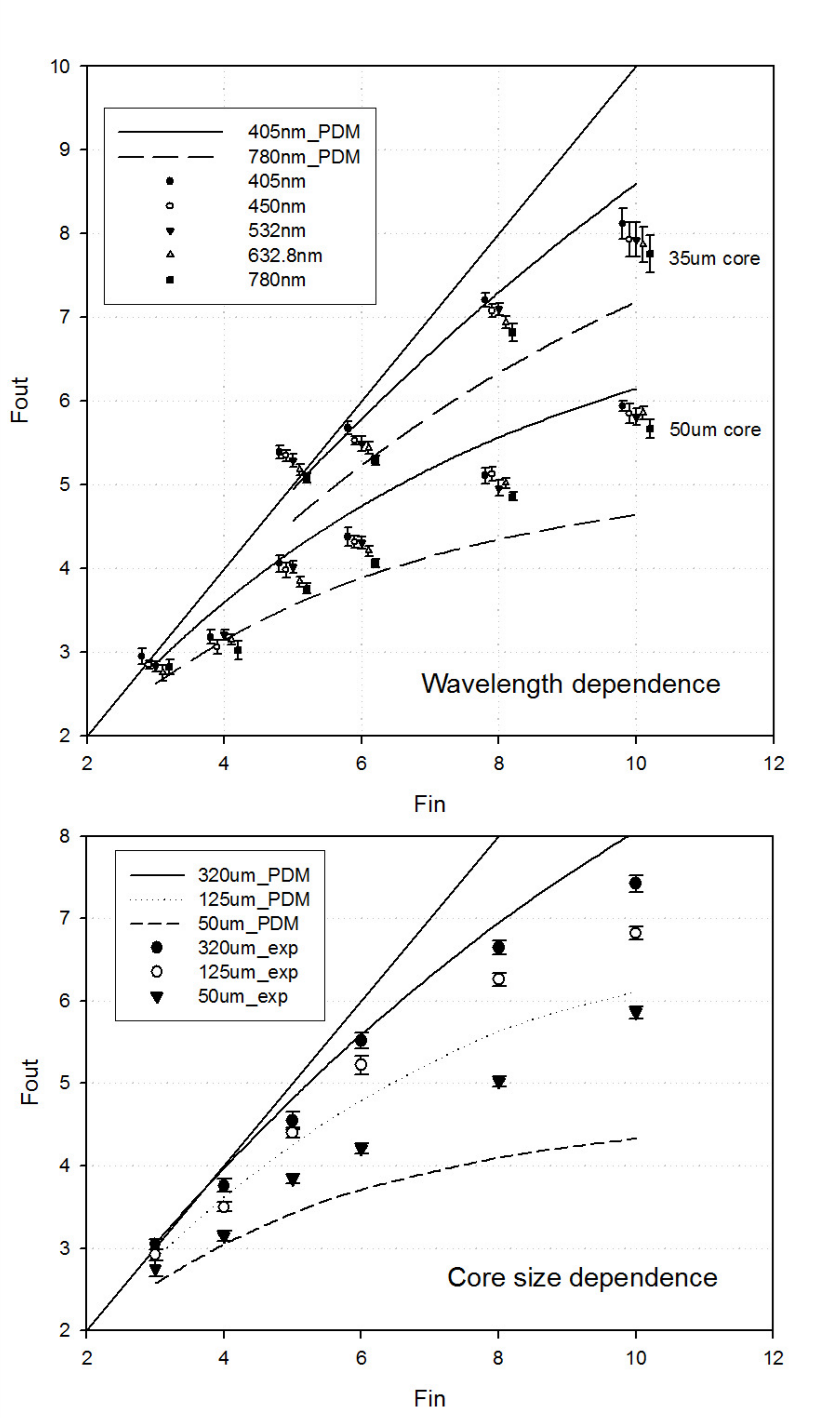}
      \caption{The wavelength prediction of PDM of 50$\mu$m and 35$\mu$m core fibre. The predicted values bracket the measured results. Comparing the dependence of the two types of fibres, the smaller core fibre (35$\mu$m) with smaller $N.A.$=0.12 has a stronger wavelength dependence. The core size dependence predicted by PDM is also stronger than the observed data. The difference between the prediction and measured value in the condition of input focal ratio $F_{in}$=3.0, 4.0 is smaller than others because the FRD in small input focal ratio is relatively unimportant compared to slower beams.}
         \label{fig:predictionofwavelength}
   \end{figure}

According to the predictions of PDM, the trend of small core fibre is broadly consistent within the wavelength from 400nm$\sim$800nm. The predictions in Fig.\ref{fig:predictionofwavelength} show that the model brackets the range of the observed data, but the FRD dependence on wavelength is ought to be stronger.

The core size dependence can be distinctly seen in Fig.\ref{fig:wavelengthdependence} that the FRD is more serious in a small core fibre. The $N.A.$ has a major contribution to the FRD. The output focal ratio of the 35$\mu$m core fibre with a small $N.A.$=0.12 is larger than that of the 50$\mu$m core fibre with $N.A.$=0.22. In this view, choosing a fibre with small $N.A.$ to improve the FRD performance can efficiently lift up the asymptote of the output focal ratio. On the other hand, the new design of the IFU with smaller core fibres should also take the wavelength dependence into consideration that FRD of the small core fibre is more sensitive to the light source of narrow band. Large FRD at the longer wavelength will decrease the throughput within the same aperture as that at the shorter wavelength.

\subsection{Subtraction of the residual light}
The residual light in CCD-IM is treated as a uniform function that it spreads radially symmetric across the whole image which is broadly in agreement with the observed data as shown in Fig.\ref{fig:residuallight} and it gives a stable FRD measurement. The difference between DEEM and CCD-IM is also suppressed within the maximum error of E$N.A.$.

Another assumption is that the residual light comes from the cladding modes and the scattering effect of the fibre end, which can also explain the variable values of the residual light in different input light. To testify the hypothesis, we need to collect more output power within an aperture larger than the limit of $N.A.$ to encircle the cladding modes and the scattering light. The total energy of output light is measured by the lens L3 and the power meter PM3 in DEEM, and the corresponding $N.A.$ of lens L3 is $N.A.$=0.32 which is much larger than the limit of the fibre of $N.A.$=0.22. The total energy in CCD-IM is measured by the integral grey value of the image of the output spot within an aperture of $N.A.$=0.22 in the previous tests. The up limits of $N.A.$ of the CCD camera is determined by the imaging position and the window size, and the range of $N.A.$ is controlled between 0.14 and 0.31 in the imaging positions from 130.0mm to 60.0mm. For example, the imaging position of CCD is placed in 100.0mm, and the maximum of the window size is 36.864mm(4096*0.009mm), then the up limit of $N.A.$ is 0.184, in which the total energy is underestimated. So the CCD camera should be placed closer to the fibre end, say 83.7mm, to cover the area of $N.A.$=0.22 at least. Therefore, in the new tests, we set the position of CCD in the range from 30.0mm to 60.0mm to measure the diameter of the output spot within the same aperture of $N.A.$=0.32 as in DEEM. Then the input light power is changed randomly to investigate the influence on the determination of the spot size of 50$\mu$m core fibre.

The results of the new tests with light source of LED show that for high input focal ratios ($F_{in}$>8.0) with high and low input light intensities, the difference between DEEM and CCD-IM is suppressed to be less than 0.62 in F-ratio and it is smaller than the difference in the previous experiments of 0.9. This infers that a larger aperture can efficiently reduce the difference between the two methods. Comparing the residual light in the annulus of $N.A.$=0.22 and $N.A.$=0.32, we notice that the average grey value remains stable when the spot is captured in the positions of 30.0mm to 40.0mm, in which the image of the output spot is very small. But the grey value increases from the annulus of $N.A.$=0.22 to $N.A.$=0.32 when the output spot is imaged in the distance of 50.0mm and 60.0mm, where the spot is larger, especially when the input focal ratio is smaller than 6.0. In this case, the residual light is no longer uniformly distributed. Then we use a wide Gaussian function to match the wings between the circles of $N.A.$=0.22 and $N.A.$=0.32 as the subtraction of the residual light. Using Gaussian function to describe the non-uniform residual light is due to the area effect that the same amount of light is scattered into the central zone of a smaller area than the outside region, making the intensity density higher in the centre area, and not due to the asymmetry in the scattering process. Then the maximum difference between DEEM and CCD-IM is significantly suppressed to 0.25 in F-ratio in the whole tested range of input focal ratio from 4.0 to 10.0. According to the maximum uncertainties in Fig. \ref{fig:ena}, the maximum error from E$N.A.$ between DEEM and CCD-IM is 0.41 in F-ratio when the output focal ratio is smaller than $F_{out}$=7.0, which is true for the whole input range being tested, and the difference of 0.25 agrees to each other within the maximum error of E$N.A.$.

But we also find out that the $f$-intercept of the fitting curve becomes large and the linearity of the fitting curve is relatively low of $r^2$<0.980 in the condition of laser sources. It is due to the short operation distance of the CCD camera that a slight offset of 0.09mm (10 pixels) in the diameter can cause an effective change of E$N.A.$=0.0018, which makes the stability of the two methods more sensitive to the alignment and the laser speckle.

Both of the uniform function and the Gaussian function can be used to model the residual light. They can efficiently reduce the difference of FRD measurements after the subtraction of the residual light. Comparison of the two kinds of tests shows that the FRD measurement of DEEM is much closer to the assumption of cladding modes and scattering light, but the stability needs to improve. If the total power of a larger aperture of $N.A.$ is needed to encircle the cladding modes and scattering light, the operation distance of CCD should be limited in a short range close to the fibre end, which will reduce the robustness and enlarge the uncertainties from the misalignment. While the model of uniform function for residual light subtraction is also an available method to be applied in the experiments and it can be done in a longer operation distance for CCD. Nevertheless, a more accurate and complete model of residual light is of great importance and required to improve the precision.

\subsection{Application}
The FASOT system requires a critical FRD performance less than 10 per cent in FRD for input focal ratio of $F_{in}$=5.0. Theoretically, the output focal ratio of 4.5 is located within the acceptable range determined by $N.A.$ ($F_{critical}$=2.3 for $N.A.$=0.22). But technically, it is difficult to guarantee other influence like stress or micro bending on the fibre which worsens the FRD performance. So in the new design, we changed the fibre with smaller $N.A.$ of 0.12 to increase the critical output focal ratio. The first generation of prototype IFU gives out lots of valuable information to help optimise the new design. The transmission efficiency is acceptable and homogeneous. In the next generation, the staggered arrangement combined with honeycomb-shaped microlens array will further increase the filling factor and the coupling efficiency.

For highly multiplexed fibre systems, small core fibres can increase the number density within a compacted structure to improve the resolving power. However, new technique or more accurate polishing process is required to improve the smoothness of the fibre end face. And the distribution of stress on fibre array should be considered. The structure of the adapter junction box, which holds the fibre cable, needs to modify to supply more free space for fibres to avoid collision and micro bending, and to release the stress. Moreover, the position precision is another challenge since the core size is much smaller.

As stated above, multiplexing in astronomical systems is increasing at a rapid rate. More and more fibres are tested during the research and the development, but also at many stages during manufacture to verify the performance. Reliable and repeatable measurements are essential to provide the performance for reference. For a single fibre, the conventional method takes about 10 minutes for the conventional FRD measurement including the positioning process, image recording, EE ratio estimation, spot size calculation and the focal ratio determination. The diameter within a certain EE ratio is acquired directly from the electric-driven diaphragm in the DEEM system, so it suppresses the total measuring time down to about half of that to improve the measurement efficiency.

\section{Conclusions}
A reliable multifunction system of DEEM has been proposed and tested. According to the practical requirement in astronomy and combining the PDM model, DEEM can unify the relation between energy usage and the spot size. So DEEM directly use the encircled energy as the indicator to measure the output focal ratio without recording the images. Many other properties, like PSF, throughput and so forth, can also be measured with DEEM. The new method will enable rapid measurement in the system of massively integral fibres system. In the data processing, the subtraction of background is essential for DEEM and CCD-IM to provide accurate FRD measurement. Three types of noise in the two systems have different contributions to the error calculations, especially for the residual light in CCD-IM, which needs to be modified to be a more accurate model. The FRD measurements of fibres of 50$\mu$m core with $N.A.$=0.22 and 35$\mu$m core with $N.A.$=0.12 show that the stress and the defects inner the quartz plate worsen the FRD performance and lower $N.A.$ can lift the asymptotic focal ratio. The output focal ratio remains larger than 5.0 in the range of input focal ratio from 5.0 to 8.0. At the same time, the FRD dependence on wavelength and core size are revealed since it is more sensitive for the small core fibres, and the simulation of PDM indicates that the dependence should be stronger than the observed data.

\section*{Acknowledgements}

We would are deeply grateful to our colleagues who have provided us with valuable guidance. Special thanks to Dr. Julia Bryant (CAASTRO) and Prof. Matthia J\"ager (Institut f\"ur Photonische Technologien) for productive discussions.




\bibliographystyle{mnras}
\bibliography{DEEM} 



\bsp	
\label{lastpage}
\end{document}